\documentstyle[aps,epsfig,floats]{revtex}

\newcommand{\simle}
{\raisebox{-0.75ex}[-1.5ex]{$\;\stackrel{<}{\sim}\;$}}
\newcommand{\simge}
{\raisebox{-0.75ex}[-1.5ex]{$\;\stackrel{>}{\sim}\;$}}
\newcommand{\dalt}
{\raisebox{-0.35ex}[-1.5ex]{$\;\stackrel{\leftrightarrow}{\partial}\!\!\!\;$}}
\def\d{\partial}
\def\s{{\sigma}}
\def\e{{\epsilon}}  
\def\k{{ { k} }}
\def\p{{ { p} }}
\def\q{{ { q} }}

\def\r{{\rho}}
\def\g{{\gamma}}
\def\w{{\omega}}
\def\a{{\alpha}}
\def\b{{\beta}}

\def\l{{\lambda}}
\def\L{{\Lambda}}
\def\G{{\Gamma}}
\def\i{{ {\rm i} }}

\begin{document}
\draft

\def\runtitle{
General Formula for the Magnetoresistance
on the Basis of the Fermi Liquid Theory
}
\def\runauthor
 {Hiroshi {\sc Kontani}}

\title{
General Formula for the Magnetoresistance \\
on the Basis of the Fermi Liquid Theory
}

\author{
Hiroshi {\sc Kontani}$^{1,2}$
}

\address{
$^1$Theoretische Physik III, Elektronische Korrelationen und Magnetismus, 
Universit\"at Augsburg, D-86135 Augsburg, Germany.
\\
$^2$Department of Physics, Saitama University,
255 Shimo-Okubo, Urawa-city, 338-8570, Japan.
}

\date{\today}

\maketitle      

\begin{abstract}
The general expression for the magnetoresistance (MR) due to the 
Lorentz force is derived by using the Fermi liquid transport theory 
based on the Kubo formula.
The obtained gauge-invariant expression is exact for any strength of
the interaction, as for the most singular term with respect to
$1/\gamma_\k^\ast$ 
($\gamma_\k^\ast$ being the quasiparticle damping rate).
By virtue of the exactness,
the conserving laws are satisfied rigorously in the present expression,
which is indispensable for avoiding unphysical solutions.
Based on the derived expression,
we can calculate the MR within the framework of
the Baym-Kadanoff type conserving approximation,
by including all the vertex corrections required by the Ward identity.
The present expression is significant
especially for strongly correlated systems 
because the current vertex corrections will be much important.
On the other hand,
if we drop all the vertex corrections in the formula,
we get the MR of the relaxation time approximation (RTA),
which is commonly used because of the simplicity.
However, the RTA is dangerous because it may give unphysical results
owing to the lack of conserving laws.
In conclusion, the present work 
enables us to study the MR with satisfying the conserving laws
which is highly demanded in strongly correlated electrons, 
such as high-$T_{\rm c}$ superconductors, organic metals,
and heavy Fermion systems.
In Appendix D, we reply to the comment by O. Narikiyo [cond-mat/0006028].
(Note that Appendix D exists only in the e-preprint version.)

\end{abstract}



\vspace{6mm}


\section{Introduction}
The transport phenomena under the magnetic field
are very important and interesting field in condensed matter physics.
Especially, the magnetoresistance (MR) 
is of current interest:
It has been attracting great attentions of many researchers
because it gives us rich information on the 
electronic structure of the system.
There are several possible mechanisms of the MR, for example,
(i) the Lorentz force on conduction electrons (orbital effect), 
(ii) the double-exchange mechanism (spin effect), and 
(iii) the Anderson localization mechanism in dirty metals,
or the Kondo effect caused by the magnetic impurities.
It is well-known that the huge negative MR due to (ii) is 
realized in Mn-oxides at the ferromagnetic transition temperature,
i.e., the colossal magnetoresistance.
The negative MR due to (iii) is observed 
in dirty semiconductors.

On the other hand, in usual pure paramagnetic metals, 
the {\it positive} MR due to (i) 
is the most dominant, which is called the orbital MR.
There is a long-history on the study of the orbital MR
based on the Boltzmann transport theory
 \cite{Ziman}.
In many theoretical studies, 
the collision term in the Boltzmann equation
is simply approximated by 
introducing the phenomenological relaxation time ($\tau_{\bf k}$), 
which is called the relaxation time approximation (RTA).
There,
the correlation effect between particle-hole pair is totally dropped.
Although RTA is a quasiclassical approximation,
various important and useful results have been obtained.
For example, the MR as a function of the angle between 
${\vec B}$ and ${\vec I}$ in various metals was reproduced 
within the RTA
 \cite{Ziman}.

Especially,
the Kohler's rule is one of the most essential relations
derived by the RTA: 
According to this rule,
the conductivity $\s_{\mu\nu}$ in the magnetic field ${\vec B}$
is represented in the functional form
$\s_{\mu\nu} = \tau \!\cdot\! F_{\mu\nu}(\tau B)$,
if the anisotropy of $\tau_{\bf k}$ is not so large.
According to the Onsager's reciprocity theorem,
$F_{\mu\nu}(x)$ is an even-function of $x$
as for the diagonal conductivity ($\mu=\nu$),
and it is an odd-function 
as for the off-diagonal conductivity ($\mu=x,\nu=y$).

By definition,
the resistance without the magnetic filed ($\rho_0$),
the Hall coefficient ($R_{\rm H}$), and
the magnetoresistance (${\mit\Delta}\rho/\rho_0$), are given by
\begin{eqnarray}
\rho_0 &=& 1/\s_0 \nonumber \\
R_{\rm H} &=& ({\mit\Delta}\s_{xy}/B)/\s_0^2 ,
 \label{eqn:coeff-def} \\
{\mit\Delta}\rho/\rho_0 &=& -{\mit\Delta}\s_{xx}/\s_0 
 - ({\mit\Delta}\s_{xy}/\s_0)^2 ,
 \nonumber 
\end{eqnarray}
where $\s_0$ is the diagonal conductivity without ${\vec B}$.
${\mit\Delta}\s_{xy}$ and ${\mit\Delta}\s_{xx}$ are components
proportional to $B$ and $B^2$, respectively.
We call ${\mit\Delta}\s_{xx}$ the magnetoconductivity (MC).
As a result, the Kohler's rule predicts the relations
$\rho_0 \propto \tau$, 
$R_{\rm H} \propto \tau^0$, and 
${\mit\Delta}\rho/\rho_0 \propto \tau^2 \!\cdot\! B^2$.
These relations are well satisfied in various ordinary metals,
which may be interpreted as nearly free electron systems.

Interestingly, however, 
the Kohler's rule is strongly violated
in several strongly correlated metals 
according to recent experimental studies.
For example, $R_{\rm H}$ in the normal state 
in high-$T_{\rm c}$ cuprates 
is approximately proportional to $T^{-1}$, and
${\mit\Delta}\rho/\rho_0 \propto \rho_0^{-2}\cdot T^{-2}$.
These {\it seemingly} non-Fermi liquid behaviors
of the transport phenomena
in high-$T_{\rm c}$ cuprates have been studied as a central issue
both theoretically and experimentally
because they should reflect the fundamental electronic property
of high-$T_{\rm c}$ cuprates.

Up to now,
several authors have studied the transport phenomena
in high-$T_{\rm c}$ cuprates
based on the RTA, and claimed that 
the strong anisotropy of $\tau_\k$ can explain the observed 
violation of the Kohler's rule.
 \cite{Pines}.
However, its effect is too insufficient 
to reproduce the prominent non-Fermi liquid behaviors
observed for wider range of temperatures.
On the other hand, 
we recently studied the $R_{\rm H}$ in high-$T_{\rm c}$ cuprates 
based on the general formula for ${\mit\Delta}\s_{xy}$
derived by Ref.\cite{Kohno},
which is formally exact of order $\tau^2$.
Then, we find out that the vertex corrections for the current,
which are dropped in the RTA,
cause the Curie-Weiss like behavior of $R_{\rm H}$
in the presence of AF fluctuations
 \cite{Kontani,Kanki,Nagoya,BEDT}.

The aim of this paper is to derive the
{\it general expression for the MC} from the Kubo formula.
The obtained expression for ${\mit\Delta}\s_{xx}$,
which is formally exact in order $\tau^3$, contains
all the vertex corrections which ensure the conserving laws.
In the system with strong correlations,
the role of the vertex corrections will qualitatively change
the behavior of the MR, 
as it does for the Hall effect.
In this respect, the RTA is unreliable
because all the current vertex corrections are neglected there.
Based on the derived expression,
we will study the MR in high-$T_{\rm c}$ cuprates 
and discuss the violation of Kohler's rule in later publications
 \cite{Future}.

In the absence of the magnetic field,
Eliashberg derived in 1961 the expression for 
the conductivity of the Fermi liquid system
from the Kubo formula 
 \cite{Eliashberg}:
\begin{eqnarray}
\s_{xx}
 = \frac{e^2}{v_B} \int_{\rm FS} \frac{dS_k}{|{\vec v}_{\bf k}|}
 v_{{\bf k} x} J_{{\bf k} x} \cdot 2\tau_{\bf k} ,
 \label{eqn:Eliashberg}
\end{eqnarray}
which is exact up to $O(\tau)$.
$\int_{\rm FS} dS_k$ means the 
two-dimensional integration on the Fermi surfaces, and
$v_B=(2\pi)^3$ in the cubic lattice.
$\tau_{\bf k} \equiv 1/( -2{\rm Im}\Sigma_{\bf k}(+\i 0) ) >0$,
where $\Sigma_{\bf k}(\w)$ is the self-energy.
$v_{{\bf k} x} \equiv (\d/\d k_x)(\e_{\bf k}^0 + {\rm Re}\Sigma_{\bf k}(0))$
is the quasiparticle velocity without the renormalization factor $z_\k$.
Moreover,
$J_{{\bf k} x}$ has the four-point vertex correction ${\cal T}_{22}$,
which is not given by the self-energy through the Ward identity.
($J_{{\bf k}x}$ is given by eq.(\ref{eqn:J2}) in this paper.)
This non-trivial current vertex correction from ${\cal T}_{22}$
is indispensable to avoid unphysical results.
For instance, Yamada and Yosida proved that
$\s_{xx}$ given by eq.(\ref{eqn:Eliashberg}) diverges
unless any Umklapp scattering processes exist,
reflecting the momentum conservation law
 \cite{Yamada}.
This physically reasonable result cannot be reproduced
if we drop the vertex correction from ${\cal T}_{22}$.

The exact expression for the Hall conductivity of order $\tau^2$
was given by Kohno-Yamada as follows:
 \cite{Kohno}
\begin{eqnarray}
{\mit\Delta}\s_{xy}
 = -B \cdot \frac{e^3}{2v_B} \int_{\rm FS} \frac{dS_\k}{|{\vec v}_{\bf k}|}
 v_{{\bf k} x} \left[ J_{{\bf k} x} \frac{\d J_{{\bf k} y}}{\d k_y}
  - J_{{\bf k} y} \frac{\d J_{{\bf k} x}}{\d k_y} \right]
 (2\tau_{\bf k})^2  ,
 \label{eqn:Kohno}
\end{eqnarray}
where the uniform magnetic field is along the $z$-axis.
Here the carrier of an electron is $-e$ ($e>0$).
Note that 
if we replace $J_{{\bf k} \mu}$ with $v_{{\bf k} \mu}$
in eqs. (\ref{eqn:Eliashberg}) and (\ref{eqn:Kohno}),
we get the results of the RTA.

As for ${\mit\Delta}\s_{xy}$,
the vertex corrections in $J_{{\bf k} x}$ also play important roles
in strongly correlated systems.
In fact, according to recent theoretical studies,
the vertex corrections in $J_{{\bf k} x}$ 
give rise to the strong enhancement of $R_{\rm H}$
in nearly AF Fermi liquids, like 
in high-$T_{\rm c}$ superconductors or in $\kappa$-(BEDT-TTF)
organic superconductors
 \cite{Kontani,Kanki,Nagoya,BEDT}.
Noteworthily,
$R_{\rm H}$ becomes negative in electron-doped compounds
irrespective of the fact that 
its Fermi surface is hole-like everywhere
 \cite{Sato}.
This long-standing mystery,
which cannot be explained within the RTA,
is naturally reproduced in Refs.
\cite{Kontani} and \cite{Kanki}
due to the fact that 
${\vec J}_{\bf k}$ is no more parallel to ${\vec v}_{\bf k}$.

The purpose of this article is to derive the conserving expression
for the magnetoconductivity ${\mit\Delta}\s_{xx}$
which is exact in order $\tau^3$.
We notice that the magnetoconductivity
within the RTA, ${\mit\Delta}\s_{xx}^{\rm RTA}$,
is given by
\begin{eqnarray}
{\mit\Delta}\s_{xx}^{\rm RTA}
 = -B^2 \cdot \frac{e^4}{4v_B} \int_{\rm FS} \frac{dS_\k}{|{\vec v}_{\bf k}|}
 \left\{ ({\vec v}_{\bf k} \times {\hat e}_z)\cdot {\vec \nabla}
 \left( 2\tau_{\bf k} v_{{\bf k} x}\right) \right\}^2 (2\tau_{\bf k})  ,
 \label{eqn:Boltzmann-Dsxx}
\end{eqnarray}
where ${\hat e}_z$ is the unit vector along the magnetic field
 \cite{Ziman}.
Note that the expression (\ref{eqn:Boltzmann-Dsxx}) 
has the contribution from the $\k$-derivative of $\tau_{\bf k}$,
while $\s_{xx}$ and ${\mit\Delta}\s_{xy}$ given by
eqs. (\ref{eqn:Eliashberg}) and (\ref{eqn:Kohno}) do not have it.
${\mit\Delta}\s_{xx}^{\rm RTA}$ given by eq. (\ref{eqn:Boltzmann-Dsxx}) 
will be unreliable in correlated electron systems
because conserving laws are violated in the sense of Baym and Kadanoff
 \cite{Baym}.
Whereas, the derived formula for ${\mit\Delta}\s_{xx}$ in this paper,
eq.(\ref{eqn:MC-final}) or eq.(\ref{eqn:MC-total-lowT}) in \S V,
is {\it conserving} and {\it exact} up to $O(\tau^3$).
In this sense, our formula for ${\mit\Delta}\s_{xx}$ 
is indispensable for the study of 
the strongly correlated Fermi liquid.
Note that our formula coincides with eq. (\ref{eqn:Boltzmann-Dsxx})
if all the vertex corrections arising from ${\cal T}_{22}$
are neglected.

The contents of this paper are the following:
In \S II,
we study the tight-binding model with Coulomb interaction,
where the vector potential is included as the Peierls phase 
in the hopping integrals.
In \S III,
the expression for the magnetoconductivity (MC) is obtained
without vertex corrections, for an instructive purpose.
It coincides with eq.(\ref{eqn:Boltzmann-Dsxx})
at lower temperatures.
In \S IV and V,
we derive the general expression for the MC 
by taking all the vertex corrections into account,
which is the main part of this work.
In the former section,
we derive all the vertexes corrections by taking account of 
the Ward identities for multi-point vertices seriously.
In the later section,
we perform the analytic continuation
and derive the {\it exact} expression for the MC of order $\tau^{3}$.
It is given by eq.(\ref{eqn:MC-final}) or eq.(\ref{eqn:MC-total-lowT}).
Based on the obtained formula, 
we present several discussions on the vertex corrections,
and give some remarks in \S VI.

\section{Kubo Formula for the Magnetoconductivity}

\subsection{Tight Binding Model in a Magnetic Field}

The tight-binding model without the magnetic field
is given by
\begin{eqnarray}
 H_{B=0}&=& H_{B=0}^0 + H_{\rm int} \nonumber \\
 H_{B=0}^0
 &=& \sum_{\langle i,j\rangle, \s} t_{i,j}^0 c_{i\s}^\dagger c_{j\s}
   \nonumber \\
 &=& \sum_{\k\s}\e_\k^0 c_{\k\s}^\dagger c_{\k\s} ,
   \label{eqn:Hamiltonian}
\end{eqnarray}
where $t_{i,j}^0$ is the hopping integral and
$\e_\k^0$ is the spectrum of the non-interacting particles,
which is given by the Fourier transformation of $t_{i,j}^0$.
$c_{i\s}^\dagger$ ($c_{\k\s}^\dagger$) is the creation
operator of the electron at ${\vec r}_i$ (with momentum ${\vec k}$)
with the spin $\s$. 
Hereafter, we may write the vectors ${\vec r}_i$ and ${\vec k}$ as 
$r$ and $k$ for simplicity.
$H_{\rm int}$ represents the two-body interaction term.
In this manuscript, we assume the density-density interactions,
like the on-site Coulomb interaction.
Note that we exclude the processes which would 
change the form of the current operator, for example,
the pair-hopping processes 
$Jc_{i\uparrow}^\dagger c_{i\downarrow}^\dagger
c_{j\downarrow} c_{j\uparrow}$.

Now, we introduce the magnetic field through the vector potential
${\vec A}_i$ on $i$-site.
We do not consider the Zeeman term in this article
because we focus only on the orbital effect, i.e., the Lorentz force.
If the magnetic field is almost uniform, 
${\vec A}_i$ is included in the Hamiltonian through the
hopping integral as the Peierls phase factor
 \cite{Fradkin}:
\begin{eqnarray}
 t_{i,j}&\equiv& t_{i,j}^0 \cdot
  {\rm exp}(\ -\i e({\vec A}_i+{\vec A}_j) \!\cdot\! 
   ({\vec r}_i-{\vec r}_j)/2 \ ) , 
  \label{eqn:hopping} \\
 t_{j,i} &=& \{t_{i,j}\}^\ast.  
  \nonumber
\end{eqnarray}
Here and hereafter, $-e$ ($e>0$) is the charge of an electron,
and we put $c=\hbar=1$.

From now on, 
we introduce the external vector potential as
\begin{eqnarray}
 {\vec A}_i^{\rm tot}= {\vec A}e^{\i {\vec q}\cdot{\vec r}_i}
  + {\vec A}'e^{\i {\vec q}^{\, \prime}\cdot{\vec r}_i} ,
\end{eqnarray}
and extract the coefficient of 
$B \!\cdot\! B'= (\i{\vec q}\times{\vec A}) \!\cdot\!
 (\i{\vec q}^{\, \prime}\times{\vec A'})$
of the conductivity. 
In the final stage, we put $A=A'$ and take the uniform limit
$\q=\q'=0$.
This procedure was originally developed in studying
the Hall conductivity in dirty metals by Fukuyama et.al.
 \cite{Fukuyama}.
In this article, we take the $z$-axis to be the direction of 
the magnetic field.

Now we expand the Hamiltonian with respect to $A$ and $A'$.
For the purpose of this paper,
we need the terms proportional to $A$, $A'$ and $AA'$.
They are given by
\begin{eqnarray}
 H_{B}= H_{B=0} 
  +\ eA_\a j_\a(-\q) + eA_\b' \cdot j_\b(-\q')
  +\ e^2A_\a A_\b' \cdot j_{\a\b}(-\q-\q')
  \ \ + \cdots  ,
  \label{eqn:Ham-B}
\end{eqnarray}
which is exact up to $O(\q,q')$ and $O(\q\q')$.
Here and hereafter, we promise that 
the summation with respect to $\a$ or $\b$
is taken implicitly.
In eq.(\ref{eqn:Ham-B}),
$j_\a(\p)$, $j_{\a\b}(\p)$ are defined as
\begin{eqnarray}
j_\a(\p) &=& \sum_\k v_\a^0(\k) c_{\k-\p/2}^\dagger c_{\k+\p/2}  ,
  \\
j_{\a\b}(\p) &=& \sum_\k v_{\a\b}^0(\k) c_{\k-\p/2}^\dagger c_{\k+\p/2}  ,
\end{eqnarray}
where $v_\a^0(\k) =\d_\a \e_\k^0$ and 
$v_{\a\b}^0(\k) =\d_{\a\b} \e_\k^0$, respectively.
(Hereafter, $\d_\a\equiv \d/\d k_\a$ and 
$\d_{\a\b}\equiv \d^2/\d k_\a \d k_\b$, etc.) 
The equation (\ref{eqn:Ham-B}) is derived in Appendix A.

Next, we examine the current operator in the magnetic field.
It is given by
${\vec j}^B = {\vec j} + e {\vec A}/m$
for a electron in vacuum as is well known.
However, it is no more valid for a tight-binding model.
By using the kinetic equation for an electron at ${\vec r}$
in the interacting representation,
the current operator of the system, ${\vec j}^B(r)$, is given by
\begin{eqnarray}
{\vec j}^B({\vec r})= \i [H_B^0, {\vec r}\r({\vec r})]  ,
 \label{eqn:jB-definition}
\end{eqnarray}
where $\rho({\vec r})$ is the density operator of the system
 \cite{Mahan}.
Based on this definition,
${\vec j}^B(\p)$ in the weak magnetic field
is derived in Appendix A.
According to the result,
\begin{eqnarray}
 j_\nu^B(\p\!=\!0) 
 &=&  j_\nu(0) + eA_\a \cdot j_{\nu\a}(-\q) 
             + eA'_\b \cdot j_{\nu\a}(-\q') 
   + e^2A_\a A'_\b \cdot j_{\nu\a\b}(-\q-\q') 
  \ + \ \cdots  , 
  \label{eqn:curr-B} \\
 j_{\nu\a\b}(\p) &=& 
 \sum_\k v_{\nu\a\b}^0(\k) c_{\k-\p/2}^\dagger c_{\k+\p/2}  ,
  \nonumber
\end{eqnarray}
where $v_{\nu\a\b}^0(\k)\equiv \d_{\nu\a\b}\e_\k^0$.
Equation (\ref{eqn:curr-B}) is exact up to $O(\q,\q')$ and $O(\q\q')$.
In eqs.(\ref{eqn:Ham-B}) and (\ref{eqn:curr-B}), 
we dropped the terms proportional to $A^2$, $A'^2$ 
because they are not required for the purpose of this paper.
In the same way, we find 
\begin{eqnarray}
 j_\mu^B(\q+\q') = j_\mu(\q+\q') + eA_\a \cdot j_{\mu\a}(\q')
 + eA'_\b \cdot j_{\mu\b}(\q) + e^2A_\a A'_\b \cdot j_{\mu\a\b}(0) 
 \ + \ \cdots  ,
 \label{eqn:JBqq}
\end{eqnarray}
up to $O(A,A')$ and $O(AA')$.

\subsection{Kubo Formula for the Conductivity}

According to Kubo formula
 \cite{Kubo},
the conductivity in the magnetic filed is given by
\begin{eqnarray}
 \s_{\mu\nu}(\q+\q',\w) 
  = \frac1{\i\w} \left[ \Phi_{\mu\nu}(\q+\q',\w+\i0)
  - \Phi_{\mu\nu}(\q+\q',\i0) \right]  ,
  \label{eqn:Kubo2}
\end{eqnarray}
\begin{eqnarray}
\Phi_{\mu\nu}(\q+\q',\w_\lambda)
 = \int_0^\b d\tau e^{-\w_\lambda\tau}
  \langle T_{\tau} j_\mu^B(\q+\q',0) j_\nu^B({0},\tau)
  \rangle_{B}  ,
  \label{eqn:Kubo}
\end{eqnarray}
where $\b=1/T$ and $\w_\lambda = \i\pi T \cdot 2\lambda$ 
($\lambda$ being the integer), respectively.
Here $T_{\tau}$ is a $\tau$ ordering operator.
$\Phi_{\mu\nu}(\q+\q',\w+\i0)$ in eq.(\ref{eqn:Kubo})
is given by the analytic continuation of 
$\Phi_{\mu\nu}(\q+\q',\w_\lambda)$ in eq.(\ref{eqn:Kubo2})
with respect to Im\,$\w_\l > 0$

The diagonal conductivity $\s_{\mu\mu}$
does not contain the term proportional to $B^{2n+1}$
because of the Onsager's reciprocity theorem.
The diagonal conductivity given by eq.(\ref{eqn:Kubo})
can be expanded with respect to $q,q'$ and $A,A'$ as 
%
\begin{eqnarray}
\s_{\mu\mu}(\q+\q',0) = \s_{\mu\mu}^0
 + 2C_{\mu\mu}^{\a\rho;\b\rho'} \cdot 
  (\i q_\rho A_\a \cdot \i q'_{\rho'} A'_\b)
 \ + \ \cdots  .
 \label{eqn:sxx-expand}
\end{eqnarray}
The coefficient $C_{\mu\mu}^{\a\rho;\b\rho'}$
should satisfy the relation 
$C_{\mu\mu}^{\a\rho;\b\rho'} = C_{\mu\mu}^{xy;xy}
 \cdot \e_{z\rho\a}\e_{z\rho'\b}$
because of the requirement from the gauge-invariance.
We call ${\mit\Delta}\s_{\mu\mu} 
\equiv C_{\mu\mu}^{\a\rho;\b\rho'} \cdot 
 (\i q_\rho A_\a \cdot \i q'_{\rho'} A'_\b) = C_{\mu\mu}^{xy;xy} \cdot B^2$ 
the magnetoconductivity (MC).
The aim of this paper is to derive the expression for 
$C_{\mu\mu}^{\a\rho;\b\rho'}$ by using the Fermi liquid theory.

In order to derive 
$C_{\mu\mu}^{\a\rho;\b\rho'}$ in eq.(\ref{eqn:sxx-expand}),
we study the term proportional to $e^2 A_\a A_\b'$
in eq.(\ref{eqn:Kubo}),
$\Phi_{\mu\nu}^{(2)}(\q+\q',\w_\lambda)$:
It is given by 
\begin{eqnarray}
 \Phi_{\mu\nu}^{(2)}(\q+\q',\w_\lambda)
 &=& \int_0^\b d\tau e^{-\w_\lambda\tau}
 \Bigl\{ \ \langle T_\tau j_\mu(\q+\q',0)
    j_{\nu\a\b}(-\q-\q',\tau) \rangle 
                 \nonumber \\
 & &\ \
  + \int_0^\b d\tau' \langle T_\tau j_\mu(\q+\q',0) 
                 j_{\nu\a}(-\q,\tau) j_\b(-\q',\tau') \rangle 
  \nonumber \\
 & &\ \
  + \int_0^\b d\tau'
   \langle(\a,\q)\!\leftrightarrow\!(\b,\q')\rangle \nonumber \\
 & &\ \
  + \int_0^\b d\tau'
  \langle T_\tau j_\mu(\q+\q',0) 
              j_{\nu}({0},\tau) j_{\a\b}(-\q-\q',\tau') \rangle 
              \nonumber \\
 & &\ \
  + \int_0^\b d\tau' d\tau''
  \langle T_\tau j_\mu(\q+\q',0) j_{\nu}({0},\tau) 
    j_{\a}(-\q,\tau') j_{\b}(-\q,\tau'') \rangle 
    \ \Bigr\} \ + \ \cdots   .
  \label{eqn:Phi-ab}
\end{eqnarray}
In later sections, we take the derivative of
$\Phi_{\mu\nu}^{(2)}(\q+\q',\w_\lambda)$ with respect to
$q_\r$ and $q'_{\r'}$, and perform the analytic continuation
with respect to $\w_\lambda$.
In the next stage, we derive the general formula for the MC,
$C_{\mu\nu}^{\a\rho;\b\rho'} 
 = \frac12 (\d^2/\d q_\rho \d q_{\rho'})
 {\rm Im}\, \Phi_{\mu\nu}^{(2)}(\q+\q',\w+\i 0)/\w |_{\w,\q,\q'\rightarrow0}$.
Hereafter, we consider $\Phi_{\mu\nu}^{(2)}(2\q+2\q',\w_\lambda)$
to simplify the calculations, and divide the final result by four.

In eq.(\ref{eqn:Phi-ab}),
the terms with vector potential(s) in $j_\mu^B(\q+\q',0)$,
given by eq.(\ref{eqn:JBqq}), were dropped.
Actually, it is easy to see that they 
make $\Phi_{\mu\nu}(\q+\q',\w)$ independent of $\q$ or $\q'$,
which lead to $C_{\mu\mu}^{\a\rho;\b\rho'}=0$

\subsection{Vertex Corrections}

The definition of the one particle Green function is
\begin{eqnarray}
G(\k,\e_n)= -\frac1\beta \int^\beta_0 d\tau d\tau'
 \e^{\e_n(\tau-\tau')} \langle T_\tau c_{\s\k}(\tau) 
 c_{\s\k}^\dagger(\tau') \rangle   ,
\end{eqnarray} 
where $\e_n = \i\pi T(2n+1)$ is the Matsubara frequency for fermion.
By introducing the self-energy $\Sigma(\k,\e_n)$,
the Green function for the interacting system is expressed as
\begin{eqnarray}
G(\k,\e_n)= (\e_n +\mu - \e_\k^0 -\Sigma(\k,\e_n))^{-1}  ,
\end{eqnarray} 
where $\mu$ is the chemical potential.

\begin{figure}
\begin{center}
\epsfig{file=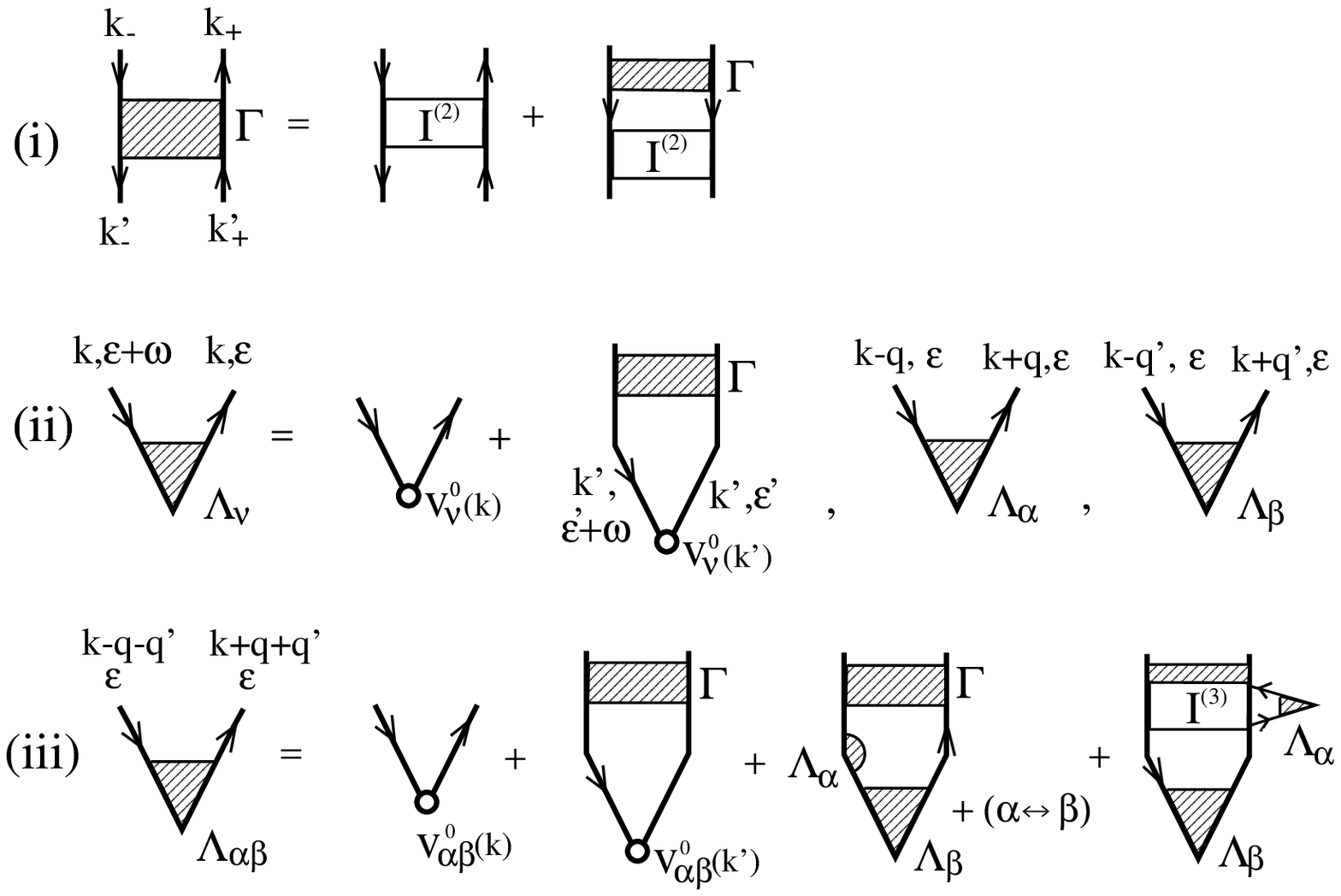,width=8cm}
\end{center}
\caption{(i) The Bethe-Salpeter integral equation for 
the four point vertex $\Gamma(\k_-;\k_+|\k'_+;\k'_-)$.
$\Gamma^I$ is the irreducible vertex.
(ii) Integral equations for $\L_\nu$, $\L_\a$ and $\L_\b$.
(iii) The integral equation for $\L_{\a\b}$.}
\label{fig:DG-init1-exp}
\end{figure}

As shown in Fig.\ref{fig:DG-init1-exp} (i),
the full four-point vertex $\Gamma$ is given by
the following integral equation:
\begin{eqnarray}
& &\Gamma(\k_-\e^+;\k_+\e|\k_+'\e';\k_-'\e'^+)=
 \Gamma^{I}(\k_-\e^+;\k_+\e|\k_+'\e';\k_-'\e'^+)
  \nonumber \\
& &\ \ \ \ \ \ \ \ \ \ \ 
 + T\sum_{\k''\e''} \Gamma^{I}(\k_-\e^+;\k_+\e|\k_+''\e'';\k_-''\e''^+)
 G(\k_-'',\e''^+)G(\k_+'',\e'')
 \Gamma(\k_-''\e''^+;\k_+''\e''|\k_+'\e';\k_-'\e'^+)  ,
  \label{eqn:full-vertex}
\end{eqnarray} 
where we put $k_{\pm}=k\pm q$ and $\e_{n}^+= \e_n+\w_\l$.
$\Gamma^{I}$ is the 'irreducible' four point vertex
with respect to the particle-hole pair.
Here and hereafter, we do not write the spin suffix explicitly,
and the suffix of the Matsubara frequencies $n$ and $\l$
are sometimes dropped for simplicity.

\begin{figure}
\begin{center}
\epsfig{file=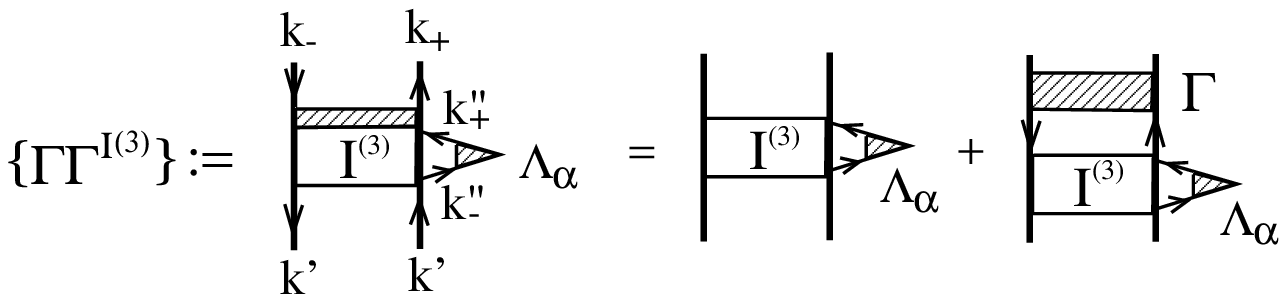,width=7cm}
\end{center}
\caption{The definition of 
$\{\Gamma\Gamma^{I(3)}\}(\k_-;\k_{+}|\k''_{+};\k''_{-}|\k';\k')$.}
\label{fig:DG-init2-add}
\end{figure}

The current vertex function $\Lambda_\nu$ is given by
\begin{eqnarray}
\Lambda_\nu(\k\e^+;\k\e)
 = v_\nu^0(\k)
 + T\sum_{\k'\e'} \Gamma(\k\e^+;\k\e|\k'\e';\k'\e'^+)
 G(\k',\e'^+)G(\k',\e') v_\nu^0(\k')   ,
\end{eqnarray} 
as is shown in Fig.\ref{fig:DG-init1-exp} (ii).
In the same way,
\begin{eqnarray}
\Lambda_\a(\k_-\e;\k_+\e)
 = v_\a^0(\k)
 + T\sum_{\k'\e'} \Gamma(\k_-\e;\k_+\e|\k_+'\e';\k_-'\e')
 G(\k_+',\e')G(\k_-',\e') v_\mu^0(\k')  .
\end{eqnarray} 
We also introduce the vertex $\Lambda_{\a\b}$
which is related to the bare vertex $v_{\a\b}^0(\k) = \d_{\a\b}\e_\k^0$
as follows:
\begin{eqnarray}
& &\Lambda_{\a\b}(\k_{--}\e;\k_{++}\e)
 = v_{\a\b}^0(\k)
 + T\sum_{\k'\e'} \Gamma(\k_{--};\k_{++}|\k'_{++};\k'_{--})
  \cdot G(\k'_{--})G(\k'_{++}) 
   \nonumber \\
& &\ \ \ \ \ \ \ 
  \times \left( \ v_{\a\b}^0(\k') 
  + \Lambda_\a(\k'_{--},\k'_{+-})
 G(\k'_{+-})\Lambda_\b(\k'_{+-};\k'_{++}) 
 + \Lambda_\b(\k'_{--};\k'_{-+})
 G(\k'_{-+})\Lambda_\a(\k'_{-+};\k'_{++}) \ \right)
   \nonumber \\
& &\ \ \ \ 
 + T^2\sum_{\k'\k''\e'\e''}
 \{\Gamma\Gamma^{I(3)}\}(\k_{--};\k_{++}|\k'_{+};\k'_{-}|\k''_{0+};\k''_{0-})
  \cdot G(\k'_{-})\Lambda_\a(\k'_{-};\k'_{+}) G(\k'_{+})
 \cdot G(\k''_{0-})\Lambda_\b(\k''_{0-};\k''_{0+}) G(\k''_{0+})  ,
  \label{eqn:Lab}
\end{eqnarray} 
where $\k_{\pm\pm}\equiv \k\pm\q\pm\q'$, $\k_{\pm\mp}\equiv \k\pm\q\mp\q'$
and $\k_{0\pm}\equiv \k\pm\q'$, respectively.
(Here frequency variables were dropped 
because it would not cause a confusion in that case.)
The eq.(\ref{eqn:Lab}) is expressed diagrammatically in 
Fig.\ref{fig:DG-init1-exp} (iii).
Here, $\Gamma^{I(3)}$ is the irreducible six-point vertex,
and the vertex $\{\Gamma \Gamma^{I(3)}\}$ is defined as
\begin{eqnarray}
& &\{\Gamma\Gamma^{I(3)}\}(\k_{--};\k_{++}|\k'_{+};\k'_{-}|\k''_{0+};\k''_{0-})
   \nonumber \\
& &\ \ \ \ \ \ \ \ \ \ \ 
= T\sum_{\p\e_m} (\ \delta_{\k,\p}/T
 + \Gamma(\k_{--};\k_{++}|\p_{++};\p_{--})G(\p_{--})G(\p_{++}) \ )
\Gamma^{I(3)}(\p_{--};\p_{++}|\k'_{+};\k'_{-}|\k''_{0+};\k''_{0-})  ,
 \label{eqn:GGI3}
\end{eqnarray} 
which is also depicted in
Fig. \ref{fig:DG-init2-add}.

In the case of $q=0$,
Ward identity assures the following relations
 \cite{Nozieres,AGD}:
\begin{eqnarray}
\Lambda_\a(\k\e_n;\k\e_n)
 &=& v_\a^0(\k) + \d_\a \Sigma(\k,\e_n)  ,  
   \label{eqn:Ward-La} \\
\Lambda_{\a\b}(\k\e_n;\k\e_n)
 &=& v_{\a\b}^0(\k) + \d_{\a\b} \Sigma(\k,\e_n)  .
   \label{eqn:Ward-Lab} 
\end{eqnarray} 

\begin{figure}
\begin{center}
\epsfig{file=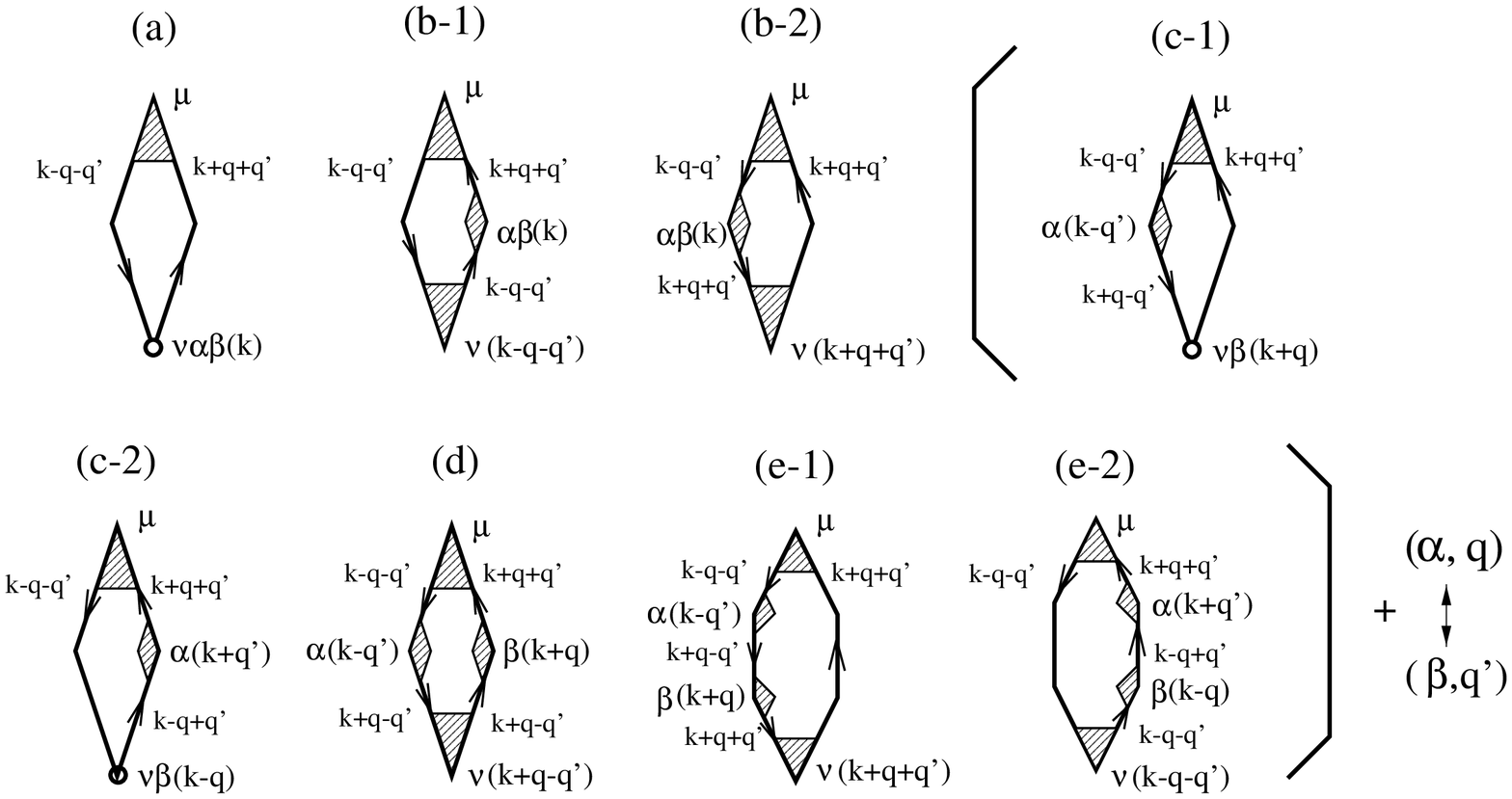,width=10cm}
\end{center}
\caption{The expressions for $\Phi_{\mu\nu}^{(2)}(2\q+2\q',\w_\l)$,
which are composed of ${\vec \L}$, $G$ and ${\vec v}^0$.}
\label{fig:DG-init1}
\end{figure}

\begin{figure}
\begin{center}
\epsfig{file=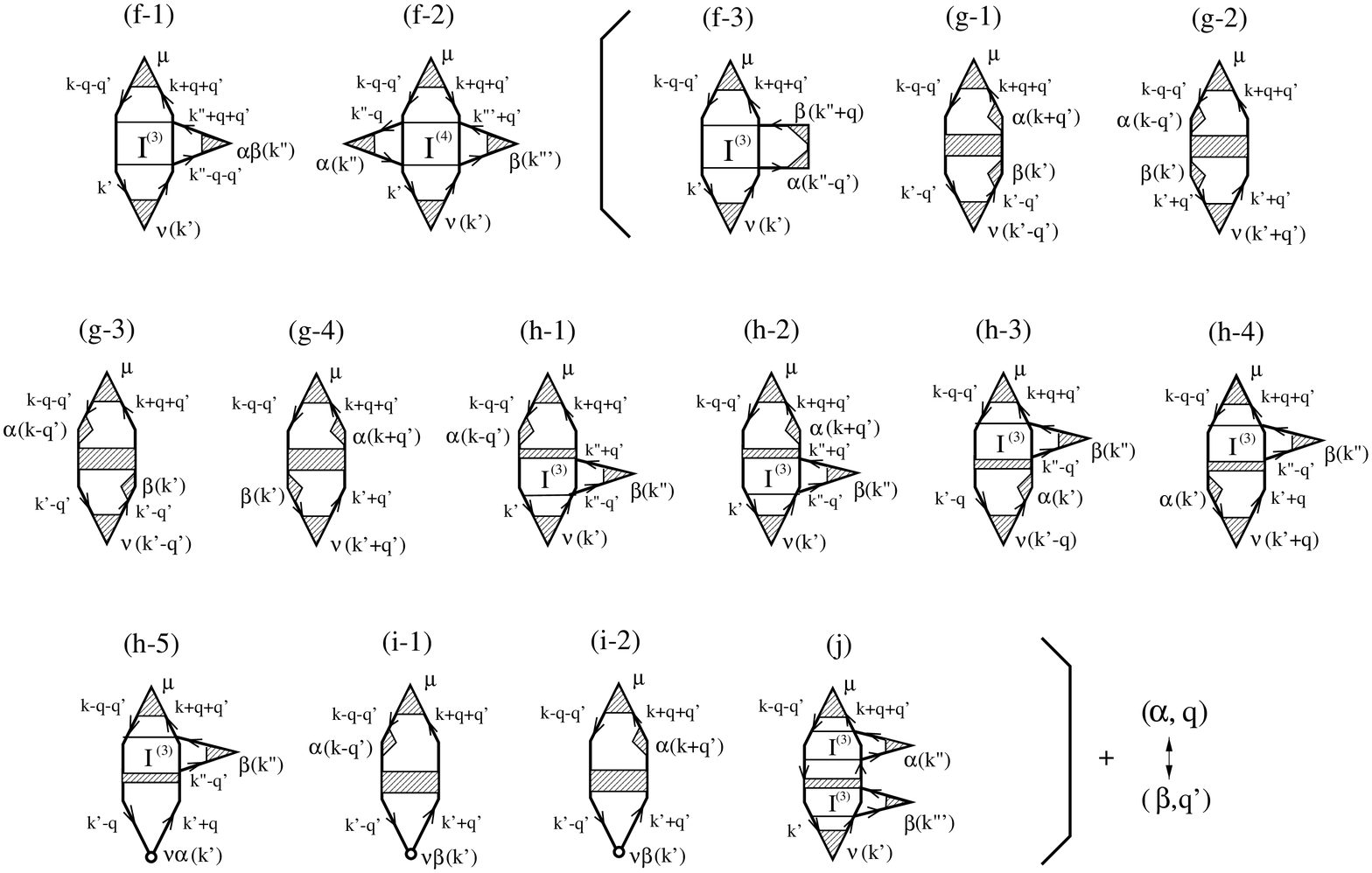,width=13cm}
\end{center}
\caption{The expressions for $\Phi_{\mu\nu}^{(2)}(2\q+2\q',\w_\l)$,
which contains at least one four (or six, eight) point vertex.}
\label{fig:DG-init2}
\end{figure}

Now we analyze $\Phi_{\mu\nu}^{(2)}(2\q+2\q',\w_\l)$
in terms of the diagrammatic technique.
They are expressed in Figs. \ref{fig:DG-init1} and \ref{fig:DG-init2}.
Below, we show analytical expressions for some of them:
\begin{eqnarray}
{\mbox{(a)}} &=& 
 -T\sum_{\k,\e} \Lambda_\mu(\k_{++};\k_{--})
 G^+(\k_{--}) v_{\nu\a\b}^0(\k) G(\k_{++})  , \\
{\mbox{(b-1)}} &=&
 -T\sum_{\k,\e} \Lambda_\mu(\k_{++};\k_{--})
 G^+(\k_{--}) \Lambda_{\nu}(\k_{--};\k_{--}) 
 G(\k_{--}) \Lambda_{\a\b}(\k_{--};\k_{++})G(\k_{++})  , \\
{\mbox{(d)}} &=&
  -T\sum_{\k,\e} \Lambda_\mu(\k_{++};\k_{--})
 G^+(\k_{--}) \Lambda_\a^+(\k_{--};\k_{+-})
 G^+(\k_{+-}) \nonumber \\
& &\times \Lambda_{\nu}(\k_{+-};\k_{+-})
 G(\k_{+-})  \Lambda_\b(\k_{+-};\k_{++})
 G(\k_{++})  ,
\end{eqnarray} 
where $v_{\nu\a\b}^0 \equiv \d_{\nu\a\b}\e_\k^0$, and
we write $G^+(\k) \equiv G(\k,\e^+)$ and
$\Lambda_\a^+(\k_{-};\k_{+}) \equiv \Lambda_\a(\k_{-}\e^+;\k_{+}\e^+)$, 
respectively.
We also give expressions for some terms in Fig. \ref{fig:DG-init2},
which include four, six and eight point vertices:
\begin{eqnarray}
{\mbox{(f-1)}} &=&  -T^3\sum_{\k\k'\k'',\e\e'\e''} 
 \Gamma^{I(3)}(\k_{--}\e^+;\k_{++}\e|
 \k'\e';\k'\e'^+|\k''_{++}\e'',\k''_{--}\e'')  
  \nonumber \\
& &\times G(\k_{++})
 \Lambda_\mu(\k_{++};\k_{--}) G^+(\k_{--}) 
  \cdot G^+(\k') \Lambda_\nu(\k';\k') G(\k') 
  \nonumber \\
& &\times G(\k''_{--})
 \Lambda_{\a\b}(\k''_{--};\k''_{++}) G(\k''_{++})  ,
  \\
{\mbox{(g-1)}} &=& -T^2\sum_{\k\k',\e\e'} 
 \Gamma(\k_{--}\e^+;\k_{++}\e|\k'_{0+}\e';\k'_{0-}\e'^+)
  \nonumber \\
& &\times G(\k_{-+}) \Lambda_\a(\k_{-+};\k_{++})
 G(\k_{++}) \Lambda_\mu(\k_{++};\k_{--}) G^+(\k_{--}) 
  \nonumber \\
& &\times G^+(\k'_{0-}) \Lambda_\nu(\k'_{0-};\k'_{0-}) 
 G(\k'_{0-}) \Lambda_\b(\k'_{0-};\k'_{0+}) G(\k'_{0+})  ,
  \\
{\mbox{(h-1)}} &=& -T^3\sum_{\k\k'\k'',\e\e'\e''}
 \{\Gamma \Gamma^{I(3)}\}(\k_{+-}\e^+;\k_{++}\e|
 k'\e';\k'\e'^+|\k''_{0+}\e'';\k''_{0-}\e'')
  \nonumber \\
& &\times G(\k_{++}) \Lambda_\mu(\k_{++};\k_{--}) 
 G^+(\k_{--})\Lambda_\a^+(\k_{--};\k_{+-}) G^+(\k_{+-}) 
  \nonumber \\
& &\times G^+(\k') 
 \Lambda_\nu(\k';\k') G(\k') 
  \cdot G(\k''_{0-})
 \Lambda_{\b}(\k''_{0-};\k''_{0+}) G(\k''_{0+})  ,
\end{eqnarray} 
where $\{\Gamma \Gamma^{I(3)}\}$ is defined in eq.(\ref{eqn:GGI3}).

\section{Magnetoconductivity: without Vertex Corrections}

In this section, 
we derive the MC by neglecting all the current vertex corrections
in the first place.
We will see that this approximation gives the consistent
result with the relaxation time approximation (RTA).
In later sections, we derive the exact formula for the MC
of order $\g^{-3}$, by taking all the current vertex corrections
into account.

\subsection{analytic continuation}
In deriving the conductivity, 
we have to perform the analytic continuation of 
$\Phi_{\mu\nu}^{(2)}(2\q+2\q',\w_\l)$.
For an instructive purpose, 
we briefly review the conductivity 
$\s_{\mu\nu}$ without the magnetic field.
If we neglect all the vertex corrections for the current,
the current-current correlation without the magnetic filed,
$\Phi_{\mu\nu}^{(0)}(\w)$, is simply given by 
\begin{eqnarray}
\Phi_{\mu\nu}^{(0)}(\w_\l) = -T\sum_{\k\e_n}G(\k,\e_{n}^+)G(\k,\e_n) \cdot
 v_\mu^0(\k)v_\nu^0(\k)  ,
\end{eqnarray}
where $\e_{n}^+ = \e_{n} + \w_\l$.
After the analytic continuation with respect to Im\,$\w_\l>0$, we get
 \cite{Eliashberg}
\begin{eqnarray}
\Phi_{\mu\nu}^{(0)}(\w) = -2\sum_\k \int\frac{d\e}{4\i\pi}
 \left[ {\rm th}\frac{\e}{2T} g_1(\k,\e;\w) 
  + \left( {\rm th}\frac{\e^+}{2T}-{\rm th}\frac{\e}{2T} \right)
  g_2(\k,\e;\w) - {\rm th}\frac{\e^+}{2T} g_3(\k,\e;\w) 
 \right] v_\mu^0(\k)v_\nu^0(\k)  ,
  \label{eqn:sigxx0}
\end{eqnarray}
where $\e^+ = \e + \w$, and
$g_1(\k,\e;\w)= G^R(\k,\e^+) G^R(\k,\e)$,
$g_2(\k,\e;\w)= G^R(\k,\e^+) G^A(\k,\e)$ and 
$g_3(\k,\e;\w)= G^A(\k,\e^+) G^A(\k,\e)$, respectively.

\begin{figure}
\begin{center}
\epsfig{file=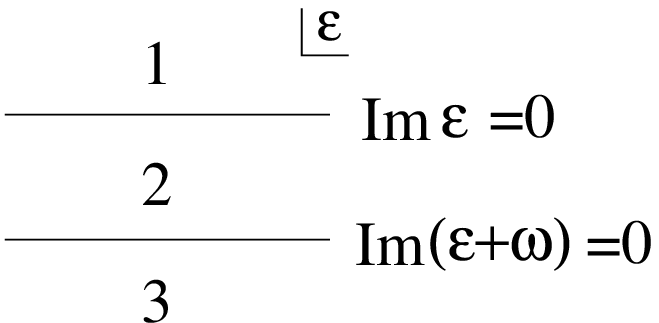,width=5cm}
\end{center}
\caption{The analytic regions 1$\sim$3 
as a function of a complex variable $\e$ (or $\e'$).
Here we put Im\,$\w_\l>0$.
From each region,
$g(\e_n;\w_\l)$, $\L_\mu(\e_{n}^+;\e_n)$ 
and $\Gamma(\e_{n}^+;\e_n|\e_{n'}^+;\e_{n'})$ 
are analytically continued to be 
$g_i(\e;\w)$, $J_\mu^{(i)}(\e^+;\e)$ 
and ${\cal T}_{ij}(\e^+;\e|\e';\e'^+)$ 
($i,j=1\sim3$), respectively.}
\label{fig:region}
\end{figure}

Here $G^R(\e)=G(\e+\i0)$ ($G^A(\e)=G(\e-\i0)$) 
is the retarded (advanced) function.
In eq.(\ref{eqn:sigxx0}), 
the first, second and third terms come from the region
1, 2 and 3 in Fig. \ref{fig:region}, respectively.
The coefficient 2 in front of the $\k$-summation in eq.(\ref{eqn:sigxx0})
comes from the spin-degeneracy.
Hereafter, we take this factor 2 correctly in all the expressions 
in this article.

In a Fermi liquid, $G^R(\k,\e)$ is expressed for $\e\sim0$ 
and $|{\vec k}|\sim k_{\rm FS}$ as
\begin{eqnarray}
G^R(\k,\e) = z_\k(\e+\mu-\e_\k^\ast +\i\g_\k^\ast )^{-1}  ,
\end{eqnarray}
where 
$z_\k = (1-(\d/\d\e){\rm Re}\Sigma(\k,\e))^{-1}|_{\e=0}$
is the renormalization factor at the Fermi energy.
We also introduced the quantities 
$\e_\k = \e_\k^0 + {\rm Re}\Sigma(\k,0)$,
$\g_\k = -{\rm Im}\Sigma^R(\k,0) >0$,
and their 'renormalized quantities'
$\e_\k^\ast = z_\k (\e_\k-\mu) +\mu$ and 
$\g_\k^\ast = z_\k \g_\k$, respectively.

At sufficiently low temperatures,
$\gamma_\k^\ast \ll T$ is satisfied in a Fermi liquid
because $\gamma_\k^\ast \propto T^2$ in 3D
and $\gamma_\k^\ast \propto T^2{\rm log}T^{-1}$ in 2D.
In this case,
\begin{eqnarray}
g_2(\k,\e;0) &\approx& \pi z_\k \delta(\e+\mu-\e_\k^\ast)/\gamma_\k,
  \label{eqn:g2} \\
g_1(\k,\e;0) &\approx& \left( \frac{z_\k}
 {\e+\mu-\e_\k^\ast+\i\delta} \right)^2  .
 \label{eqn:g1} 
\end{eqnarray}
After the integration with respect to $\e$,
eq.(\ref{eqn:g2}) gives the singular term of order 
$\gamma^{-1} \sim \tau^{1}$.
Whereas eq.(\ref{eqn:g1}), which have no pole in the region
Im$\e\ge0$, give no singularities after the $\e$-integration.

In conclusion, the conductivity 
$\s_{\mu\nu}= \lim_{\w\rightarrow0} {\rm Im} \Phi_{\mu\nu}(\w)/\w$
comes from the term possessing $g_2$-section 
in eq.(\ref{eqn:sigxx0}) as follows:
\begin{eqnarray}
\s_{\mu\nu}= e^2 \sum_\k 
 \left(-\frac{\d f}{\d \e}\right)_{\e_\k^\ast-\mu}
 z_\k \cdot \frac{1}{\gamma_\k} v_\mu^0(\k)v_\nu^0(\k)  ,
\end{eqnarray}
which is proportional to $\gamma^{-1}$.
Note that we get the exact expression (\ref{eqn:Eliashberg})
if we take account of all the current vertex correction completely.
It is still of order $\gamma^{-1}$ because $J_\k$ is of order $\gamma^{0}$.
In the next subsection, we perform the analytic continuation for 
the MC in a similar way,
and we also see that only the term which 
comes from the region 2 in Fig. \ref{fig:region}
gives the most singular contribution with respect to $\gamma^{-1}$.

\subsection{Magnetoconductivity}

In this subsection,
we perform the analytic continuation
and derive the expression for the MC 
without vertex corrections for the current.
In this approximation,
(i) all the diagrams in Fig. \ref{fig:DG-init2} are dropped 
because each of them contains more than one 
four (or six, eight) point vertices, and
(ii) the three-point vertices $\Lambda_{\mu(\nu)}$ 
in Fig \ref{fig:DG-init1}
are replaced with the bare ones, $v_{\mu(\nu)}^0(\k)$.
On the other hand,
we notice the Ward identities for $\Lambda_{\a}$ and $\Lambda_{\a\b}$
given by eqs.~(\ref{eqn:Ward-La}) and (\ref{eqn:Ward-Lab}),
in order to get the gauge-invariant expression for the MR
as shown below.

\begin{figure}
\begin{center}
\epsfig{file=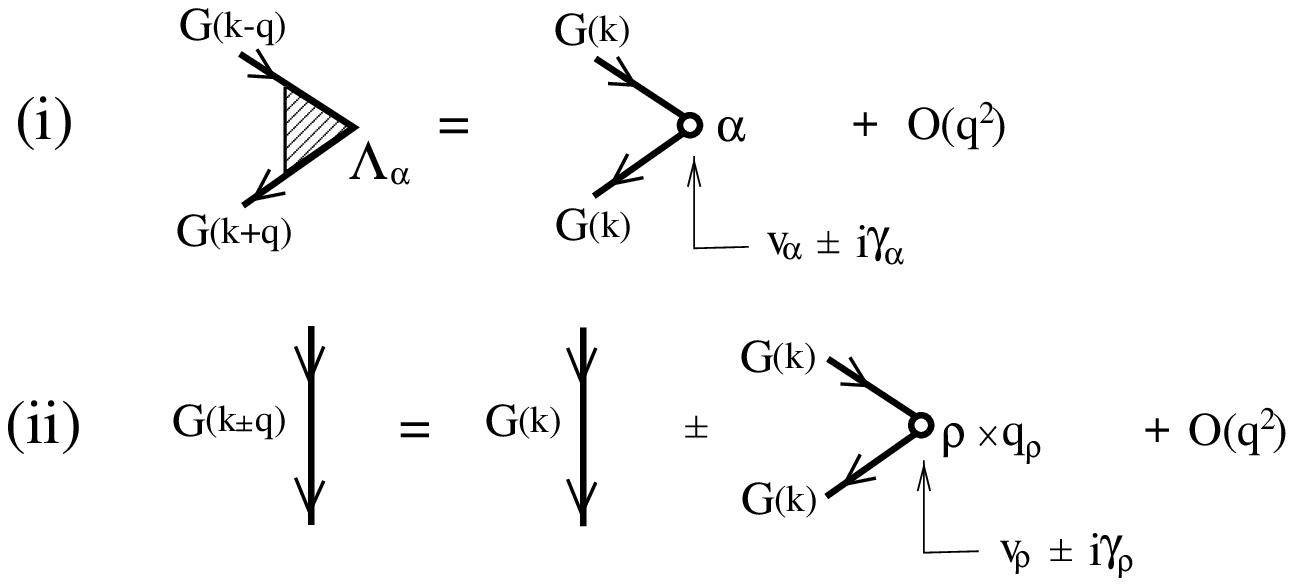,width=7cm}
\end{center}
\caption{
The expansion of 
(i) $\L_\a(\k_-\e;\k_+\e)$, and 
(ii) $G(\k_\pm)$ with respect to $q$.
}
\label{fig:DG-qexpand-G}
\end{figure}

As explained in the previous subsection,
the most singular terms with respect to $\g^{-1}$
come from the region 2 in Fig. \ref{fig:region}.
Thus, we replace $G(\k,\e^+) \rightarrow G^R(\k,\e)$ and 
$G(\k,\e)\rightarrow G^A(\k,\e)$, respectively.
In this subsection, 
we take the $\q_\rho,{q'}_{\rho'}$-derivative of 
$\Phi^{(2)}(2\q+2\q',\w_\l)$ 
in order to derive the MC.
For this purpose, we use the relation
\begin{eqnarray}
\L_\a^R(\k_-\e;\k_+\e) &=& V_\a^R(\k,\e) 
 \ \ + \ O(q^2), 
  \label{eqn:L-der} \\
V_\a^R(\k,\e) &\equiv& v_\a^0(\k)+\d_\a\Sigma(\k,\e+\i\delta) \nonumber \\
  & &\ \ = v_\a(\k,\e) - \i\gamma_\a(\k,\e)  ,
\end{eqnarray}
where $v_\a(\k,\e) = v_\a^0(\k)+\d_\a{\rm Re}\Sigma(\k,\e)$
and $\gamma_\a(\k,\e) = -\d_\a{\rm Im}\Sigma(\k,\e+\i\delta)$.
They are shown in Fig.\ref{fig:DG-qexpand-G} (i).
As a result, $\w$-linear term of $\Phi^{(2)}(2\q+2\q',\w+\i 0)$ 
without the vertex corrections is given by
\begin{eqnarray}
& &\lim_{\w\rightarrow0}\Phi^{(2)}(2\q+2\q',\w+\i0)/\w
 \nonumber \\
& &\ \ 
 = -\sum_\k \int\frac{d\e}{\pi}\left(-\frac{\d f}{\d\e}\right)
 \cdot v_{\mu}^0(\k) G^R(\k_{--}) G^A(\k_{++}) 
 \nonumber \\
& &\ \ \ \ \ \times
 \Bigl\{ v_{\nu\a\b}^0(\k)
 + \left( V_{\a\b}^R(\k) G^R(\k_{++}) v_{\nu}^0(\k_{++}) 
      + v_{\nu}^0(\k_{--}) G^A(\k_{--}) V_{\a\b}^A(\k) \right)
 \nonumber \\
& &\ \ \ \ \ \ \ + \Bigl[
 \left( V_\a^R(\k_{0-}) G^R(\k_{+-}) v_{\nu\b}^0(\k_{+}) 
      + v_{\nu\b}^0(\k_{-}) G^A(\k_{-+}) V_\a^A(\k_{0+}) \right)
 \nonumber \\
& &\ \ \ \ \ \ \ \ \ \ + 
  V_{\a}^R(\k_{0-}) G^R(\k_{+-}) v_{\nu}^0(\k_{+-})
        G^A(\k_{+-}) V_{\b}^A(\k_{+}) 
 \nonumber \\
& &\ \ \ \ \ \ \ \ \ \ +
 \left(V_{\a}^R(\k_{0-}) G^R(\k_{+-}) V_{\b}^R(\k_{+}) 
        G^R(\k_{++}) v_{\nu}^0(\k_{++})
      + v_{\nu}^0(\k_{--}) G^R(\k_{--}) V_{\a}^R(\k_{0-}) 
        G^A(\k_{+-}) V_{\b}^A(\k_{+}) \right) 
        \ \Bigr]
 \nonumber \\
& &\ \ \ \ \ \ \
 + \ \Bigl[ (\a,\q)\leftrightarrow (\b,\q') \Bigr] \ 
   \Bigr\}   .
   \label{eqn:Phi2-0}
\end{eqnarray}
%

To take the $\q,\q'$-derivative of eq.(\ref{eqn:Phi2-0}),
we employ the diagrammatic technique
for the preparation of deriving the exact formula
in the next section.
Then, we use the followings:
\begin{eqnarray}
G^{R}(\k_\pm,\e) &=& G^{R}(\k,\e) 
 \pm \q_\rho \cdot V_\rho^R(\k,\e) { G^{R}(\k,\e) }^2 
 \ \ + \ O(q^2),  
  \label{eqn:G-der}\\
V_\b^{R}(\k_\pm,\e) &=& _\b^{R}(\k,\e) 
 \pm \q_{\rho} \cdot V_{\b\rho}^{R}(\k,\e)  
 \ \ + \ O({q}^2),  
 \label{eqn:V-der} 
\end{eqnarray}
where $V_{\a\b}^{R}(\k,\e)\equiv \d_\b V_\a^{R}(\k,\e)
 = \d_\a V_\b^{R}(\k,\e)$.
There, we took the summations of $\rho$ and $\rho'$ implicitly.
Eq.(\ref{eqn:G-der}) is shown in Fig.\ref{fig:DG-qexpand-G} (ii).

\begin{figure}
\begin{center}
\epsfig{file=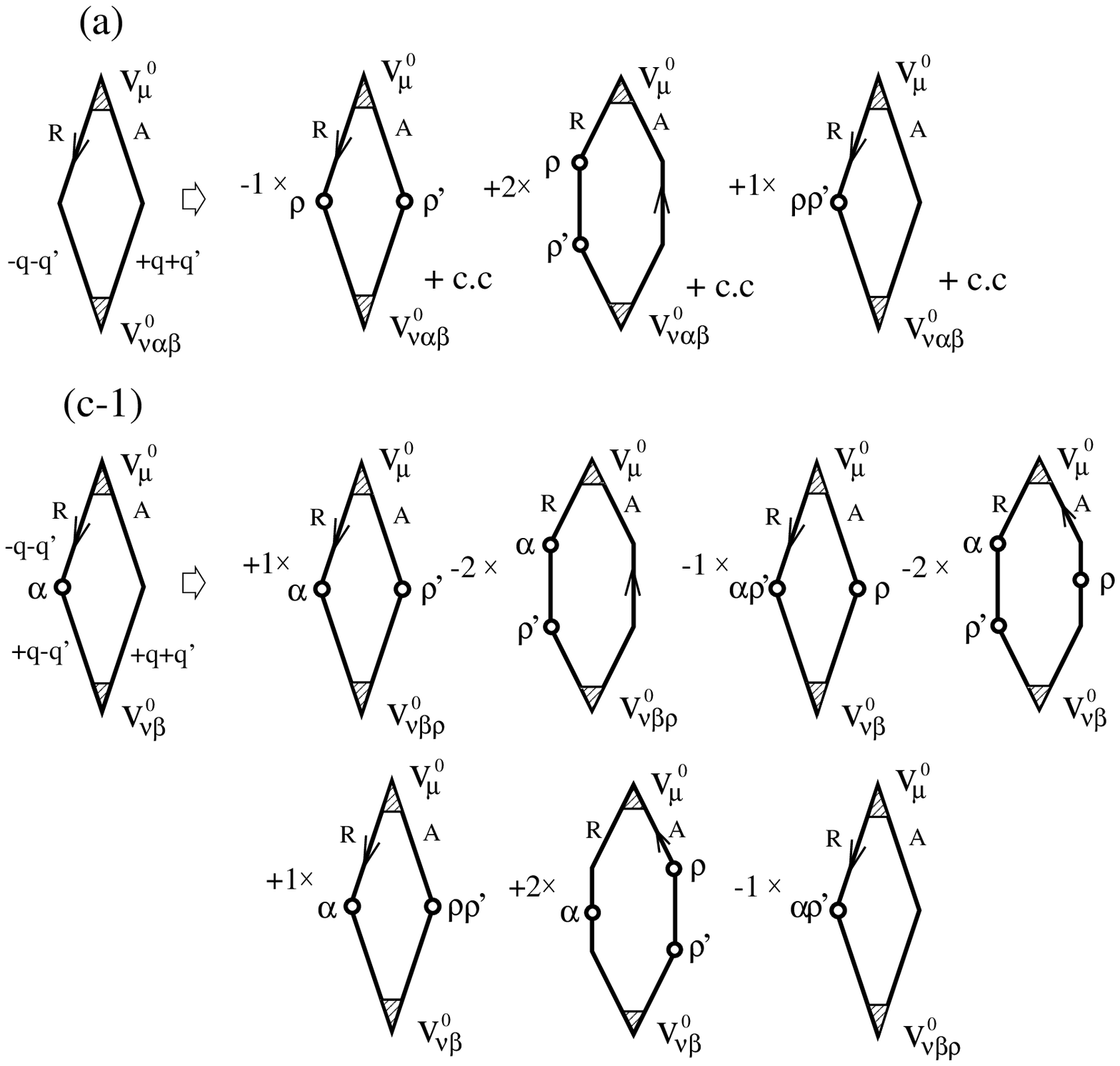,width=8cm}
\end{center}
\caption{The $\q,\q'$-derivative
of (a) and (c-1) in Fig.\ref{fig:DG-init1}
without vertex corrections.
'R' ('A') represent retarded (advanced) Green functions.
'$\circ\rho$' on retarded (advanced) Green functions
means $v_\rho - \i\g_\rho$ ($v_\rho + \i\g_\rho$).
Note that the sign of each diagrams
comes from $\d/\d q$ and $\d/\d q'$.
}
\label{fig:DG-noVC1-exm}
\end{figure}

Figure \ref{fig:DG-noVC1-exm}
shows the diagrammatic expressions for the $\q,\q'$-derivative
of (a) and (c-1) in Fig.\ref{fig:DG-init1}.
The $\q$-derivative of the Green functions
increases the number of $G$'s by one, 
which increases the singularity of 
the conductivity by $\gamma^{-1}$.
In Fig. \ref{fig:DG-noVC1-exm},
the coefficient of a diagram and that of a diagram
which differ by $\a\leftrightarrow\r$ are opposite in sign. 
This fact ensures the gauge-invariance of the expression.

\begin{figure}
\begin{center}
\epsfig{file=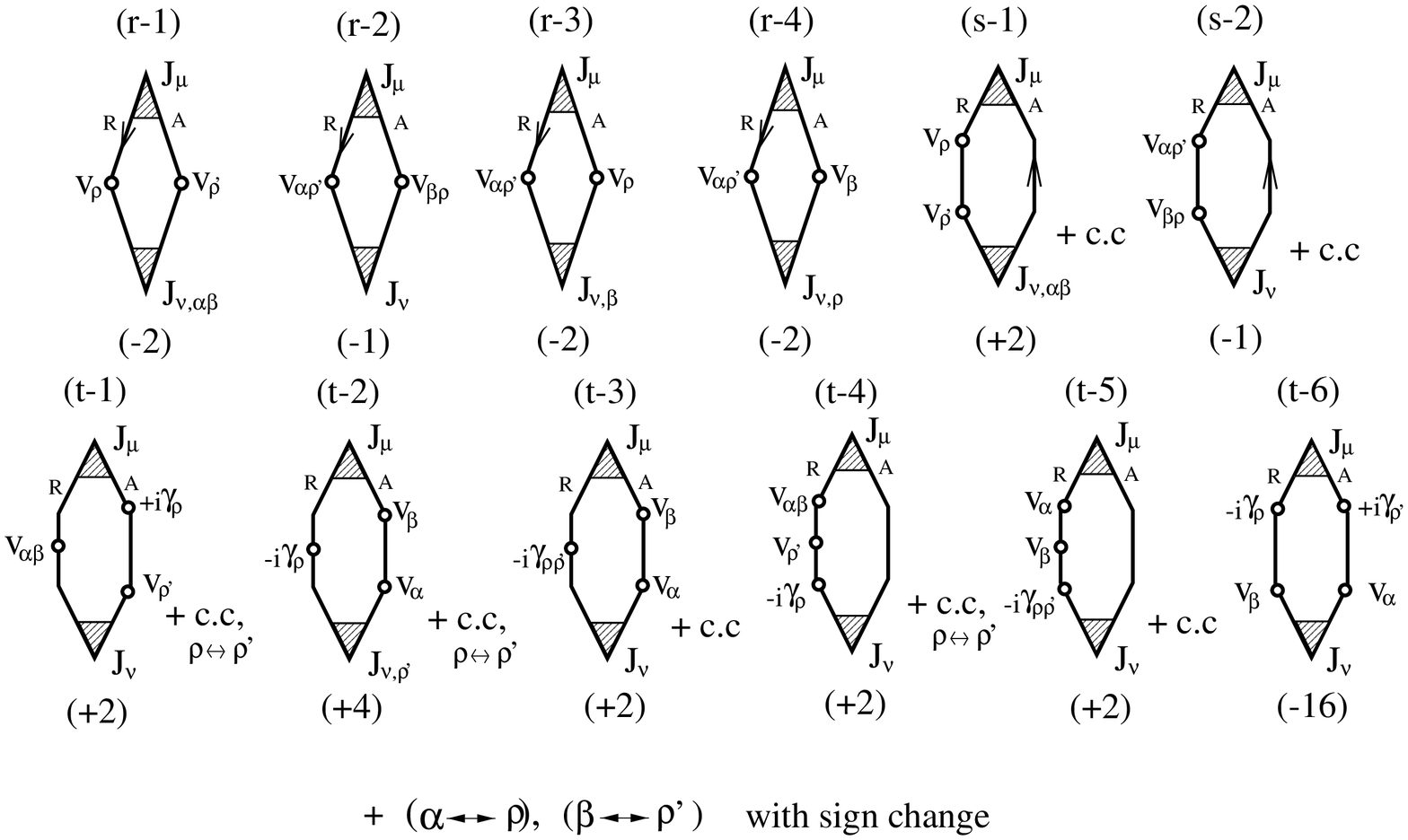,width=12cm}
\end{center}
\caption{}
{\small
The diagrams for the MC of order $\g^{-3}$
which are composed of only $G$, ${\vec v}$ and ${\vec J}$.
(r-1$\sim$4) corresponds to $A^{(1)}$ in eq.(\ref{eqn:MC-part1}).
In the same way,
(s-1,2) corresponds to $A^{(2)}$, 
(t-1$\sim$3) corresponds to $A^{(3)}$, 
(t-4,5) corresponds to $A^{(4)}$, and 
(t-6) corresponds to $A^{(6)}$, 
respectively.
(r-1)$\sim$(s-2) contain neither $\g_\a$ nor $\g_{\a\b}$, but
(t-1)$\sim$(t-6) contain $\g_\a$ and/or $\g_{\a\b}$.
}
\label{fig:DG-noVC1}
\end{figure}

To derive the formula for MC, 
we take the $\q,\q'$-derivative
of the all other terms eq.(\ref{eqn:Phi2-0}).
Figure \ref{fig:DG-noVC1} 
shows the obtained terms for the MC 
of order $\gamma^{-3}$,
which are the most singular terms as we will see later.
Note that the terms (t-1$\sim$6) 
become zero in an isotropic system
because they contain the momentum derivative
of Im$\Sigma_\k$ along the Fermi surface.
[see eq.(\ref{eqn:MC-noVC-lowT}) in this paper.]
However, they will play important roles in rather anisotropic 
system like high-$T_{\rm c}$ superconductors.


\begin{figure}
\begin{center}
\epsfig{file=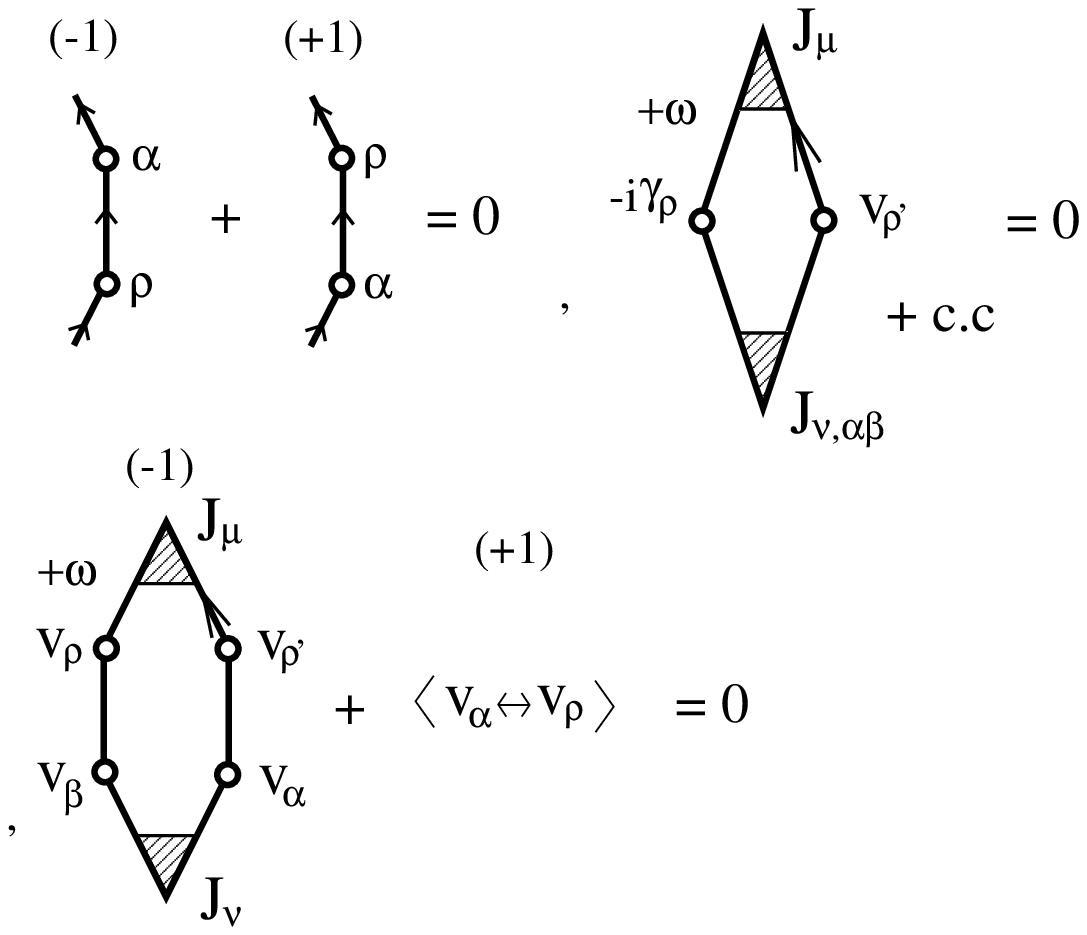,width=7cm}
\end{center}
\caption{Examples of the diagrams which vanish identically.}
\label{fig:DG-vanish}
\end{figure}

\begin{figure}
\begin{center}
\epsfig{file=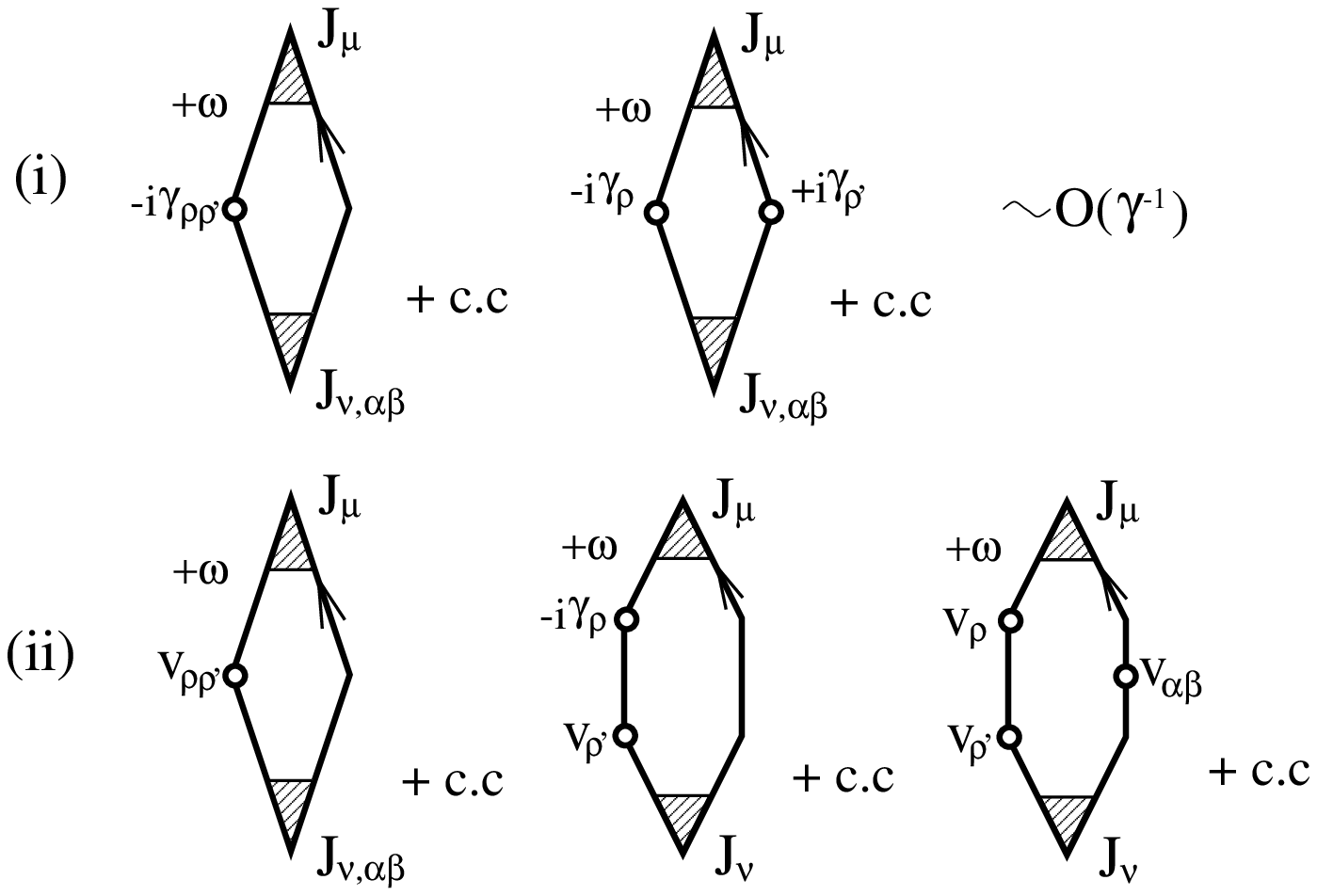,width=7.5cm}
\end{center}
\caption{(i) The diagrams of order $\g^{-1}$,
which are negligible compared with the terms $O(\g^{-3})$.
(ii) The diagrams which are odd with respect to 
$(\e+\mu-\e_\k^\ast)$:
They will be vanishingly small after the integrations
with respect to $\e$ and ${\vec k}$.}
\label{fig:DG-noVC1-add}
\end{figure}

In Fig. \ref{fig:DG-noVC1},
we have dropped some diagrams which cancel out exactly 
with their counterparts.
We show the example of the cancellation in Fig. \ref{fig:DG-vanish}.
Also, the less singular terms with respect to $\g^{-1}$,
some examples are shown in Fig. \ref{fig:DG-noVC1-add} (i),
are discarded.
Furthermore, we also dropped the terms proportional to 
$|G(\k,\e)|^m \!\cdot\! {\rm Im}G^{2n}(\k,\e)$ or to 
$|G(\k,\e)|^m \!\cdot\! {\rm Re}G^{2n+1}(\k,\e)$:
Some of the them are shown in Fig. \ref{fig:DG-noVC1-add} (ii).
Because they are odd-functions of $(\e+\mu-\e_\k^\ast)$,
their contribution for the MC should be vanishingly small
after the $\e$ and ${\k}$-integrations,
except that the density of states is extraordinarily
asymmetric around the chemical potential.


In this paper, 
we assume that the magnetic filed is along the $z$-axis.
We can verify that 
the obtained expression for the MC,
which is given in Fig. \ref{fig:DG-noVC1},
changes its sign with the replacement $\a \leftrightarrow \rho$, or
$\b \leftrightarrow \rho'$.
Thus, the gauge-invariant result of the MC,
$C_{\mu\nu}^{\a\rho;\b\rho'} 
 = C_{\mu\nu}^{xy;xy}\cdot \e_{\a\rho z}\e_{\b\rho' z}$ 
in eq.(\ref{eqn:sxx-expand}), is assured.

Below, we give the analytical expression for Fig.\ref{fig:DG-noVC1}.
\begin{eqnarray}
 {\mit\Delta}\s_{\mu\nu}^a
 &=& B^2 \cdot \frac{e^4}{8v_B} \sum_{\k,i=1,\cdots,5} \int\frac{d\e}{\pi}
 \left(-\frac{\d f}{\d\e}\right) J_{\mu}(\k,\e) 
 \cdot A^{(i)}(\k,\e)  ,
  \label{eqn:MC-part1} \\
 A^{(1)}(\k,\e) 
 &=& |G(\k,\e)|^4 \cdot 2 \left[
  (v_{xy}^2 - v_{xx}v_{yy})J_{\nu}
  + 2(v_yv_{xy}-v_xv_{yy})J_{\nu,x}
  + 2(v_xv_{xy}-v_yv_{xx})J_{\nu,y} \right.
  \nonumber \\
 & &+ \left. (v_y^2 J_{\nu,xx} -2v_xv_y J_{\nu,xy} +v_x^2 J_{\nu,yy})
  \right]_{(\k,\e)}   , \nonumber \\
 A^{(2)}(\k,\e) 
 &=& |G(\k,\e)|^2{\rm Re}\{ G_\k^2(\e) \} \cdot 4 \left[
  (v_{xy}^2 - v_{xx}v_{yy})J_{\nu} 
  -(v_y^2 J_{\nu,xx} -2v_xv_y J_{\nu,xy} +v_x^2 J_{\nu,yy})
  \right]_{(\k,\e)}  , \nonumber \\
 A^{(3)}(\k,\e) 
 &=& -|G(\k,\e)|^4{\rm Im}\{ G^{\rm R}(\k,\e) \} 
  \cdot 4 \left[ 
  2(v_xv_{yy}\g_x +v_yv_{xx}\g_y 
 -v_xv_{xy}\g_y -v_yv_{xy}\g_x)J_{\nu} \right.
   \nonumber \\
& &  +(-v_x^2 \g_{yy} + 2v_xv_y\g_{xy} -v_y^2 \g_{xx})J_{\nu}
  \nonumber \\
 & &\left. +4(v_xv_y\g_y -v_y^2\g_x)J_{\nu,x}
           +4(v_xv_y\g_x -v_x^2\g_y)J_{\nu,y}
  \right]_{(\k,\e)}  , \nonumber \\
 A^{(4)}(\k,\e) 
 &=& -|G(\k,\e)|^2{\rm Im}\{ {G^{\rm R}}^3(\k,\e) \} \cdot 4 \left[
  2(v_xv_{yy}\g_x +v_yv_{xx}\g_y -v_xv_{xy}\g_y -v_yv_{xy}\g_x)J_{\nu} 
   \right. \nonumber \\
 & &\left. +(v_x^2 \g_{yy} - 2v_xv_y\g_{xy} 
  +v_y^2 \g_{xx})J_{\nu} \right]_{(\k,\e)}  , \nonumber \\
 A^{(5)}(\k,\e)  
 &=& |G(\k,\e)|^6 \cdot 16 \left[
   (v_x^2\g_y^2 +v_y^2 \g_x^2 -2v_xv_y\g_x\g_y) J_{\nu}
  \right]_{(\k,\e)}  ,  \nonumber 
\end{eqnarray}
where 
$v_{xy} \equiv \d_x v_y = \d_{xy}(\e_\k^0+{\rm Re}\Sigma_\k(0))$
and 
$\g_{xy} \equiv \d_x \g_y = \d_{xy}{\rm Im}\Sigma_\k(-\i\delta)$.
In eq. (\ref{eqn:MC-part1}), 
we replaced $v_{\mu(\nu)}^0(\k)$ in eq.(\ref{eqn:Phi2-0}) with
$J_{\mu(\nu)}(\k,\e)= v_{\mu(\nu)}(\k,\e)$, and
$J_{\nu,x}\equiv v_{\nu x}(\k,\e)$, $J_{\nu,xy}\equiv v_{\nu xy}(\k,\e)$.
Note that we have checked eq.(\ref{eqn:MC-part1})
by taking the $q,q'$-derivative of eq.(\ref{eqn:Phi2-0}) directly
by using MATHEMATICA.
In later sections,
we study all the vertex corrections 
which are dropped here,
and find that each $J_{\mu}$ and $J_{\nu}$ in eq.(\ref{eqn:MC-part1})
have a vertex correction by ${\cal T}_{22}$.
(see eq.(\ref{eqn:J2}).)

In eq.(\ref{eqn:MC-part1}),
$A^{(1)}$ and $A^{(2)}$
are expressed diagrammatically as (r-1)$\sim$(r-4) and (s-1,2) 
in Fig.\ref{fig:DG-noVC1},
and $A^{(3)}$, $A^{(4)}$, and $A^{(5)}$
are expressed as (t-1$\sim$6).
The latter contains the momentum derivative of
$\gamma(\k,\e)$.
In eq.(\ref{eqn:MC-part1}), the factor 2 due to the 
spin degeneracy is taken into account.

Finally, we perform the $\e$-integration in 
eq. (\ref{eqn:MC-part1}) 
under the assumption that $\gamma^\ast \ll T$.
In this case, the expression for the MR, eq.~(\ref{eqn:MC-part1}),
becomes much simpler as follows:
In the case of $\gamma_\k^\ast \ll T$, 
the following replacements are allowed in eq.~(\ref{eqn:MC-part1}).
\begin{eqnarray}
& &|G|^4 \rightarrow \pi z_\k \delta(\e+\mu-\e_\k^\ast) 
 \frac1{2\g_\k^{3}}  , \nonumber \\
& &|G|^2 {\rm Re}\{G^2\}
 \rightarrow \pi z_\k \delta(\e+\mu-\e_\k^\ast) 
 \frac{-1}{4\g_\k^{3}}  , \nonumber \\
& &|G|^4 {\rm Im}\{G^R\}
 \rightarrow \pi z_\k \delta(\e+\mu-\e_\k^\ast) 
 \frac{-3}{8\g_\k^{4}} 
 \label{eqn:qp}  , \\
& &|G|^2 {\rm Im}\{{G^R}^3\}
 \rightarrow \pi z_\k \delta(\e+\mu-\e_\k^\ast) 
 \frac1{8\g_\k^{4}}  , \nonumber \\
& &|G|^6  \rightarrow \pi z_\k \delta(\e+\mu-\e_\k^\ast) 
 \frac3{8\g_\k^{5}}  , \nonumber 
\end{eqnarray}
By using these relations (\ref{eqn:qp}),
we see that 
(r-1) and (s-1) in Fig. \ref{fig:DG-noVC1} are equal, whereas
(r-2) and (s-2) cancel out completely 
in the case of $\gamma_\k^\ast \ll T$.
After the long but straightforward calculation,
we get the following simple result for the MC.
\begin{eqnarray}
 {\mit\Delta}\s_{\mu\mu}^a
 &=& B^2 \cdot \frac{e^4}{4v_B} \sum_{\bf k} \int d\e
  \left(-\frac{\d f}{\d\e}\right) z_{\bf k} 
  \delta(\e+\mu-\e_{\bf k}^\ast)
   \cdot \frac{J_{{\bf k}\mu}}{\g_{\bf k}}
  ({\hat e}_z\times{\vec v}_{\bf k})\cdot{\vec \nabla} 
  \left\{ \frac1{\gamma_{\bf k}}
  ({\hat e}_z\times{\vec v}_{\bf k})\cdot{\vec \nabla} 
  \left( \frac{J_{{\bf k}\mu}}{\g_{\bf k}} \right) \right\}
  \nonumber \\
 &=& B^2 \cdot \frac{e^4}{4v_B} \oint_{\rm FS} dS_\k
  \frac{J_{{\bf k}\mu}}{\g_{\bf k}}
  \frac{\d}{\d k_\parallel}
  \left\{ \frac{|{\vec v}_{\bf k}|}{\gamma_{\bf k}}
  \frac{\d}{\d k_\parallel}
  \left( \frac{J_{{\bf k}\mu}}{\g_{\bf k}} \right) \right\}
  \nonumber \\
 &=& -B^2 \cdot \frac{e^4}{4v_B} \oint_{\rm FS} dS_\k
 \frac{|{\vec v}_{\bf k}|}{\g_{\bf k}} \cdot
 \left( \frac{\d}{\d k_\parallel} 
 \left( \frac{J_{{\bf k}\mu}}{\g_{\bf k}} \right) \right)^2  ,
  \label{eqn:MC-noVC-lowT}
\end{eqnarray}
where we used the relation
$({\hat e}_z\times{\vec v}_\k)\!\cdot\!{\vec \nabla} 
= v_x \d_y -v_y \d_x = |{\vec v}_\k| \frac{\d}{\d k_\parallel}$.
Here $k_\parallel$ is the momentum along 
${\hat e}_\parallel \equiv ({\hat e}_z\times{\vec v}_\k)$, 
that is, $k_\parallel$ is along the Fermi surface
on the $xy$-plane.
We also used the replacement:
\begin{eqnarray}
\sum_\k z_\k\delta(\mu-\e_\k^\ast) \ \rightarrow \
\int_{\rm FS} \frac1{|{\vec v}_\k|} dS_\k  
\ \ \ \ \ \ \ \ \ \ \ \ 
{\mbox{in 3D}}  .
\end{eqnarray}
On the other hand, in the two-dimensional case,
\begin{eqnarray}
\sum_\k z_\k\delta(\mu-\e_\k^\ast) \ \rightarrow \
\int_{\rm FS} \frac1{|{\vec v}_\k|} dk_\parallel
\ \ \ \ \ \ \ \ \ \ \ \ 
{\mbox{in 2D}}  .
\end{eqnarray}
The result (\ref{eqn:MC-noVC-lowT})
is nothing but the result of the relaxation time
approximation given by eq.(\ref{eqn:Boltzmann-Dsxx}),
if we replace $2\tau_\k\rightarrow \g_k^{-1}$.
In conclusion, the MC derived in this section,
${\mit\Delta}\s_{xx}^a$,
is of order $\gamma_\k^{-3}$.


\section{Magnetoconductivity: with Full Vertex Corrections}

In the previous section, we derived the MC by neglecting
the vertex corrections as the first step.
The obtained expression becomes equal to the result
of the relaxation time approximation.
However, it will not be applicable for correlated electron systems
because the vertex corrections
are indispensable to satisfy the conservation laws.
In this section, we derive the {\it exact expression}
for the MC in the Fermi liquid with arbitrary interaction, 
up to the most divergent contribution with respect to $\gamma^{-1}$.

At first, 
we take the $q_\rho q'_{\rho'}$-derivative of 
$\Phi^{(2)}(2\q+2\q',\w_\l)$ 
which is shown in Figs. \ref{fig:DG-init1} and \ref{fig:DG-init2}.
Now, we should remind that 
all the diagrams are composed of (a) the bare vertices
$v_\nu^0(\k)$, $v_{\mu\a}^0(\k)$, etc, 
(b) the two point Green functions $G(\k,\e)$, and
(c) the irreducible four, six and eight vertices $\Gamma^I$,
$\Gamma^{I(3)}$ and $\Gamma^{I(4)}$.
(see Fig.\ref{fig:DG-init1-exp}.)
To obtain the exact result,
we have to take the $q_\rho,q'_{\rho'}$-derivative of the
{\it every element of (a)-(c)} included in 
$\Phi^{(2)}(2\q+2\q',\w_\l)$ once, 
by taking care of the momentum suffixes.

\begin{figure}
\begin{center}
\epsfig{file=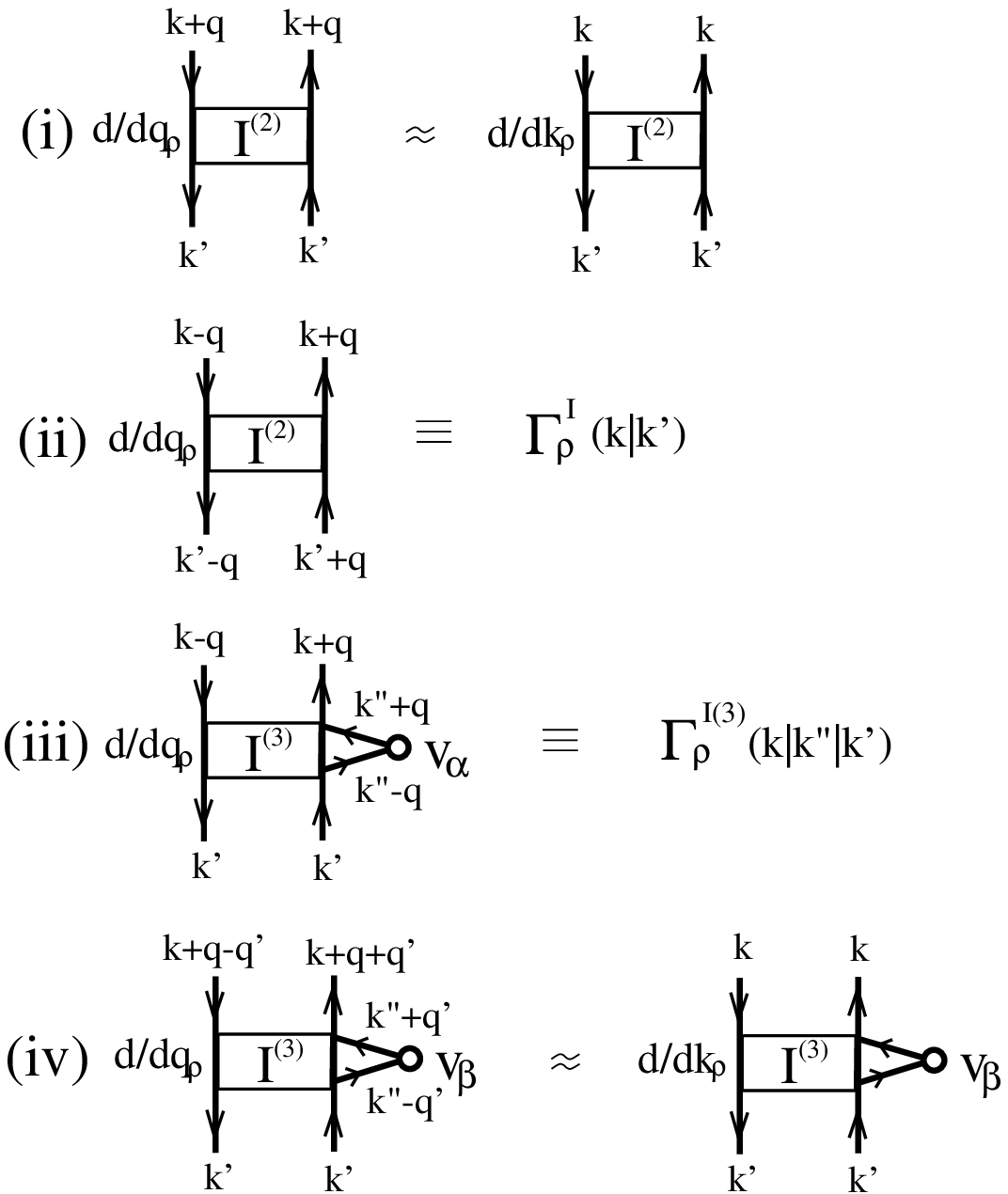,width=7cm}
\end{center}
\caption{$\q$-derivative of several types of the 
irreducible vertices contained in $\Phi_{\mu\nu}^{(2)}(2\q+2\q';\w_\l)$:
(i) $\G^I(\k_+;\k_+|\k';\k')$,
(ii) $\G^I(\k_-;\k_+|\k'_+;\k'_-)$, 
(iii) $\G^{I(3)}(\k_-;\k_+|\k';\k'|\k''_+;\k''_-) \cdot \d''_\a G(\k'')$,
and (iv) 
$\G^{I(3)}(\k_{+-};\k_{++}|\k';\k'|\k''_{0+};\k''_{0-}) \cdot \d''_\b G$.
Only the $\q$-derivative of (i) and (iv) can give the
$O(\g^{-3})$ contributions for the MC. 
Whereas (ii) and (iii) give at most
$O(\g^{-2})$, which will be explained in Appendix B.}
\label{fig:DG-qexpand}
\end{figure}

The $q_\rho(q'_{\rho'})$-derivative of (b)
is already shown in Fig. \ref{fig:DG-qexpand-G}.
So we consider on the derivative of (c):
Figure \ref{fig:DG-qexpand} 
shows several types of the irreducible vertices
which are included in $\Phi^{(2)}(2\q+2\q',\w_\l)$.
In the case of (i) and (iv),  
$\d/\d q_\rho$ can be replaced by $\d/\d k_\rho$,
while it is impossible for (ii) and (iii).
In this section, 
we consider only the $q_\rho$-derivative of (i) and (iv)
and neglect that of (ii) and (iii), 
because they turn out to 
give less singular contributions with respect to $\gamma^{-1}$: 
We will give the explanation for this fact in Appendix B.

\begin{figure}
\begin{center}
\epsfig{file=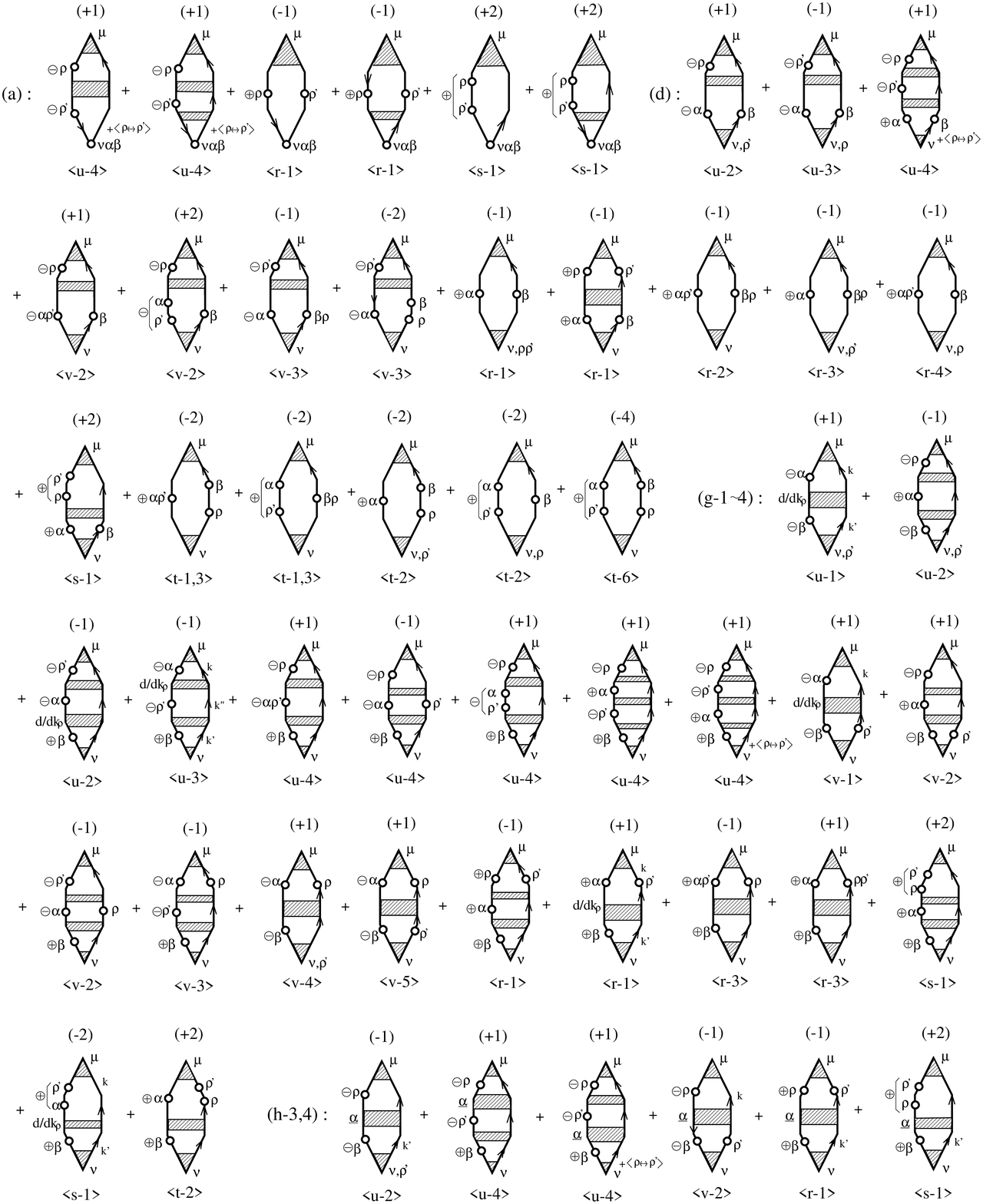,width=15cm}
\end{center}
\caption{}
{\small 
Some examples of 
$(\d^2/\d q_\rho \d q'_{\rho'}) \Phi^{(2)}(2\q+2\q',\w_\l)$
which come from  
(a) and (d) in Fig.\ref{fig:DG-init1}, and
(g-1$\sim$4) and (h-3,4) in Fig.\ref{fig:DG-init2}.
We note that the diagrams which are less singular than 
$O(\g^{-3})$, e.g., the first term of (i) and (ii)
in Fig.\ref{fig:DG-noVC1-add}, are dropped in this figure.
Note that the diagrams with the replacement
$(\a,\rho)\leftrightarrow(\b,\rho')$
also exist for (g-1$\sim$4) and (h-3,4).
Other examples are shown in Appendix C.
}
\label{fig:DG-withVC-exp1} 
\end{figure}

\begin{figure}
\begin{center}
\epsfig{file=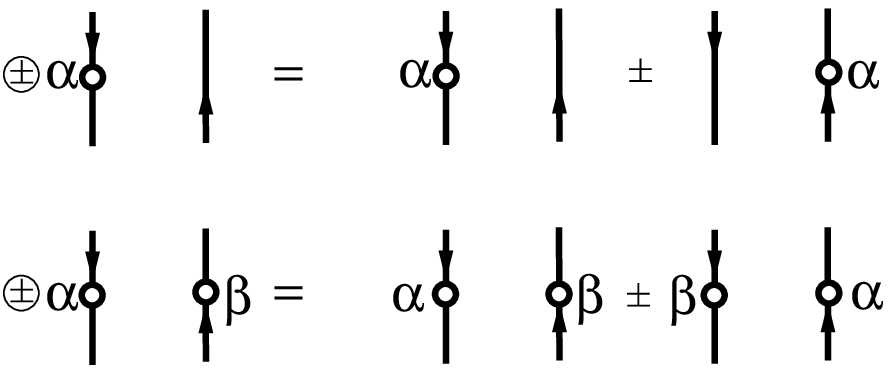,width=5cm}
\end{center}
\caption{
The definition of the operation $\oplus$ and $\ominus$.}
\label{fig:DG-pm} 
\end{figure}

From now on, 
we derive $(\d^2/\d q_\rho\d q'_{\rho'}) \Phi^{(2)}(2\q+2\q',\w_\l)$
by taking care of the definitions of $\G$ and $\L$
given in Fig. \ref{fig:DG-init1-exp}.
This procedure is not easy because we meet hundreds of 
diagrams in total.
For example, in Fig. \ref{fig:DG-withVC-exp1},
we show 'all the terms' 
come from the $\q\q'$-derivative of 
(a) and (d) in Fig. \ref{fig:DG-init1},
and (g-1$\sim$4), (h-3,4) in Fig. \ref{fig:DG-init2}.
The definition of the operation $\oplus$ and $\ominus$ 
in each diagram is given in Fig. \ref{fig:DG-pm}.
As for the $\q,\q'$-derivative of $\G$,
only (i) and (iv) in Fig.\ref{fig:DG-qexpand} are taken into account.
In Fig. \ref{fig:DG-withVC-exp1},
we write its sign and the factor above each diagram.

\begin{figure}
\begin{center}
\epsfig{file=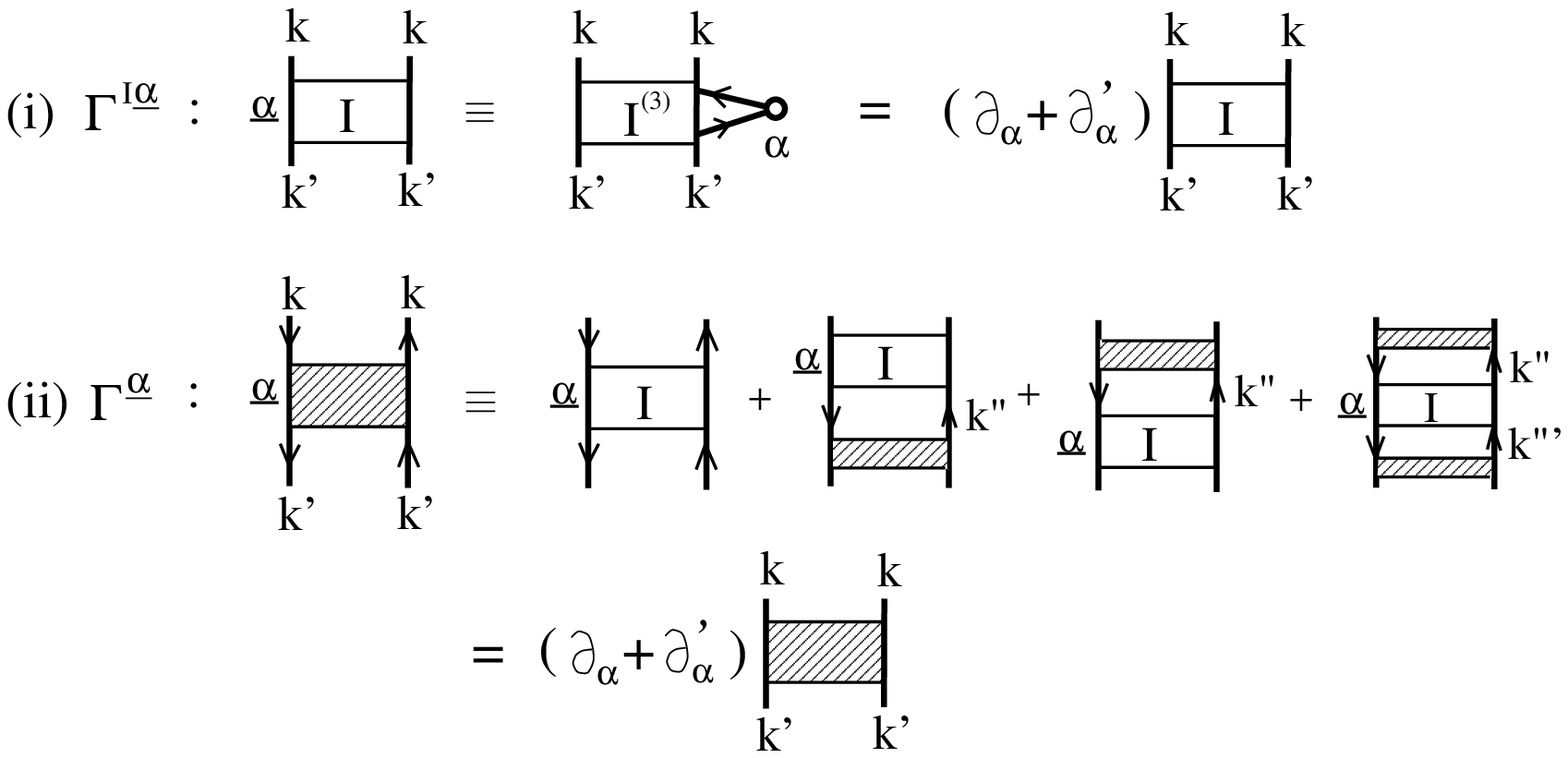,width=8cm}
\end{center}
\caption{
The definitions of $\G^{I{\underline{\a}}}(\k|\k')$
and $\G^{\underline{\a}}(\k|\k')$.
According to the Ward identity,
they are equal to $(\d_\a+\d'_\a)\G^{I}$
and $(\d_\a+\d'_\a)\G^{}$,
respectively.
}
\label{fig:DG-Ward}
\end{figure}

In Fig. \ref{fig:DG-withVC-exp1}, 
(a) does not have the momentum derivative of $\G$
because we drop the terms with (ii) and (iii) in
Fig.\ref{fig:DG-qexpand}.
On the other hand,
(d) contains $\d'_{\rho'}\L_\mu$, 
(g-1$\sim$4) contains $\d_{\rho}\G$, and
(h-3,4) have the six-point vertex $\G^{\underline{\a}}$, 
respectively.
Hereafter, we write $\d_{\a}\L_\nu(\k) \equiv \L_{\nu,\a}(\k)$
and $\d_{\a\b}\L_\nu(\k) \equiv \L_{\nu,\a\b}(\k)$, respectively.
Some of the diagrams in Fig. \ref{fig:DG-withVC-exp1} are 
expressed as
\begin{eqnarray}
{\mbox{1st of (a)}}
&=& -T^2 \sum_{\k\k',\e\e'}
 \L_\mu(\k) [G \dalt_\rho G^+ ] \cdot \G(\k|\k') \cdot
 [G {\dalt}_{\rho'}^{\> \prime} G^+] \L_{\nu,\a\b}(\k')  ,
   \\
{\mbox{3rd of (a)}}
&=& +T \sum_{\k,\e}
 \L_\mu(\k) [ \d_{\rho'}G \cdot \d_\rho G^+
  + \langle \rho \leftrightarrow \rho' \rangle ] 
  \L_{\nu,\a\b}(\k)  ,
   \\
{\mbox{5th of (a)}}
&=& -T \sum_{\k,\e}
 \L_\mu(\k) \{ G [\d_{\rho\rho'}G^+ ]_{(3)} 
  + [\d_{\rho\rho'} G ]_{(3)} G^+ \}
  \L_{\nu,\a\b}(\k)  ,
   \\
{\mbox{1st of (g-1$\sim$4)}}
&=& -T^2 \sum_{\k\k',\e\e'}
 \L_\mu(\k) [G \dalt_\a G^+ ] \cdot 
 \d_\rho \G(\k|\k') \cdot [G {\dalt}_\b^{\> \prime} G^+ ] 
  \L_{\nu,\rho'}(\k')  ,
   \\
{\mbox{2nd of (g-1$\sim$4)}}
&=& +T^3 \sum_{\k\k'\k'',\e\e'\e''}
 \L_\mu(\k) [G \dalt_\rho G^+ ] \cdot 
 \G(\k|\k') \cdot [ \d'_\a \{GG^+\} ] 
 \nonumber \\
& &\times \G(\k'|\k'') \cdot [G {\dalt}_\b^{\> \prime\prime} G^+ ] 
  \L_{\nu,\rho'}(\k'')  ,
   \\
{\mbox{1st of (h-3,4)}}
&=& +T^2 \sum_{\k\k',\e\e'}
 \L_\mu(\k) [G \dalt_\rho G^+ ] \cdot 
 \G^{\underline{\a}}(\k|\k') \cdot [G \dalt_\b^{\> \prime} G^+ ] 
  \L_{\nu,\rho'}(\k')  ,
   \label{eqn:h34}
\end{eqnarray}
where 
$\d'_x = \d/\d k'_x$ and $\d''_x = \d/\d k''_x$.
$[A\dalt_\a B] \equiv A \!\cdot\! \d_\a B - B \!\cdot\! \d_\a A$, and 
$[ \cdots ]_{(3)}$ means to pick up
only the terms which is proportional to $G^3$.
Moreover, we wrote 
$\G(\k\e^+;\k\e|\k'\e',\k'\e'^+) \equiv \G(\k|\k')$, and 
$\G^{\underline{\a}}(\k|\k')$ in eq.(\ref{eqn:h34})
is defined as
\begin{eqnarray}
 \Gamma^{\underline{\a}}(\k|\k')
 \equiv T^2\sum_{\p\p'} (1/T + \G\cdot GG^+)_{(\k;\p)}
 \cdot \Gamma^{I{\underline{\a}}}(\p|\p')
 \cdot (1/T + GG^+\cdot\G)_{(\p';\k')}  ,
 \label{eqn:Ward3}
\end{eqnarray}
where $\Gamma^{I{\underline{\a}}}$ is introduced as
\begin{eqnarray}
\Gamma^{I{\underline{\a}}}(\k|\k') =
  T\sum_{\k''\e''}\Gamma^{I(3)}(\k|\k'|\k'')
  \cdot \d''_\a G(\k'')  .
  \label{eqn:Ward1}
\end{eqnarray}
Both $\Gamma^{\underline{\a}}$ and $\Gamma^{I{\underline{\a}}}$
are shown by Feynman diagrams
in Fig. \ref{fig:DG-Ward} (i) and (ii), respectively.


\begin{figure}
\begin{center}
\epsfig{file=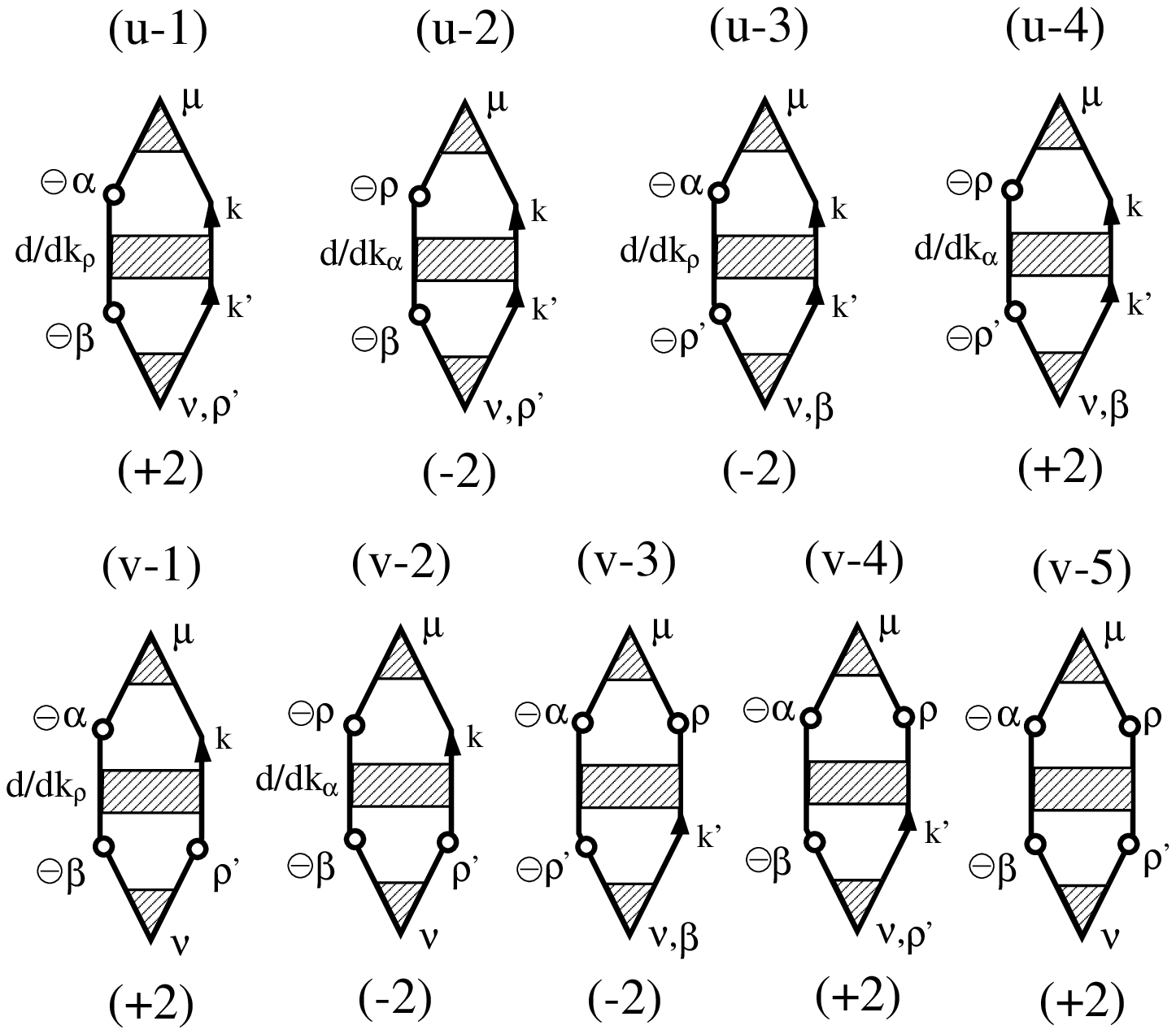,width=8cm}
\end{center}
\caption{The diagrams for the MC
with the $q$-derivative of four-point vertices.
The exact MC of order $\g^{-3}$ is given by these diagrams 
together with Fig.\ref{fig:DG-noVC1}.
Here, the coefficient $\pm 2$ comes from the 
exchange between $(\a,\rho)\leftrightarrow(\b,\rho')$.}
\label{fig:DG-withVC}
\end{figure}

To derive 
$(\d^2/\d q_\rho \d q'_{\rho'}) \Phi^{(2)}(2\q+2q',\w_\l)$,
we also have to take the $\q,\q'$-derivative
for all other diagrams 
in Figs.\ref{fig:DG-init1} and \ref{fig:DG-init2}.
After all, we get hundreds of complicated diagrams for the MC,
which is not a practical formula for applications.
Fortunately, all the diagrams
turn out to be collected into small number of
simpler diagrams.
From now on, we will prove that they are equivalent to
Fig. \ref{fig:DG-withVC} together with Fig. \ref{fig:DG-noVC1} 
if $J_{\mu(\nu)}$ is replaced with $\Lambda_{\mu(\nu)}$.
The expressions for the diagrams in 
Fig. \ref{fig:DG-withVC} are given by
\begin{eqnarray}
{\mbox{(u-1)}} &=&
 -2 T^2\sum_{\k\k'\e\e'}
  \L_\mu(\k) [G\dalt_{\a} G^+] \cdot \d_\rho
  \G(\k|\k') \cdot [G\dalt_{\b}^{\> \prime} G^+] 
  \L_{\nu,\rho'}(\k')
   \label{eqn:u-1}  , \\
{\mbox{(v-1)}} &=&
 -2 T^2\sum_{\k\k'\e\e'}
  \L_\mu(\k) [G\dalt_{\a} G^+] \cdot \d_\rho
  \G(\k|\k') \cdot 
  [\d'_\b G^+ \cdot \d'_{\rho'} G
  -\d'_{\rho'} G^+ \cdot \d'_\b G] \L_\nu(\k')  .
\end{eqnarray}
In the same way, we get
(u-2)$= -{\mbox{(u-1)}}|_{\a\leftrightarrow\rho}$,
(u-3)$= -{\mbox{(u-1)}}|_{\b\leftrightarrow\rho'}$, and
(u-4)$= -{\mbox{(u-2)}}|_{\b\leftrightarrow\rho'}$.

The expressions for (r-1) and (s-1) in Fig. \ref{fig:DG-noVC1}
after the replacement $J_{\mu(\nu)}\rightarrow \L_{\mu(\nu)}$
are given by
\begin{eqnarray}
{\mbox{(r-1)}} &=&
  T\sum_{\k\e}
  \L_\mu(\k) [ \d_{\rho'} G \cdot \d_{\rho} G^+ ] \L_{\nu,\a\b}(\k) ,
   \\
{\mbox{(s-1)}} &=&
 -T\sum_{\k\e}
  \L_\mu(\k) \{\ [\d_{\rho\rho'} G^+]_{(3)} \cdot G 
  + \langle G^+ \leftrightarrow G \rangle \ \} \L_\nu(\k)  .
\end{eqnarray}
The procedure of proof is that 
(i) we decompose the diagrams with 
$\d_{\a(\b)}$ or $\d_{\a\b}$ in 
Figs.\ref{fig:DG-noVC1} and \ref{fig:DG-withVC}
by taking several types of Ward identities, and
(ii) recognize the one to one correspondence between them and
$(\d^2/\d q_\rho \d q'_{\rho'}) \Phi^{(2)}(2\q+2q',\w_\l)$.

\begin{figure}
\begin{center}
\epsfig{file=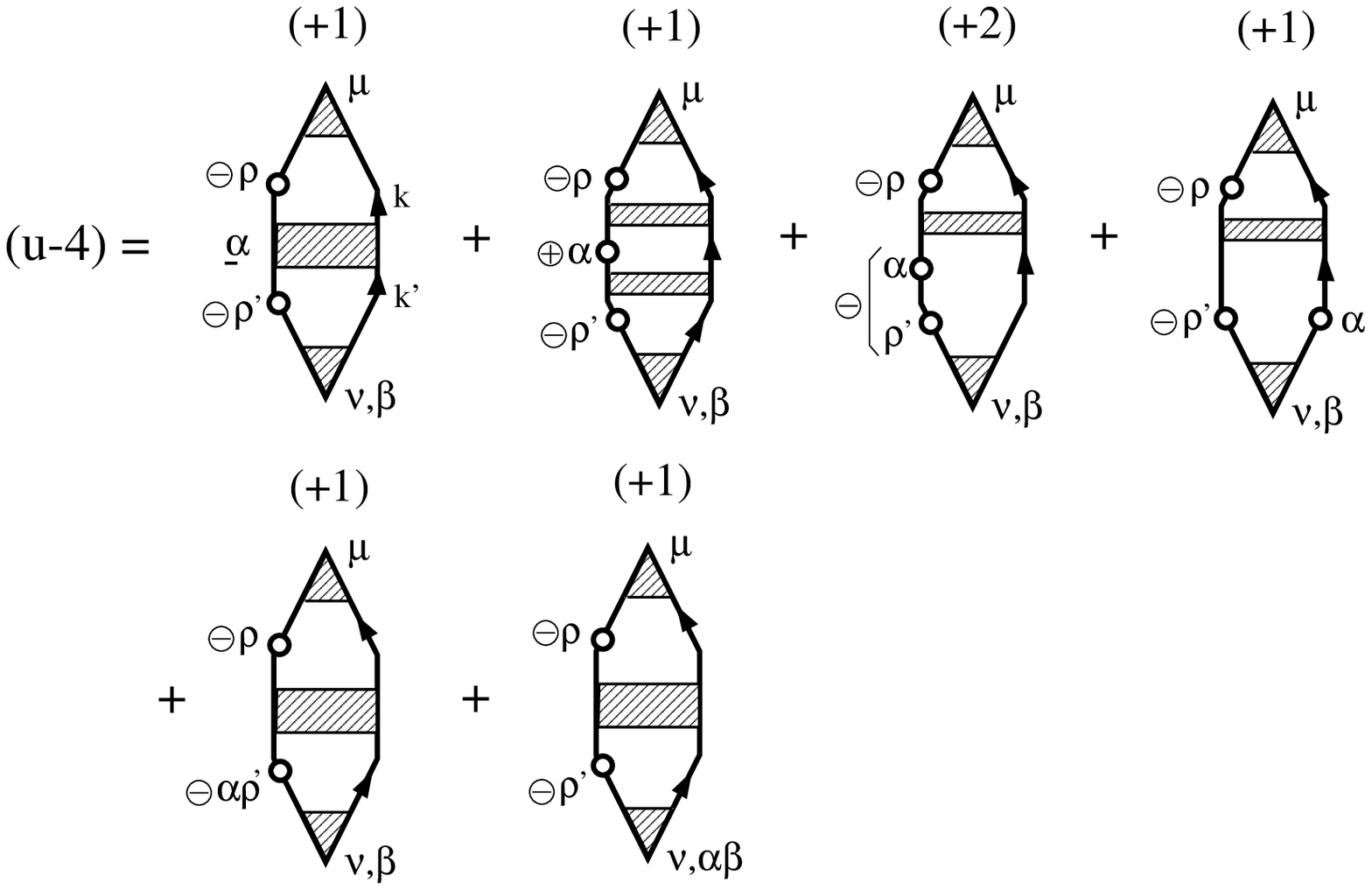,width=8cm}
\end{center}
\caption{}
{\small 
The digram for (u-4) in Fig.\ref{fig:DG-withVC} is expanded
by using a Ward identity, eq.(\ref{eqn:Ward2}).
}
\label{fig:DG-withVC-add1}
\end{figure}

\begin{figure}
\begin{center}
\epsfig{file=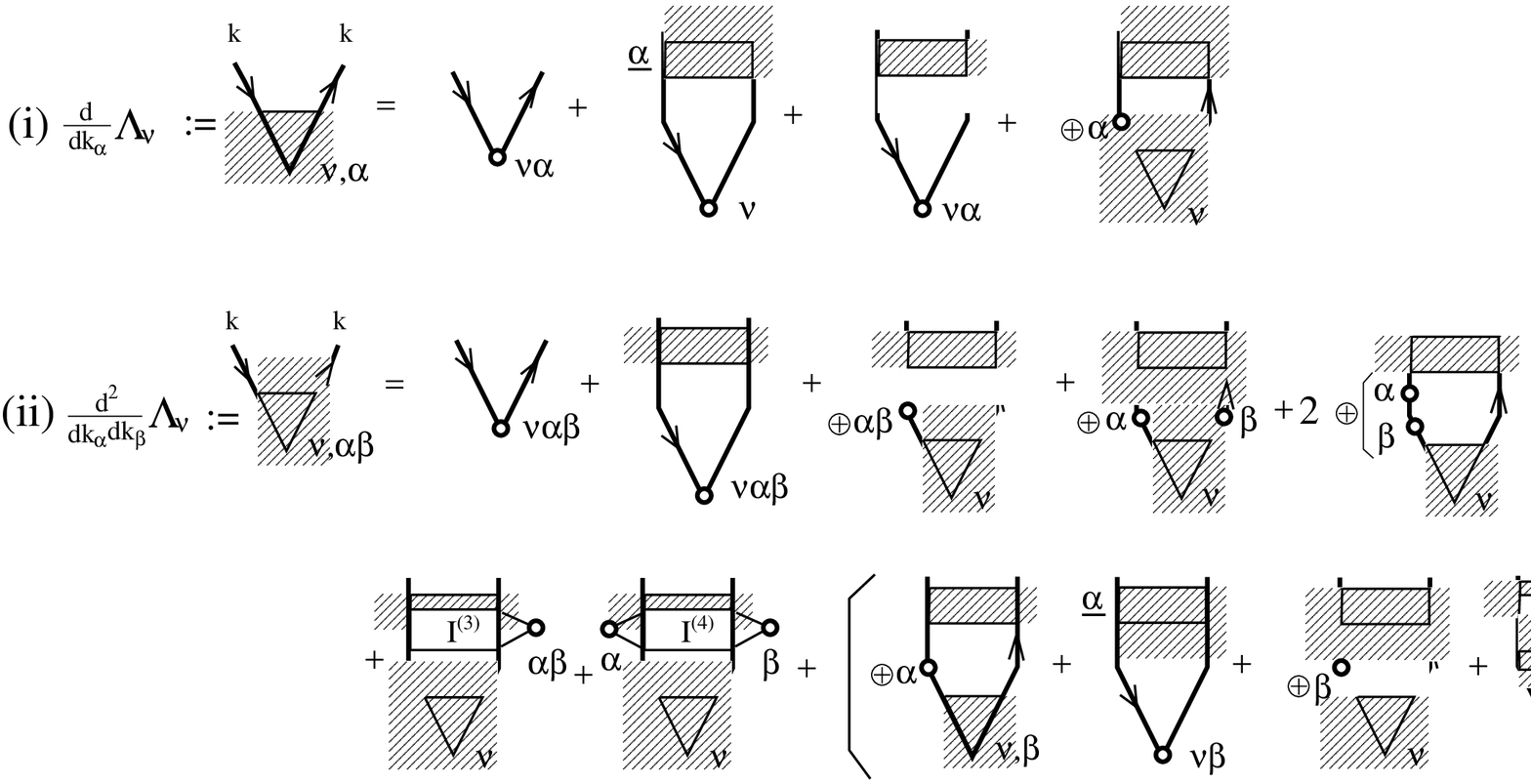,width=12cm}
\end{center}
\caption{The diagrammatic expression for 
(i) $\L_{\nu,\a}(\k) = \d_\a \L_\nu(\k)$ and 
(ii) $\L_{\nu,\a\b}(\k) = \d_{\a\b} \L_{\nu}(\k)$, respectively.
}
\label{fig:DG-withVC-add2}
\end{figure}

For example, we can show that
(u-2) in Fig. \ref{fig:DG-withVC},
is decomposed as Fig. \ref{fig:DG-withVC-add1}
by using a kind of Ward identity as follows:
\begin{eqnarray}
\d_\a \Gamma^I(\k|\k')
 &=& \Gamma^{I{\underline{\a}}}(\k|\k')
   \ - \ \d'_\a \Gamma^I(\k|\k'),
   \nonumber \\
\d_\a \Gamma(\k|\k')
 &=& \Gamma^{\underline{\a}}(\k|\k')
 +T\sum_{\k''\e''} \Gamma(\k|\k'')
  \cdot \d''_\a \{GG\} \cdot \Gamma(\k''|\k')
   \ - \ \d'_\a \Gamma(\k|\k')  ,
   \label{eqn:Ward2}
\end{eqnarray}
which are shown in Fig.\ref{fig:DG-Ward}.
To prove these identities,
we consider the $\k$-derivative of a function
$F(\k)\equiv \sum_{\k'}\Gamma(\k|\k')G(\k')$.
Then, $(\d/\d k_\a)F(\k)$ is given by 
taking the $\k$-derivative of every $G$
(i) on the 'connected line' between two $\k$-points in $F(\k)$
supposing that $\k$ runs only on the connected line, 
and (ii) on all the 'closed loops' after shifting 
all the momenta by $\k$ at the same time virtually.
After all, we see that $(\d/\d k_\a)F(\k)$ is given by
replacing each one of $G(\p)$'s in $F(\k)$ with $(\d/\d\p_\a)G$
and taking the summation of them.
Thus, eq.(\ref{eqn:Ward2}) is proved identically.
Note that the contribution from (ii) vanishes after all
 \cite{Nozieres}.
In the same way, we can derive the Ward identities for 
$\d_\a \L_\nu(\k;\k) \equiv \L_{\nu,\a}$ and 
$\d_{\a\b} \L_\nu(\k;\k) \equiv \L_{\nu,\a\b}$, which 
are expressed as Fig.\ref{fig:DG-withVC-add2}.
$\L_{\nu,\a\b}$ contains the irreducible
eight-point vertex $\G^{I(4)}$.

We note that 
$\L_{x,y}(\k;\k) \neq \L_{y,x}(\k;\k)$ 
whereas $\L_{\a\b}(\k;\k) = \L_{\a\b}(\k;\k)$.
As for $\L_{\a}$,
the diagrams with $v_\a^0$ on a closed loop cancel out
identically because of the Ward identity,
$\L_{\a} = \d_\a(\e_\k^0+\Sigma_\k)$.
On the other hand,
$\L_{\nu} \neq \d_\nu(\e_\k^0+\Sigma_\k)$
because the external frequency $\w$ is finite in the 
current study.
Because of this fact,
the diagrams with $v_\nu^0$ on a closed loop
in $\L_{\nu}$ do not vanish.


By using the identity in Fig. \ref{fig:DG-withVC-add2},
(u-4) in Fig.~\ref{fig:DG-withVC-add1} 
(with $\d_{\a}\G$ or $\d_{\a\b}\G$)
is decomposed into 42 diagrams which are consist of four, six and
eight-point vertices (without $\d_{\a}\G$ and $\d_{\a\b}\G$).
In the same way,
we can decompose all other terms in Fig.\ref{fig:DG-withVC}.
At the final stage of the proof, we verify the
one to one correspondence between them 
and all the terms of
$(\d^2/\d q_\rho \d q'_{\rho'}) \Phi^{(2)}(2\q+2q',\w_\l)$,
with signs and factors.
Note that each term in Fig.\ref{fig:DG-withVC-exp1}
is included in one of the diagrams (r-1)$\sim$(v-5)
in Figs. \ref{fig:DG-noVC1} and \ref{fig:DG-withVC},
as denoted under each term.
Especially, all the diagrams for 
(u-1), (v-1), (v-4) and (v-5) are already
appeared in Fig.\ref{fig:DG-withVC-exp1}.
In Appendix C,
we present the group of the terms in
$(\d^2/\d q_\rho \d q'_{\rho'}) \Phi^{(2)}(2\q+2q',\w_\l)$
which gives (u-2$\sim$4) and (v-2,3) precisely.
In the same way, one can check that the rest of terms in
$(\d^2/\d q_\rho \d q'_{\rho'}) \Phi^{(2)}(2\q+2q',\w_\l)$
give (r-1$\sim$4), (s-1,2) and (t-1$\sim$6) precisely.

In conclusion, we have proved that 
$(\d^2/\d q_\rho \d q'_{\rho'}) \Phi^{(2)}(2\q+2q',\w_\l)$
is exactly given by 
Figs. \ref{fig:DG-noVC1} and \ref{fig:DG-withVC}.
In the next section,
we perform its analytic continuation with respect to $\w_\l$,
and derive the exact expression for the MC of order $\tau^3$.

\section{Analytic continuation}
In this section, we perform the analytic continuation of 
$(\d^2/\d q_\rho \d q'_{\rho'})\Phi^{(2)}(2\q+2\q',\w_\l)$
obtained in the previous sections, and derive the coefficient 
of the $\w$-linear term, i.e., $C_{\mu\nu}^{\a\rho,\b\rho'}$
in eq.(\ref{eqn:sxx-expand}).
For this purpose,
we have to study the analytic continuation of the
following two types of functions:
\begin{eqnarray}
L_{\mu\nu}^{a}(\w_\l) &=& T\sum_{\e_n} \L_\mu(\e_n;\w_\l) 
 \cdot G(\e_{n}^+)G(\e_n) \cdot \L_\nu(\e_n;\w_\l)  ,
 \label{eqn:L11} \\
L_{\mu\nu}^{b}(\w_\l) &=& T^2\sum_{\e_n,\e_{n'}} 
 \L_\mu(\e_n;\w_\l) \cdot G(\e_{n}^+)G(\e_n)
 \cdot \G(\e_n^+,\e_n|\e_{n'},\e_{n'}^+) \cdot G(\e_{n'}^+)G(\e_{n'}) 
 \cdot \L_\nu(\e_{n'};\w_\l)  ,
  \label{eqn:L12}
\end{eqnarray}
where we dropped the momentum suffix for simplicity.
Figs. \ref{fig:DG-noVC1} and \ref{fig:DG-withVC} correspond to
$L_{\mu\nu}^{a}(\w_\l)$ and $L_{\mu\nu}^{b}(\w_\l)$, respectively.
We promise that several symbols for the momentum derivative of
Green functions are implicit in eqs.(\ref{eqn:L11}) and (\ref{eqn:L12}).

After the analytic continuation with respect to $\w_\l (\l>0)$,
the $\e_n$ and $\e_{n'}$-summations
are replaced with the integrations, depending on the region
1$\sim$3 in Fig. \ref{fig:region}.
As a result, we get ($x=a,b$)
\begin{eqnarray}
L_{\mu\nu}^{x}(\w) &=& \int\frac{d\e}{4\i\pi}
 \left[ {\rm th}\frac{\e}{2T} K_{\mu\nu}^{x,(1)}(\e;\w)
 + \left( {\rm th}\frac{\e^+}{2T} -{\rm th}\frac{\e}{2T} \right)
 K_{\mu\nu}^{x,(2)}(\e;\w) 
 - {\rm th}\frac{\e^+}{2T} K_{\mu\nu}^{x,(3)}(\e;\w) \right]
 \label{eqn:L-1}  , \\
K_{\mu\nu}^{a,(i)}(\e;\w)
 &=& J_\mu^{(i)}(\e;\w) g_i(\e;\w) J_\mu^{(i)}(\e;\w)
 \label{eqn:K-1}  , \\
K_{\mu\nu}^{b,(i)}(\e;\w)
 &=& J_\mu^{(i)}(\e;\w) g_i(\e;\w) \cdot T\sum_{\k',j=1,2,3} 
 \int\frac{d\e'}{4\pi\i} {\cal T}_{i,j}(\e^+;\e|\e';\e'^+)
  g_j(\e';\w) J_\mu^{(j)}(\e';\w)  ,
 \label{eqn:K-2} 
\end{eqnarray}
where we put $i,j=1,2,3$,
and ${\cal T}_{i,j}(\e^+;\e|\e';\e'^+)$
is given by the analytic continuation of 
$\G(\e_{n}^+;\e_n|\e_{n'};\e_{n'}^+)$
from the region $i$ for $\e_n$ 
and the region $j$ for $\e_{n'}$, respectively.
Its explicit expression is given in eq.(12) of Ref. 
 \cite{Eliashberg}.
Moreover, $J_{\mu}^{(i)}(\k,\e;\w)$ is defined as
\begin{eqnarray}
J_{\mu}^{(i)}(\k,\e;\w) = v_\mu^0(\k)+
 \sum_{\k',j=1,2,3} \int \frac{d\e'}{4\pi\i} 
 {\cal T}_{ij}(\k\e^+;\k\e|\k'\e';\k'\e'^+)
 g_j(\k',\e';\w) v_\mu^0(\k')  .
 \label{eqn:J-def}
\end{eqnarray}
Now we put $\w=0$ in eq.(\ref{eqn:J-def})
because we study the static conductivity 
$\s(\w\rightarrow0)$.
In this case, according to the Ward identity
 \cite{Eliashberg},
\begin{eqnarray}
J_{\mu}^{(1)}(\k,\e;0) &=& v_\mu(\k,\e) - \i\g_\mu(\k,\e)  ,
  \\
J_{\mu}^{(3)}(\k,\e;0) &=& v_\mu(\k,\e) + \i\g_\mu(\k,\e)  .
\end{eqnarray}
On the other hand, 
$J_{\mu}(\k,\e) \equiv J_{\mu}^{(2)}(\k,\e;0)$ can be rewritten as
\begin{eqnarray}
J_{\mu}(\k,\e) = v_\mu(\k,\e)+
 \sum_{\k'} \int\frac{d\e'}{4\pi\i} {\cal T}_{22}(\k\e|\k'\e')
 |G(\k',\e')|^2 v_\mu(\k',\e')  ,
 \label{eqn:J2}
\end{eqnarray}
which is derived by using the following relation
satisfied for $|\e|\sim T$:
\begin{eqnarray}
 v_\mu^0(\k) + \sum_{\k',j=1,3} \int \frac{d\e'}{4\pi\i} 
 {\cal T}_{2j}^{(0)}(\k\e|\k'\e') g_j(\k',\e';0) v_\mu^0(\k') 
 \ \approx \ v_\mu(\k,\e)  ,
\end{eqnarray}
where ${\cal T}_{2j}^{(0)}$ is the 'irreducible'
vertex with respect to $g_{2}$-section
 \cite{Eliashberg}.
Here we denote 
${\cal T}_{ij}(\k\e;\k\e|\k'\e';\k'\e')$ as
${\cal T}_{ij}(\k\e|\k'\e')$ for simplicity.

Based on the above argument,
we derive the general expression for the MC
up to the most singular contribution with respect to 
$\gamma^{-1}$.
At first, we consider the terms in Fig. \ref{fig:DG-noVC1},
which correspond to $L_{\mu\nu}^{a}(\w)$ in eq.(\ref{eqn:L-1}).
By doing the same argument in the previous sections,
the term $K_{\mu\nu}^{a,(2)}$ in eq.(\ref{eqn:L-1})
gives the most singular term, $O(\gamma^{-3})$.
Thus,  the expression for the MC
derived from $K_{\mu\nu}^{a,(2)}$,
${\mit\Delta}\s_{\mu\nu}^a$,
is given by eq.(\ref{eqn:MC-part1})
where $J_\mu(\k,\e)$ is replaced with eq.(\ref{eqn:J2}).
This expression is still of order $\g^{-3}$ 
because (the momentum derivative of) 
$J_{\mu(\nu)}$ is of order $\g^0$.

Next, we study the terms in Fig. \ref{fig:DG-withVC},
which are expressed as $L_{\mu\nu}^{b}(\w)$ in eq.(\ref{eqn:L-1}).
In this case, 
all the terms except $i=j=2$ turn out to vanish identically 
because eq.(\ref{eqn:K-2}) contains the following momentum derivatives 
of $G$ implicitly:
\begin{eqnarray}
& &[G^{R}\dalt_\a G^{R}] = 0  , \\
& &(\d_\a G^{R})\cdot( \d_\rho G^{R} )
 - \langle \a\leftrightarrow\rho\rangle = 0  .
\end{eqnarray}
In conclusion, we get the expression for MC 
which comes from Fig.\ref{fig:DG-withVC}, ${\mit\Delta}\s_{\mu\nu}^b$, 
as follows:
\begin{eqnarray}
 {\mit\Delta}\s_{\mu\nu}^b
 &=& B^2 \cdot \frac{e^4}{4v_B} \sum_\k \int \frac{d\e}{\pi}
  \left(-\frac{\d f}{\d\e}\right) 
   2\Bigl\{ 
   |G|^2 |{\rm Im}G| 
   J_\mu(v_x\d_y - v_y\d_x) 
     +|G|^4 W_\mu^{(2)} \Bigr\}_{(\k,\e)}
     \nonumber \\
 & &\times \sum_{\k'} \int \frac{d\e'}{4\pi\i}
  {\cal T}_{22}(\k\e|\k'\e')
  2\Bigl\{
     |G|^2 |{\rm Im}G| W_\nu^{(1)}
  +|G|^4 W_\nu^{(2)} \Bigr\}_{(\k',\e')}   ,
    \label{eqn:MC-withVC-pt2} 
\end{eqnarray}
where
\begin{eqnarray}
 & &W_\mu^{(1)}(\k,\e) = v_x(\k,\e) J_{\mu, y}(\k,\e)
             -v_{y}(\k,\e) J_{\mu, x}(\k,\e)  ,
    \nonumber \\
 & &W_\mu^{(2)}(\k,\e) = ( \ -v_x(\k,\e) \g_{y}(\k,\e)
             +v_{y}(\k,\e) \g_{x}(\k,\e) \ )J_\mu(\k,\e)  ,
\end{eqnarray}
where $J_\mu(\k,\e)$ is given by eq.(\ref{eqn:J2}),
and $J_{\mu,x}\equiv \d_x J_\mu$ and $J_{\mu,xy}\equiv \d_{xy} J_\mu$.
Moreover, $\d_x(\e_\k^0 + \Sigma^R(\k,\e))\equiv v_x(\k,\e) -\i\g_x(\k,\e)$.
Note that $J_{x,y}\neq J_{y,x}$.
In eq.(\ref{eqn:MC-withVC-pt2}), 
the factor 2 due to the spin degeneracy
has been included.
$v_B=(2\pi)^3$ in the cubic lattice.

\begin{figure}
\begin{center}
\epsfig{file=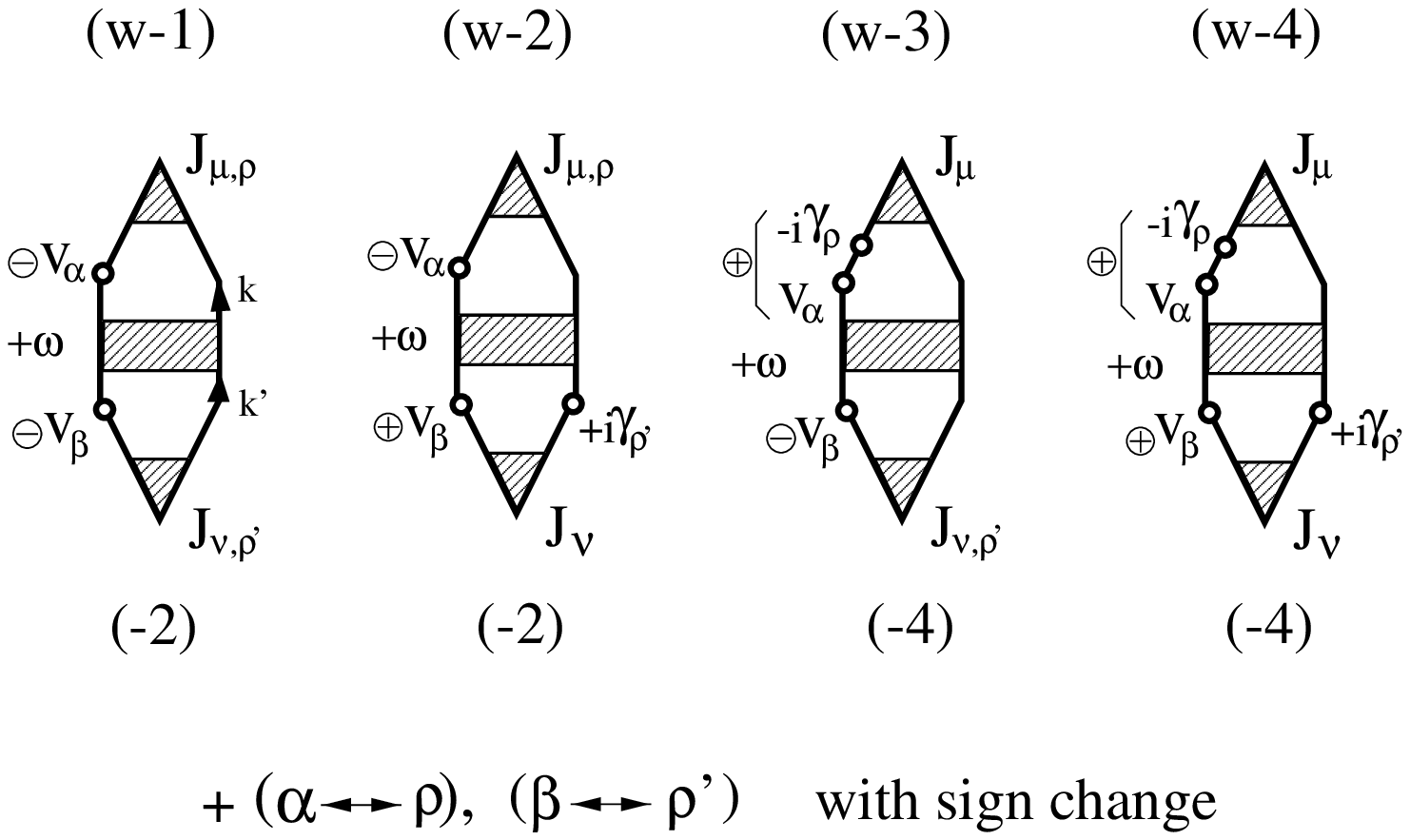,width=8cm}
\end{center}
\caption{}
\begin{center}
{\small 
The expressions for ${\mit\Delta}\s_{\mu\nu}^b$
in eq.(\ref{eqn:MC-withVC}).
They are of order $\g^{-3}$.
}
\end{center}
\label{fig:DG-withVC2}
\end{figure}

Equation (\ref{eqn:MC-withVC-pt2}) can be rewritten 
by performing the partial integration with respect to $\k$.
After the reformation of the equation,
we get the following simpler form:
\begin{eqnarray}
 {\mit\Delta}\s_{\mu\nu}^b
 &=& -B^2 \cdot \frac{e^4}{4v_B} \sum_\k \int \frac{d\e}{\pi}
  \left(-\frac{\d f}{\d\e}\right) 
  2\left\{ |G|^2 |{\rm Im}G| W_\mu^{(1)}
  - 2|G|^2 {\rm Re}G^2 W_\mu^{(2)} \right\}_{(\k,\e)}
     \nonumber \\
 & &\times \sum_{\k'} \int \frac{d\e'}{4\pi\i}
  {\cal T}_{22}(\k\e|\k'\e')
  2\left\{ |G|^2 |{\rm Im}G| W_\nu^{(1)} 
  +|G|^4 W_\nu^{(2)} \right\}_{(\k',\e')}   .
    \label{eqn:MC-withVC} 
\end{eqnarray}
The diagrammatic expression is shown in Fig.\ref{fig:DG-withVC2}.
They are of order $\g^{-3}$
because the factor $\g^{-2}$ comes from each $g_2$-lines,
and ${\cal T}_{22}$ is of order $\g$.

In conclusion,
the general expression for the MC in the Fermi liquid system,
which is exact in order $\g^{-3}$, is given by
\begin{eqnarray}
 {\mit\Delta}\s_{xx}
 = {\mit\Delta}\s_{xx}^a + {\mit\Delta}\s_{xx}^b  ,
    \label{eqn:MC-final} 
\end{eqnarray}
where ${\mit\Delta}\s_{\mu\nu}^a$ is given by eq.(\ref{eqn:MC-part1})
where $J_{\mu(\nu)}$ is given by eq.(\ref{eqn:J2}).

Next, we derived the simplified the expression of eq.(\ref{eqn:MC-final}) 
under the condition $\g^\ast \ll T$.
In this case, 
${\mit\Delta}\s_{xx}^a$ is given by eq.(\ref{eqn:MC-noVC-lowT})
where $J_{{\bf k}\mu(\nu)}$ is replaced with $J_{\mu(\nu)}(\k,\e=0)$ 
in eq.(\ref{eqn:J2}).
As for eq.(\ref{eqn:MC-withVC}), 
the following replacement are allowed:
$|G|^2|{\rm Im}G| \ \rightarrow \ \pi z_\k\delta(\e+\mu-\e_\k^\ast)/2\g_\k^2$,
$|G|^2{\rm Re}G^2 \ \rightarrow \ -\pi z_\k\delta(\e+\mu-\e_\k^\ast)/4\g_\k^3$,
and $|G|^4 \ \rightarrow \ \pi z_\k\delta(\e+\mu-\e_\k^\ast)/2\g_\k^3$.
Then, we can perform 
the energy integrations in eq.(\ref{eqn:MC-withVC}) easily.

After the long but straightforward calculation,
we obtain the following much compact expression for the MC
for $\g^\ast \ll T$:
\begin{eqnarray}
 {\mit\Delta}\s_{\mu\nu}
 &=& -B^2 \cdot \frac{e^4}{4v_B} \oint_{\rm FS} 
 \frac{dS_\k}{|{\vec v}_\k|} \frac{1}{\g_\k} d_\mu(k)D_\nu(k),
 \label{eqn:MC-total-lowT}   \\
 d_\mu(\k)
  &=& \left(v_{x}(\k,0)\frac{\d}{\d k_y}
           -v_{y}(\k,0)\frac{\d}{\d k_x} \right)
  \cdot \left(\frac{J_\mu(\k,0)}{\g(\k,0)} \right)
  \nonumber \\
  &=& W_\mu^{(1)}(\k,0)\g^{-1}(\k,0) + W_\mu^{(2)}(\k,0)\g^{-2}(\k,0)  ,
  \nonumber \\
 D_\mu(\k)
  &=& \frac1{v_B} \sum_{\k'}\int\frac{d\e'}{4\pi\i}
  {\cal T}_{22}(\k0|\k'\e') |G_{\k'}(\e')|^2 d_\mu(\k') 
  + d_\mu(\k)
   \nonumber \\
  &=& \frac1{v_B} \oint_{\rm FS} \frac{dS_{\k'}}{|{\vec v}_{\k'}|} 
    \int\frac{d\e'}{4\i}
  {\cal T}_{22}(\k0|\k'\e') 
   \frac1{\g_{\k'}} d_\mu(\k') + d_\mu(\k)  ,
   \nonumber 
\end{eqnarray}
where $\int_{\rm FS} dS_{\k}$ represents 
the two-dimensional integration on the Fermi surface.
Both $D_\mu(\k)$ and $d_\mu(\k)$ are real
because ${\cal T}_{22}$ is pure imaginary for $\w=0$.

On the other hand, the MC
within the relaxation time approximation is given by
\begin{eqnarray}
 {\mit\Delta}\s_{\mu\nu}^{\rm RTA}
 &=& -B^2 \cdot \frac{e^4}{4v_B} \oint_{\rm FS} 
 \frac{dS_\k}{|{\vec v}_\k|} \frac{1}{\g_\k} d_\mu^0(k)d_\nu^0(k),
 \label{eqn:MCRTA-total-lowT} \\
 d_\mu^0(\k)
 &=& \left(v_{x}(\k,0)\frac{\d}{\d k_y}
          -v_{y}(\k,0)\frac{\d}{\d k_x} \right)
 \cdot \left(\frac{v_\mu(\k,0)}{\g(\k,0)} \right).
 \nonumber
\end{eqnarray}
(see eq.(\ref{eqn:Boltzmann-Dsxx}).)
By comparing the exact formula (\ref{eqn:MC-total-lowT}) with
eq. (\ref{eqn:MCRTA-total-lowT}),
we see that $v_{\mu,\nu}$ is replaced with $J_{\mu,\nu}$
due to the vertex correction by ${\cal T}_{22}$.
Moreover, $d_{\mu}$ is replaced with $D_{\mu}$
due to another vertex ${\cal T}_{22}$.
Conversely, we get ${\mit\Delta}\s_{xx}^{\rm RTA}$ 
if we drop all ${\cal T}_{22}$'s in the exact expression for the MC,
eq.(\ref{eqn:MC-total-lowT}).
Note that $D_\mu(\k)$ is of order $\g^{-1}$,
as $d_\mu(\k)$ and $d_\mu^0(\k)$ are,
because ${\cal T}_{22}$ is of order $\g^{1}$.
This fact means that ${\mit\Delta}\s_{xx}$ 
given by eq.(\ref{eqn:MC-total-lowT})
is of order $\g^{-3}$.

Before concluding this section,
we consider a system with four-fold symmetry around the $z$-axis 
(${\vec e}_z\parallel {\vec B}$).
In this case, we get the expression 
\begin{eqnarray}
 {\mit\Delta}\s_{xx}
 = -B^2 \cdot \frac{e^4}{4v_B} \oint_{\rm FS} 
 \frac{dS_\k}{|{\vec v}_\k|} \frac{1}{\g_\k} 
 {\vec d}_\perp(k)\cdot {\vec D}_\perp(k) /2  ,
  \label{eqn:MC-sym}
\end{eqnarray}
where ${\vec d}_\perp = (d_x,d_y)$ and 
${\vec D}_\perp = (D_x,D_y)$.
Equation (\ref{eqn:MC-sym}) means that ${\mit\Delta}\s_{xx}$
is independent of the choice of the $x$-axis 
on the $xy$-plane.

%
%
\section{Discussion and Conclusions}

\subsection{The Role of the Vertex Corrections}

In this subsection, we analyze the obtained exact formula
for MC, eq.(\ref{eqn:MC-total-lowT}),
and discuss the role of the vertex corrections (VC's) by ${\cal T}_{22}$.
As shown in eqs. (\ref{eqn:Eliashberg}), (\ref{eqn:Kohno})
and (\ref{eqn:MC-total-lowT}),
$\s_{xx}$, ${\mit\Delta}\s_{xy}$ and ${\mit\Delta}\s_{xx}$ 
have one, two and three ${\cal T}_{22}$'s, respectively.

At first, we discuss an isotropic correlated electron system 
without any Umklapp processes.
In this case, 
Yamada and Yosida proved that 
$J_{\mu}^{(2)}(\k,\e;\w)$ given by eq.(\ref{eqn:J-def})
diverges in the static limit ($\w=0$)
 \cite{Yamada}.
This fact brings the divergence of the static conductivities
$\s_{xx}$, ${\mit\Delta}\s_{xy}$ and ${\mit\Delta}\s_{xx}$ 
even at finite temperatures,
reflecting the momentum conservation law.
On the other hand, 
$J_{\mu}^{(2)}(\k,\e;\w)$ gives the total current
with the 'backflow' in the case of $\g_k^\ast\ll\w\ll T$,
or in the 'zero-sound regime'
 \cite{Nozieres,AGD}.
In this case ${\vec J}_{{\bf k}}$ is finite.
According to our formula (\ref{eqn:MC-total-lowT}),
${\vec J}_{{\bf k}} = C_{\rm U} {\vec v}_{{\bf k}}$,
${\vec d}_{{\bf k}} = C_{\rm U} {\vec d}_{{\bf k}}^{\, 0}$, and
${\vec D}_{{\bf k}} = C_{\rm U}^2 {\vec d}_{{\bf k}}^{\, 0}$,
where $C_{\rm U}$ is the 'Umklapp coefficient' 
coming from ${\cal T}_{22}$
 \cite{Yamada}.
$C_{\rm U}$ is finite for finite $\w$.
This fact means that
${\mit\Delta}\rho(\w) = 0$ up to $B^2$
even in the zero-sound regime
in an isotropic system.

We also discuss the conductivity
of an isotropic free electron system with weak disorders,
within the self-consistent Born approximation.
Then, it is known that
$C_{\rm U}$ coming from ${\cal T}_{22}$ 
becomes the 'geometric factor'.
In this approximation,
$\gamma = n_{\rm imp}\sum_{\k'}|V_{\k-\k'}|^2\pi\rho_{\k'}(0)$
where $V_{\k}$ is the Fourier transformation of the impurity potential.
According to the Ward identity, 
${\cal T}_{22}$ is given by the ladder-type vertex corrections.
By solving the Bethe-Salpeter equation, we get
$C_{\rm U}^{-1} = n_{\rm imp}\sum_{\k'}|V_{\k-\k'}|^2\pi\rho_{\k'}(0)
 (1-{\rm cos}\theta_{\k-\k'})/2\g$.
In conclusion, the conductivities are given by
$\s_{xx}= C_{\rm U} \cdot ({ne^2}/{2m\g})$, 
${\mit\Delta}\s_{xy}= C_{\rm U}^2 \cdot ({ne^3}/{4m^2\g^2})$ and 
${\mit\Delta}\s_{xx}= C_{\rm U}^3 \cdot ({ne^4}/{8m^3\g^3})$,
respectively.
As a result, the relations
$R_{\rm H}=1/ne$ and ${\mit\Delta}\rho=0$ up to $B^2$, 
which are equal to the results the RTA,
hold in the self-consistent Born approximation
if we take the VC's into account correctly,
because of the cancellation of $C_{\rm U}$'s.

Finally,
we discuss the role of the VC's
in anisotropic systems,
which will be much important.
For example,
the momentum dependence of ${\vec J}_{\bf k}$
in nearly AF Fermi liquids
becomes singular singular due to the VC's by ${\cal T}_{22}$
 \cite{Kontani,Kanki,Nagoya,BEDT}.
This fact explains the {\it seemingly} non-Fermi liquid behavior 
of the Hall coefficient in high-$T_{\rm c}$ cuprates successfully,
without assuming any non-Fermi liquid ground state
 \cite{Anderson}.
 
It is natural to expect that the VC's by 
${\cal T}_{22}$ are also important for the MR.
In high-$T_{\rm c}$ cuprates, 
the Kohler's rule ${\mit\Delta}\rho/\rho \propto \rho^{-2}$
is strongly violated in a wide range of temperatures
 \cite{Kimura}.
Based on the exact expression derived in this paper,
we recently studied this long-standing problem 
by using the spin-fluctuation theory, 
with including all the VC's to keep the conserving laws
 \cite{Future}.
We found the approximate relation
${\mit\Delta}\rho/\rho \propto \xi_{\rm AF}^4 \cdot \rho^{-2}$ 
due to the VC's by ${\cal T}_{22}$
in the presence of AF fluctuations
($\xi_{\rm AF}$ being the AF correlation length).
We confirmed that the factor $\xi_{\rm AF}^4$,
which comes from the VC's,
well explains the violation of the Kohler's rule
in high-$T_{\rm c}$ cuprates.

\subsection{Concluding Remarks}

In this paper,
we derived the general expression for the MC
based on the Fermi liquid theory from the Kubo formula.
We treated the vertex corrections in the exact way
by following the Ward identities.
The obtained expression is given by 
eq.(\ref{eqn:MC-final}), which is exact up to $O(\g^{-3})$.
This expression can be simplified as 
eq.(\ref{eqn:MC-total-lowT}) in the case of $\g^\ast \ll T$.
However, we should use
eq.(\ref{eqn:MC-final}) to obtain the reliable result
when $\gamma_\k^\ast \simle T$ is realized like in
high-$T_{\rm c}$ cuprates.
In all these expressions,
we put the factor 2 due to the spin-degeneracy,
assuming the paramagnetic state without the magnetic field.
Fortunately, 
these obtained expressions do not contain
six and eight-point vertices, although
the original diagram for $\Phi^{(2)}$ given by Fig. \ref{fig:DG-init2}
have them.
This fact makes the numerical calculation easier.

It is noteworthy that we can calculate the MC in the conserving way
by virtue of the obtained formula 
(\ref{eqn:MC-final}) or (\ref{eqn:MC-total-lowT}):
In the purpose, 
we get at first the proper 'irreducible vertex'
through the functional derivative;
${\hat \G}^I \equiv \delta{\hat \Sigma}/\delta{\hat G}$.
Next, we perform the analytic continuation
and get the irreducible vertex ${\cal T}_{22}^{\rm I}$.
Finally, we calculate the full vertex correction ${\cal T}_{22}$
through the integral equation
like eq.(\ref{eqn:full-vertex}),
and input it in eq.(\ref{eqn:J2}) for $J_\mu(\k,\e)$ and
in eq.(\ref{eqn:MC-total-lowT}) for $D_\mu(\k)$.
This procedure is classified as the 'conserving approximation'
developed by Baym and Kadanoff,
which is indispensable to derive reliable transport phenomena
 \cite{Baym}.
In future, we can apply the present work 
to the study of interesting magnetotransport phenomena
in various kinds of strongly correlated electron systems.

In the obtained expression,
the $z$-axis is taken to be the direction of the magnetic field ${\vec B}$.
We obtain the 'transverse' MC
if we put $\mu=\nu$ on the $x,y$-plane.
The transverse MR is given by eq.(\ref{eqn:coeff-def}).
In the same way, we get the 'longitudinal' MC
if we put $\mu=\nu=z$.
In this case, the longitudinal MR is given by
${\mit\Delta}\rho/\rho = -{\mit\Delta}\s_{zz}/\s_{zz}^0$
because of ${\mit\Delta}\s_{zx} = 0$.
In both cases, ${\mit\Delta}\s \sim O(\tau^3)$
and ${\mit\Delta}\rho/\rho \sim O(\tau^2)$, respectively.
In usual three-dimensional systems,
transverse MR and longitudinal MR are of the same order
 \cite{Ziman}.
In quasi-two dimensional system like high-$T_{\rm c}$ cuprates,
on the other hand, the longitudinal MR on the $ab$-plane
in the case of ${\vec B}\parallel{\vec a}$,
$({\mit\Delta}\rho_a/\rho_a)_\parallel$,
is much smaller than the transverse one 
in the case of ${\vec B}\parallel{\vec c}$,
$({\mit\Delta}\rho_a/\rho_a)_\perp$.
Their ratio will be about $O(t_c^2/t_a^2)$, where $t_c$($t_a$)
is the hopping integral along the $c$($a$)-axis.

Because we discussed the MC (MR) up to $O(B^2)$,
our formula will be valid under the weak magnetic field,
$\max_\k \{\w_c/\g_\k^\ast\}\ll 1$,
where $\w_c \equiv eB/mc$ is the cyclotron frequency.
In general, this condition is satisfied for $B\ll1$Tesla
in good metals at low temperatures.
In optimally-doped high-$T_{\rm c}$ cuprates, however,
$\max_\k \{\w_c/\g_\k^\ast\} \ll 1$ will be satisfied
even for $B\sim1$Tesla because 
$\max_\k \{1/\g_\k^\ast\} \simle T$ is expected above $T_{\rm c}$.
In this article, we did not consider the MR caused by the 
spin-dependence of $\g_{{\bf k}\s}^\ast$ due to the Zeeman effect, 
because it is negligible in high-$T_{\rm c}$ cuprates
 \cite{Kimura}.
However, this spin-effect may cause the negative MR 
in the vicinity of the ferromagnetic instability
 \cite{Arita}.

Finally, we make short comments on the reliableness of the
calculation: We used the MATHEMATICA for the check of some parts
of the calculation.  
Moreover,
(1) The gauge invariance of the obtained expression for 
$\Delta\s_{xx}$ is a severe verification of its correctness,
because the gauge invariance is violated if we modify only
one of the coefficient of diagrams for $\Delta\s_{xx}$.
We have to mistake at least four coefficients of diagrams
accidentally to recover the gauge invariance.
(2) The obtained $\Delta\s_{xx}$ coincides with 
$\Delta\s_{xx}^{\rm RTA}$ if we drop all the vertex corrections.
(3) The exact expression for $\Delta\s_{xy}$ up to $O(\tau^2)$,
which was originally obtained in Ref. \cite{Kohno},
can be derived much easier by using the technique 
developed in this article.


\vspace{5mm}
\begin{center}
{\bf acknowledgment}
\end{center}

The author are grateful to 
D. Vollhardt for enlightening discussions.
He is also thankful to K. Yamada, P. W{\"o}lfle, W. Metzner 
and W. Hofstatter for useful comments and discussions.
Finally, he thank many participants 
in 'XII Workshop on Strongly Correlated Electron Systems'
at ICTP in Trieste for stimulating discussions.

\appendix
\section{The Hamiltonian and the Current in the Magnetic Field}
In this appendix,
we derive the expression for 
eqs. (\ref{eqn:Ham-B}), (\ref{eqn:curr-B}) and (\ref{eqn:JBqq})
in \S II.
Here we study the one-dimensional system
with only the nearest neighbor hopping for simplicity.
It is easy to verify that 
the obtained results (\ref{eqn:A-Ham-B})-(\ref{eqn:A-curr-B})
are valid for any dimensional systems
with arbitrary range of hoppings.

In the case of $eA\ll1$,
the Peierls phase factor for $t_{l,l+1}$,
eq.(\ref{eqn:hopping}), can be expanded as

\begin{eqnarray}
 {\rm exp}(\pm\i e(A_{l+1}^{\rm tot}+A_l^{\rm tot})/2)
 &=& 1 \pm \i\frac{e}{2}A(e^{\i q}+1)e^{\i lq}
       \pm \i\frac{e}{2}A'(e^{\i q'}+1)e^{\i lq'}
    \nonumber \\
 & &- \frac{e^2}{4} AA'(e^{\i q}+1)(e^{\i q'}+1)e^{\i l(\q+\q')}
    \nonumber \\
 & &+ \ \cdots  .
\end{eqnarray}
By expanding $t_{ij}$ and $t_{ij}^\ast$
in eq.(\ref{eqn:hopping}) in the same way,
and by performing the Fourier transformation,
we get the following result:
\begin{eqnarray}
H_B &=& H_{B=0} 
 + eA_\a \sum_\k \frac12 \left(v_{\a,\k+\q/2}^0+v_{\a,\k-\q/2}^0 \right)
 c_{\k+\q/2}^\dagger c_{\k-\q/2}
 + \langle (\a,\q)\rightarrow(\b,\q')\rangle
   \nonumber \\
& & + e^2 A_\a A'_\b \sum_\k 
  \left( \frac14 \sum_{s,s'}^{-1,1} v_{\a\b,\k+(s\q+s'\q')/2}^0 \right)
 c_{\k+(\q+\q')/2}^\dagger c_{\k-(\q+\q')/2}  .
  \label{eqn:A-Ham-B}
\end{eqnarray}
By expanding $v_{\a,\k\pm\q/2}^0$ and $v_{\a\b,\k\pm(\q\pm\q')/2}^0$
in eq.(\ref{eqn:A-Ham-B})
with respect to $q_\rho$ and $q'_{\rho'}$,
we see that eq.(\ref{eqn:Ham-B}) in \S II
is exact up to $O(q,q')$ and $O(qq')$.

Next, we consider the current operator in the magnetic field,
whose definition is given by eq.(\ref{eqn:jB-definition}).
We can derive that 
\begin{eqnarray}
 j(m) &=& \i[H_B^0,mc_m^\dagger c_m]
   \nonumber \\
  &=& \i t \cdot m \left\{
   e^{\i e(A_m+A_{m-1})/2}c_{m-1}^\dagger c_m
  -e^{\i e(A_m+A_{m+1})/2}c_{m}^\dagger c_{m+1} \right.
   \nonumber \\
  & &+ \left. e^{-\i e(A_m+A_{m+1})/2}c_{m+1}^\dagger c_m
     -e^{-\i e(A_m+A_{m-1})/2}c_{m}^\dagger c_{m-1} \right\}  ,
\end{eqnarray}
where the integer $m$ denotes the site index.
By performing the Fourier transformation, we get
\begin{eqnarray}
 j(p) &=& \sum_m j(m)e^{\i pm}
   \nonumber \\
  &=& \frac{\d}{\d p} \cdot t \sum_{m\k\k'}
   c_\k^\dagger c_{\k'} e^{\i(-\k+\k'+p)m} \left\{
   e^{\i e(A_m+A_{m-1})/2}e^{\i\k}
  -e^{\i e(A_{m+1}+A_{m})/2}e^{\i\k'}
   \right. \nonumber \\
   & &\left.
  +e^{-\i e(A_{m+1}+A_{m})/2}e^{-\i\k}  
  -e^{-\i e(A_m+A_{m-1})/2}e^{-\i\k'}  \right\}  .
  \label{eqn:A-j1}
\end{eqnarray}
After the straightforward calculation from eq.(\ref{eqn:A-j1}),
we obtain the following expression for $j_\nu(p)$:
\begin{eqnarray}
 j_\nu(p) &=& \frac{\d}{\d p_\nu} \sum_{\k} 
  c_{\k+\p/2}^\dagger c_{\k-\p/2} 
  (\e_{\k+\p/2}^0 - \e_{\k-\p/2}^0) 
    \nonumber \\
 & &+ \frac{\d}{\d p_\nu} eA_\a \sum_{\k} 
  c_{\k+(\p+\q)/2}^\dagger c_{\k-(\p+\q)/2} \cdot \frac12 
  \sum_{s}^{-1,1}
  \left( v_{\a\b,\k+(\p+s\q)/2}^0 - v_{\a\b,\k-(\p+s\q)/2}^0 \right)
    \nonumber \\
 & &+ \langle(\a,\q)\rightarrow(\b,\q')\rangle
     \nonumber \\
 & &+ \frac{\d}{\d p_\nu} e^2A_\a A'_\b \sum_{\k} 
  c_{\k+(\p+\q+\q')/2}^\dagger c_{\k-(\p+\q+\q')/2} \cdot 
  \frac14 \sum_{s,s'}^{-1,1}
  \left( v_{\a\b,\k+(\p+s\q+s'\q')/2}^0 
       - v_{\a\b,\k-(\p+s\q+s'\q')/2}^0 \right)
     \nonumber \\
 & &\ + \ \cdots ,
   \label{eqn:A-curr-B}
\end{eqnarray}
By expanding $v_{\a,\k\pm\q/2}^0$, $v_{\a\b,\k\pm(\q\pm\q')/2}^0$
and $v_{\a\b,\k\pm(\p\pm\q\pm\q')/2}^0$
in eq.(\ref{eqn:A-curr-B})
with respect to $q_\rho$ and $q'_{\rho'}$,
we can check that eq.(\ref{eqn:curr-B}) in \S II
is exact up to $O(q,q')$ and $O(qq')$.

Finally, we comment that 
another unquestionable definition for the current operator 
is ${\vec j} = \d H_B^0 / \d {\vec A}$
 \cite{Fradkin}.
Based on this definition, we can also derive 
eqs.(\ref{eqn:curr-B}) and (\ref{eqn:JBqq})
exactly up to $O(q,q')$ and $O(qq')$.

\section{The MC of Order $\g^{-2}$: The Next Singular Contribution}
\begin{figure}
\begin{center}
\epsfig{file=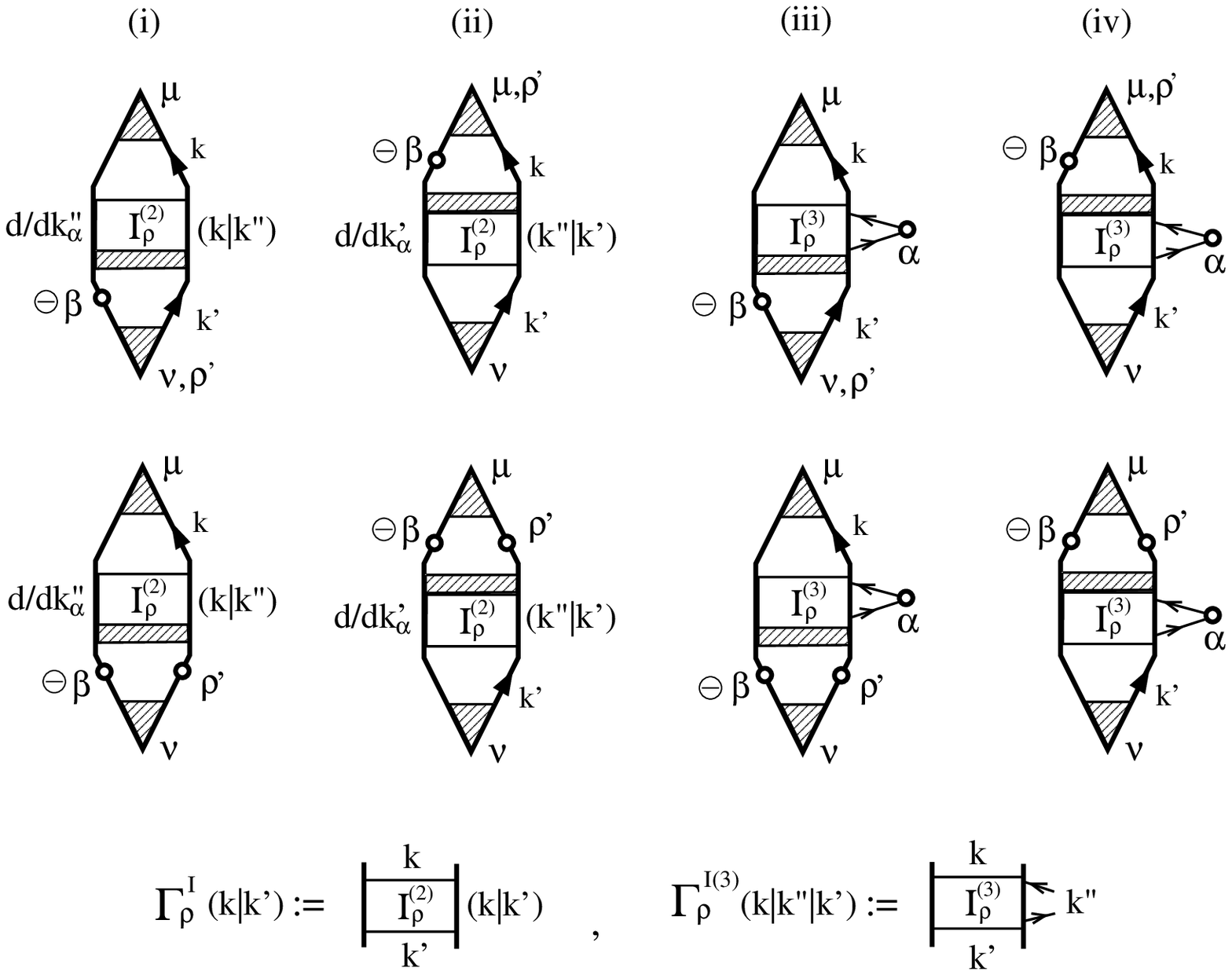,width=8cm}
\end{center}
\caption{The diagrams for the MC of order $\g^{-2}$,
all of which contain the $\q$-derivative of four-point 
vertices shown in Fig.\ref{fig:DG-qexpand}
(ii) or (iii).
The terms with the exchanges $\b\leftrightarrow\rho'$  
with negative signs also exist.}
\label{fig:DG-withVC-correction}
\end{figure}

In this appendix,
we study the most divergent terms which come from the 
$q$-derivative of the irreducible vertices
(ii) and (iii) in Fig. \ref{fig:DG-qexpand};
$\G_\rho^I$ and $\G_\rho^{I(3)}$.
For this purpose,
we take the $\q\q'$-derivative of 
$\Phi^{(2)}(2\q+2\q')$ given in 
Figs.\ref{fig:DG-init1} and \ref{fig:DG-init2},
and gather all the diagrams which contain 
$\G_\rho^I$ or $\G_\rho^{I(3)}$,
which have been neglected in the discussion in \S IV.
This procedure produces a huge number of complicated diagrams.
Fortunately, we find that they are collected into
eight diagrams shown in Fig. \ref{fig:DG-withVC-correction},
by using the Ward identity.
In Fig. \ref{fig:DG-withVC-correction},
only the most divergent terms are presented.
This procedure is similar to that developed in \S III.

For example, the expressions for (i) and (iii) of 
Fig. \ref{fig:DG-withVC-correction} are given by,
\begin{eqnarray}
{\mbox{(i)}} &=& T^3\sum_{\k\k'\k''}
\L_\mu(\k) GG^+\cdot \d''_\a\G_{\rho}^I(\k|\k'')
 \cdot (\delta_{\k'',\k'}/T+GG^+\cdot \G) 
  \nonumber \\
& & \cdot 
([G{\dalt}_\b^{\> \prime} G^+] \L_{\mu,\rho'}(\k') + 
(\d'_{\rho'}G\cdot\d'_{\b}G^+ - \d'_{\b}G\cdot\d'_{\rho'}G^+ )
 \L_{\mu}(\k') )  ,
 \label{eqn:B-1} \\
{\mbox{(iii)}} &=& T^4\sum_{\k\k'\k''\k'''}
\L_\mu(\k) GG^+\cdot \G_{\rho}^{I(3)}(\k|\k''|\k''') 
 \cdot \d'''_\a G \cdot 
(\delta_{\k'',\k'}/T+GG^+\cdot \G) 
\nonumber \\
& & \cdot
([G{\dalt}_\b^{\> \prime} G^+] \L_{\mu,\rho'} + 
(\d'_{\rho'}G\cdot\d'_{\b}G^+ - \d'_{\b}G\cdot\d'_{\rho'}G^+ ) 
 \L_{\mu} )  ,
 \label{eqn:B-2}
\end{eqnarray}
where 
$\G_{\rho}^I$ and $\G_{\rho}^{I(3)}$ 
are introduced in Fig. \ref{fig:DG-qexpand}.
After performing the analytic continuation of them
in the same way as $L_{\mu\nu}^b$ in eq.(\ref{eqn:L-1}),
the term $G^+G \rightarrow G^R G^A$ gives the most singular contribution.
By using the relation (\ref{eqn:qp})
together with the fact that ${\cal T}_{22}$ is of order $\g$,
we can clearly show that 
eqs.(\ref{eqn:B-1}) and (\ref{eqn:B-2}) are 
of order $\g^{-2}$
In conclusion, we can drop 
Fig.\ref{fig:DG-withVC-correction} safely in usual metals,
where $\g_\k^\ast$ is much smaller than the Fermi energy.

We can also show that 
Fig.\ref{fig:DG-withVC-correction} gives the exact
$O(\g^{-2})$-terms of the MC, 
and any diagrams which are not included in
Figs.\ref{fig:DG-noVC1}, \ref{fig:DG-withVC} and 
\ref{fig:DG-withVC-correction} are at most $O(\g^{-1})$.
We note that the expression in Fig.\ref{fig:DG-withVC-correction}
should be gauge-invariant because we take all the terms 
$O(\gamma^{-2})$ into account,
although it is not written in a gauge-invariant form
explicitly.

\def\FIGA{{\ref{fig:DG-noVC1}}}
\def\FIGB{{\ref{fig:DG-withVC}}}
\section{Derivation of Figs. \FIGA \ and \FIGB}

In order to derive 
$(\d^2/\d q_\rho \d q'_{\rho'}) \Phi^{(2)}(2\q+2\q',\w)$,
we have to take the $\q,\q'$-derivative of all the terms
Figs. \ref{fig:DG-init1} and \ref{fig:DG-init2}.
In this appendix,
we present the part of them which give 
(u-1$\sim$4) and (v-1$\sim$5) in Fig.\ref{fig:DG-withVC} explicitly.
Note that (u-1), (v-1), (v-4) and (v-5)
have already been included in Fig.\ref{fig:DG-withVC-exp1}.

\begin{figure}
\begin{center}
\epsfig{file=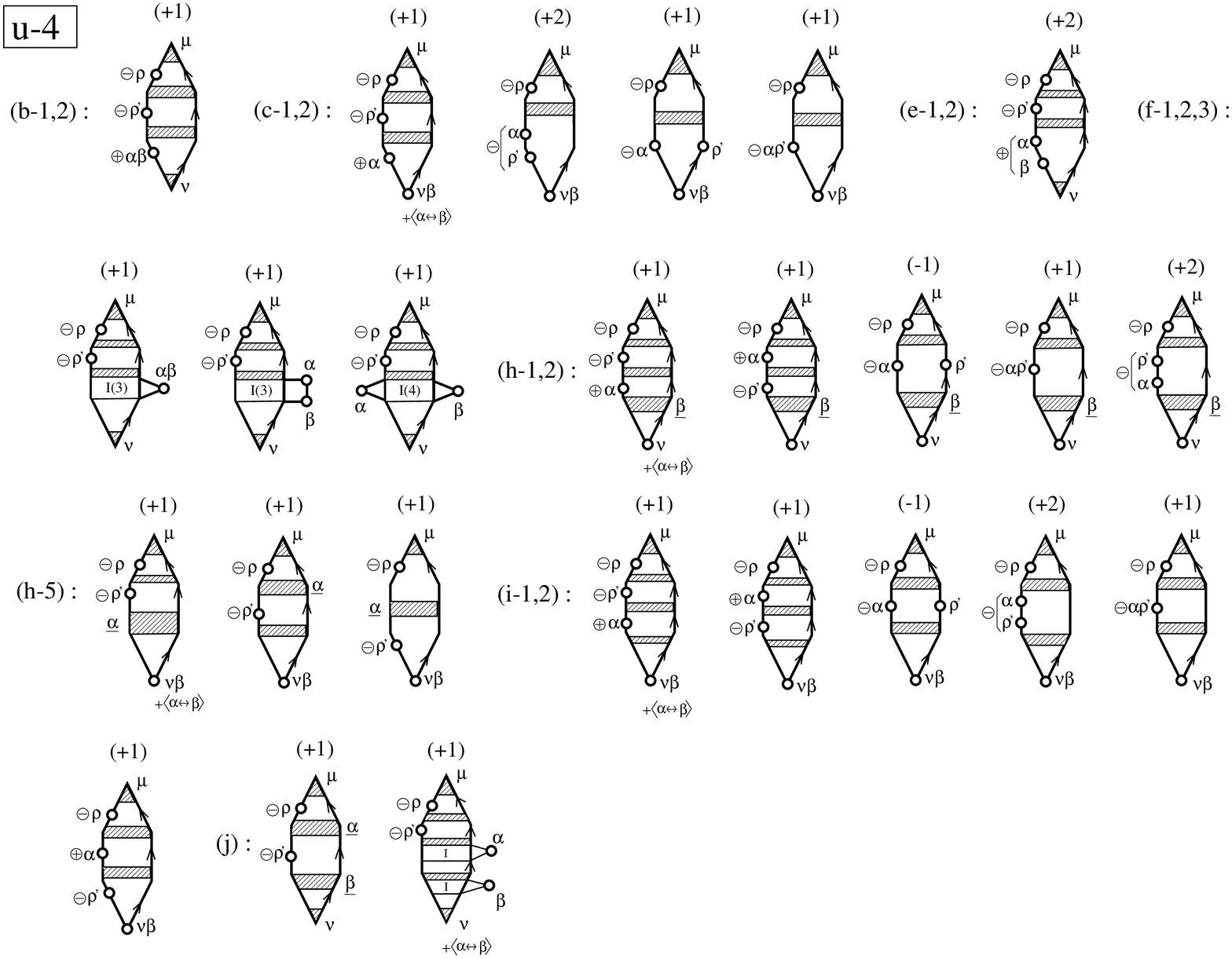,width=14cm}
\end{center}
\caption{
The part of the diagrams in 
$(\d^2/\d q_\rho \d q'_{\rho'}) \Phi^{(2)}(2\q+2\q',\w)$.
These diagrams together with the terms with the notation
'$\langle\mbox{u-4}\rangle$' in Fig.\ref{fig:DG-withVC-exp1}
give (u-4) in Fig.\ref{fig:DG-withVC}.
}
\label{fig:DG-u4-append}
\end{figure}

At first,
we concentrate on (u-4) in Fig.\ref{fig:DG-withVC}.
By using the Ward identity, eq.(\ref{eqn:Ward2}),
(u-4) is decomposed into Fig. \ref{fig:DG-withVC-add1}.
There, $\L_{\nu,\a}$ and $\L_{\nu,\a\b}$ 
are expressed as Fig. \ref{fig:DG-withVC-add2},
which are also kind of the Ward identities.
Thus, (u-4) is decomposed into 42 diagrams in total.
On the other hand,
Fig. \ref{fig:DG-u4-append}
show the part of the terms coming from 
$(\d^2/\d q_\rho \d q'_{\rho'}) \Phi^{(2)}(2\q+2\q',\w)$.
Then,
we can recognize that the diagrams in Fig.\ref{fig:DG-u4-append}
together with the terms with the notation 
'$\langle\mbox{u-4}\rangle$' in Fig.\ref{fig:DG-withVC-exp1}
give (u-4) precisely,
by checking the one-to-one correspondence between them.
In the procedure, the symmetry 
$(\a,\rho)\leftrightarrow(\b,\rho')$ 
should be taken into account carefully.

\begin{figure}
\begin{center}
\epsfig{file=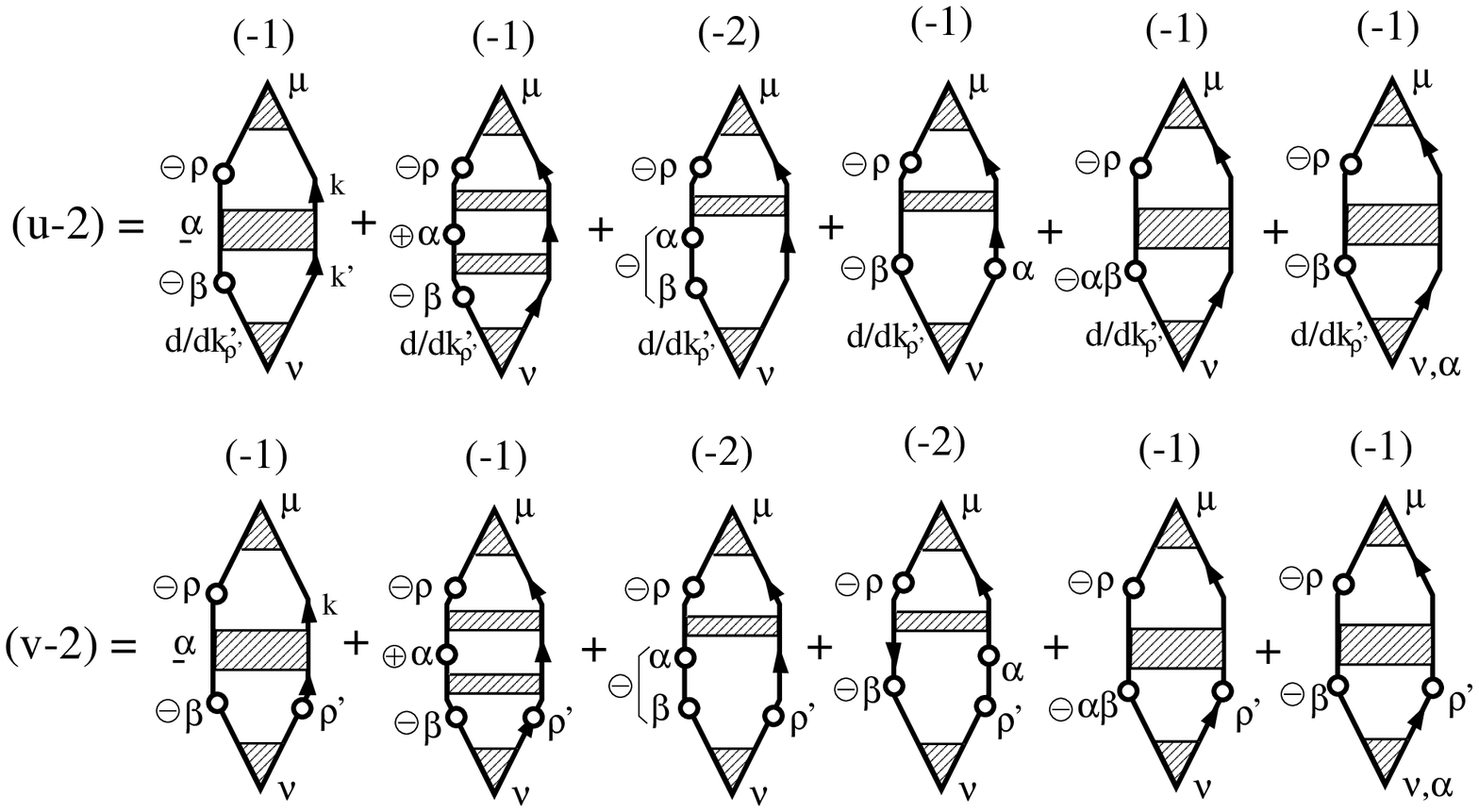,width=8cm}
\end{center}
\caption{
The expanded expressions for (u-2) and (v-2) 
given in Fig.\ref{fig:DG-withVC}.
They are derived by using a Ward identity,
eq.(\ref{eqn:Ward2}).
}
\label{fig:DG-u2-append2}
\end{figure}
\begin{figure}
\begin{center}
\epsfig{file=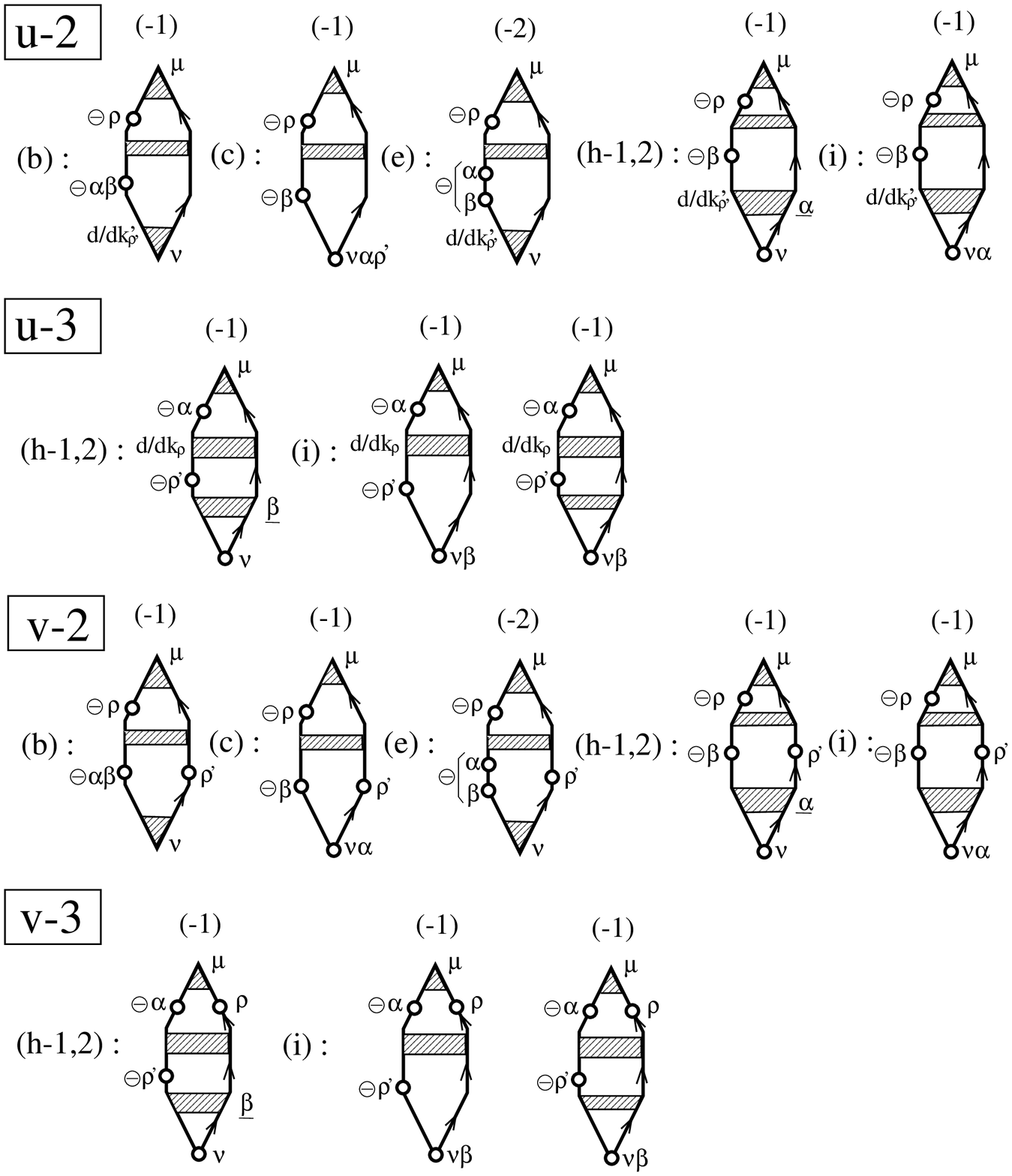,width=8cm}
\end{center}
\caption{
The part of the diagrams in
$(\d^2/\d q_\rho \d q'_{\rho'}) \Phi^{(2)}(2\q+2\q',\w)$.
These diagrams together with the corresponding terms in
Fig.\ref{fig:DG-withVC-exp1}
give (u-2), (u-3), (v-2) and (v-3) in Fig.\ref{fig:DG-withVC}.
}
\label{fig:DG-u2-append}
\end{figure}
\begin{figure}
\begin{center}
\epsfig{file=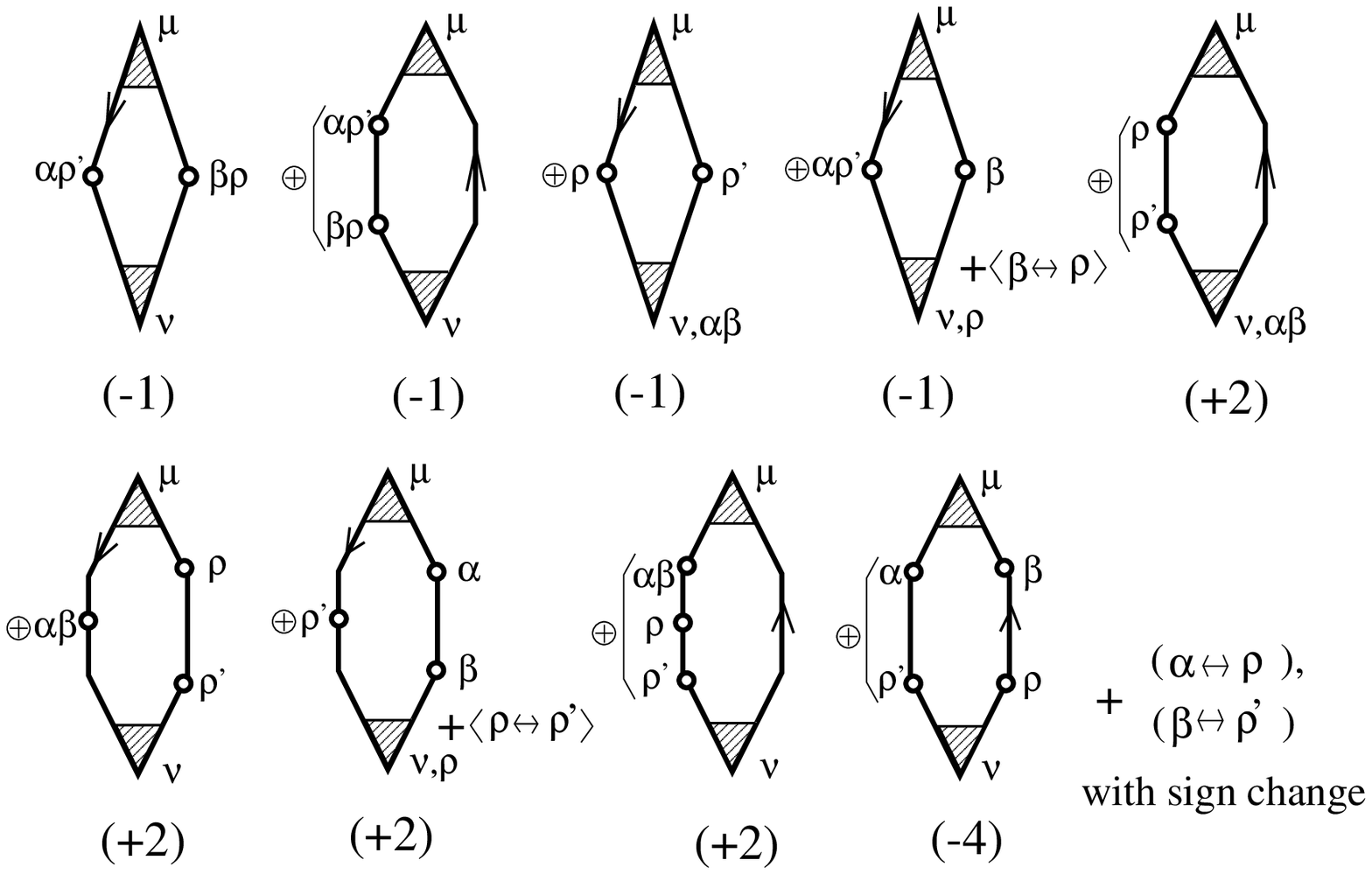,width=8cm}
\end{center}
\caption{
The rest of the diagrams which come from
$(\d^2/\d q_\rho \d q'_{\rho'}) \Phi^{(2)}(2\q+2\q',\w_\l)$.
After the analytic continuation of these terms,
we get (r-1$\sim$4) (s-1,2) and (t-1$\sim$6) 
in Fig.\ref{fig:DG-noVC1}.
}
\label{fig:DG-rst-append}
\end{figure}

In the next stage, 
we study (u-2), (u-3), (v-2) and (v-3).
As for (u-2) and (v-2), they
are decomposed into Fig.\ref{fig:DG-u2-append2}
by using the Ward identity, Eq.(\ref{eqn:Ward2}).
Moreover,
$\L_{\mu,\a(\b)}(\k)= \d_{\a(\b)} \L_\mu(\k)$ 
is expressed diagrammatically in Fig.\ref{fig:DG-withVC-add2}.
Thus, each (u-2) and (v-2) is decomposed into 9 terms.
In the same way, each (u-3) and (v-3) is given by 4 terms.
On the other hand, 
Fig. \ref{fig:DG-u2-append}
show the part of the terms coming from 
$(\d^2/\d q_\rho \d q'_{\rho'}) \Phi^{(2)}(2\q+2\q',\w)$.
Then,
we can also see that Fig.\ref{fig:DG-u2-append},
together with the diagrams with the notations
'$\langle\mbox{u-2}\rangle$', '$\langle\mbox{u-3}\rangle$',
'$\langle\mbox{v-2}\rangle$' and '$\langle\mbox{v-3}\rangle$'
in Fig.\ref{fig:DG-withVC-exp1},
give (u-2), (u-3), (v-2) and (v-3) in 
Fig.\ref{fig:DG-withVC} precisely.
Then, we have to take the symmetry 
$(\a,\rho)\leftrightarrow(\b,\rho')$ into account carefully.

In the same way,
we can write down all the rest of the terms in
$(\d^2/\d q_\rho \d q'_{\rho'}) \Phi^{(2)}(2\q+2\q',\w)$
which are of order $\g^{-3}$.
It is rather straightforward to see that 
they are collected into the 9 terms in
Fig.\ref{fig:DG-rst-append} precisely.
We will not show the detail to avoid redundancy.
Next, we perform the analytic continuation 
of Fig.\ref{fig:DG-rst-append}
to obtain the expression for the MC.
By dropping several terms which turn out to be zero,
we get the results
(r-1$\sim$4) (s-1,2) and (t-1$\sim$6) 
in Fig.\ref{fig:DG-noVC1} which is of order $\g^{-3}$.

In conclusion, $C_{\mu\nu}^{\a\rho;\b\rho'} = 
 (\d^2/\d q_\rho \d q'_{\rho'}) \Phi^{(2)}(2\q+2q',\w+\i0)/8$
is exactly given by Figs.\ref{fig:DG-noVC1} and \ref{fig:DG-withVC}
as for the most singular terms with respect to $\gamma^{-1}$.
Because the MC is given by
${\mit\Delta}\s_{xx}= C_{xx}^{xy;xy}$,
the derived expression for the MC in the present work,
eq.(\ref{eqn:MC-final}) or eq.(\ref{eqn:MC-total-lowT}),
is exact up to $O(\g^{-3})$.

\section{Comment on cond-mat/0006028}

In recent works, we find the following relations
based on the conserving approximation,
${\Delta}\s_{xy} \propto \xi^2_{\rm AF}\gamma^{-2}$ 
 \cite{Kontani} and
${\Delta}\s_{xx} \propto \xi^4_{\rm AF}\gamma^{-3}$
 \cite{Future},
where $\xi_{\rm AF}$ is the AF correlation length,
and $\gamma={\rm Im}\Sigma_k(-\i\delta)$.
They are the most divergent terms with respect to $\gamma^{-1}$.
However,
in a recent e-preprint [O. Narikiyo, cond-mat/0006028],
the author strongly claimed that 
the above results are inadequate for 
explaining the non-Fermi liquid behaviors 
in high-$T_{\rm c}$ cuprates.
His claims will be summarized as follows: \ 
(i) The conductivity at finite temperatures
given by the conserving approximation remains finite
even if no Umklapp processes exist,
which is a serious unphysical result. \ 
(ii) In the conserving approximation,
the Maki-Thompson process violates the Fermi liquid behavior
$\rho\propto T^2$. \ \
(iii) The contribution for ${\mit\Delta}\s_{xy}$
from the momentum derivative of $\gamma_\k$ 
is overlooked in Ref.\cite{Kontani}
although it is necessary. \\
In this appendix,
we show that the statements (i)-(iii) are inappropriate,
so the conserving approximation gives reliable results
for high-$T_{\rm c}$ cuprates as follows.

First, we show that the statement (i),
which is the main claim of the author,
is just false.
Actually, we can show that
the conserving approximation based on the FLEX method produces
{\it infinite conductivity}
in the spherical model without Umklapp process:
We can prove it straightforwadly by following the
Yamada and Yosida's transport theory given by Ref.\cite{Yamada},
by taking account of both the Maki-Thompson (MT) 
and the Aslamazov-Larkin (AL) processes.
On the other hand,
only the MT-process becomes crucial if the AF fluctuations are strong
like in high-$T_{\rm c}$ cuprates,
as shown in Ref.\cite{Kontani}.)
In this case, 
the Bethe-Salpeter eq. (6$\cdot$14) of Ref.\cite{Yamada} is also
satisfied by replacing eq.(6$\cdot$14) with 
\begin{eqnarray}
{\mit\Delta}_0(k,k';k'+q,k-q)
= \pi\rho_{k-q}(0)\rho_{k'+q}(0)\rho_{k'}(0) [(\pi T)^2+\e^2] W_q(0),
\end{eqnarray}
where $W_q(0)$ is given in eq.(A6) of Ref.\cite{Kontani}.
Moreover, 
${\rm Im}\Sigma_\k(-\i\delta)= 
 \frac12\sum_{k'q} {\mit\Delta}_0(k,k';k'+q,k-q)$
in three dimension.
As a result, by following the discussion in Ref.\cite{Yamada},
we see that the FLEX approximation can produce the fact that
$\s_{xx}=\infty$ even for $T>0$ unless the Umklapp processes exist,
as a consequence of the momentum conservation law.

It is a commonsense that the conserving laws are quite important 
for the study of transport phenomena; 
see Ref.\cite{Baym}.
We found that it is also the case in high-$T_{\rm c}$ cuprates.


The statement (ii) is inappropriate:
The MT process does not prevent the 
low temperature Fermi liquid behavior $\rho\propto T^2$
because the MT process does not give 
a singular temperature dependence of ${\vec J}_\k$ 
at the cold spot.
In fact, according to the analysis in Ref.\cite{Kontani},
${\vec J}_k \simge {\vec v}_k/2$
at the cold spot due to the MT process
is the presence of the strong AF fluctuations.
Secondly, Im$\Sigma_k(-\i\delta)\propto T^2$ in 3D
is ensured in the FLEX approximation at sufficiently 
low temperatures unless its ground state has a AF order,
because the FLEX is classified as the conserving approximation.
As a result, the MT process in the conserving approximation 
does not violate the relation $\rho\propto T^2$ 
in the Fermi liquid regime.
(In the numerical results of Ref.\cite{Kontani},
the lowest temperature is too high to see
the behavior $\rho\propto T^2$.)

On the other hand, the Hall coefficient is proportional to 
the coefficient $(\d\theta_J/\d k_{\parallel})$ at the cold spot.
In Refs. \cite{Kontani,Kanki}, we find that
the MT process produces a {\it seemingly} non-Fermi liquid
behavior of $R_{\rm H}$ at higher temperatures 
where $\xi_{\rm AF}$ is strongly temperature dependent.
On the other hand,
$R_{\rm H}$ should be nearly constant in the Fermi liquid regime
where $\xi_{\rm AF}$ becomes nearly constant.

The statement (iii) is false;
actually, ${\mit\Delta}\s_{xy}$
within the relaxation time approximation
is given by
\begin{eqnarray}
{\mit\Delta}\s_{xy}
 \propto \oint_{\rm FS} dk_\parallel
 \left( {\vec l}_\k \times \frac{\d}{\d k_\parallel}{\vec l}_\k \right),
 \label{eqn:ddd}
\end{eqnarray}
where ${\vec l}_\k = {\vec v}_\k\gamma_\k^{-1}$.
If we take the vertex corrections into account,
${\vec v}_k$ is replaced with ${\vec J}_k$ in eq.(\ref{eqn:ddd}).
Thus, ${\mit\Delta}\s_{xy}$ seems to contain
the momentum derivative of $\gamma_\k$.
However, eq.(\ref{eqn:ddd}) (${\vec l}_\k = {\vec J}_\k\gamma_\k^{-1}$)
is rewritten as
\begin{eqnarray}
{\mit\Delta}\s_{xy}
 \propto \oint_{\rm FS} dk_\parallel
 |{\vec l}_k|^2 \left( \frac{\d \theta_J(k)}{\d k_\parallel} \right),
  \label{eqn:DD}
\end{eqnarray}
where $\theta_J(k)= \tan^{-1}(J_{x}/J_{y})_k 
= \tan^{-1}(l_{x}/l_{y})_k.$
(see Ref.\cite{Kontani}.)
Thus, the momentum derivative of $\gamma_k$ does not apprear
in eq. (\ref{eqn:DD}).
As a result, the calculations of $R_{\rm H}$ in Refs. \cite{Kontani,Kanki} 
based on eq. (\ref{eqn:DD}) are correct.

In conclusion,
our results of the Hall effect and the magnetorestance
based on the conserving approximation 
are reliable and consistent with experiments.
We find that only the MT-type process
plays an important role in the transport phenomema 
in nearly AF Fermi liquid.
We note that some vertex corrections 
are dropped in eq.(3) in Ref.[26] of cond-mat/0006028,
which is for quasiprticle contribution to ${\mit\Delta}\s_{xx}$.
The correct expression for the MC of order $O(\g_k^{-3})$ 
is given by eq.(\ref{eqn:MC-final}) in this article,
or by eq.(\ref{eqn:MC-total-lowT}) in the case of $\g_k^\ast \ll T$.


\end{document}